\documentclass[pra,notitlepage,nofootinbib,superscriptaddress,longbibliography,twocolumn]{revtex4-1}

\usepackage{color}
\usepackage{amssymb}
\usepackage{amsmath}
\usepackage{subfig}
\usepackage{graphicx}
\usepackage[latin1]{inputenc}
\definecolor{myblue}{rgb}{0.1,0.15,0.6}
\usepackage[colorlinks=true, citecolor=myblue, linkcolor= myblue, urlcolor=myblue,bookmarks=true]{hyperref}

\def \bs{\boldsymbol}
\def \csq{cm$^{ \text{-} 2}$}
\def \Hel{H_{\mathrm{e}  \text{--}  \mathrm{l}}}
\def \Hee{H_{\mathrm{e}  \text{--}  \mathrm{e}}}

\newcommand{\expect}[1]{\langle#1\rangle}
\newcommand{\mypar}{{\mkern3mu\vphantom{\perp}\vrule depth 0pt\mkern2mu\vrule depth 0pt\mkern3mu}}

\begin{document}

\title{Generalized Wigner crystallization in moir\'e materials}
\author{Bikash Padhi}
\affiliation{Department of Physics and Institute for Condensed Matter Theory, University of Illinois at Urbana-Champaign, 1110 W. Green Street, Urbana, IL 61801, USA.}
\author{R. Chitra}
\affiliation{ Institute for Theoretical Physics, ETH Z\"urich, Wolfgang-Pauli-Stra{\ss}e 27, 8093 Zürich, Switzerland.}
\author{Philip W. Phillips}
\affiliation{Department of Physics and Institute for Condensed Matter Theory, University of Illinois at Urbana-Champaign, 1110 W. Green Street, Urbana, IL 61801, USA.}

\begin{abstract}
Recent experiments on the twisted transition metal dichalcogenide (TMD) material, $\rm WSe_2/WS_2$, have observed insulating states at fractional occupancy of the moir\'e bands. Such states were conceived as generalized Wigner crystals (GWCs). 
In this article, we investigate the problem of Wigner crystallization in the presence of an underlying (moir\'e) lattice. 
Based on the best estimates of the system parameters, we find a variety of homobilayer and heterobilayer TMDs to be excellent candidates for realizing GWCs. In particular, our analysis based on $r_{s}$ indicates that $\rm MoSe_{2}$ (among the homobilayers) and  $\rm MoSe_2/WSe_2$ or $\rm MoS_2/ WS_2$ (among the heterobilayers) are the best candidates for realizing GWCs. We also establish that due to larger effective mass of the valence bands, in general, hole-crystals are easier to realize that electron-crystals as seen experimentally.  For completeness, we show that satisfying the Mott criterion $n_{\rm Mott}^{1/2} a_{\ast} = 1$ requires densities  nearly three orders of magnitude larger than the  maximal density for  GWC formation. This indicates that for the typical density of operation, HoM or HeM systems are far from the Mott insulating regime. These crystals realized on a moir\'e lattice, unlike the conventional Wigner crystals, are incompressible due the gap arising from pinning with the lattice.
Finally, we capture this many-body gap by variationally renormalizing the dispersion of the vibration modes.
We show these low-energy modes, arising from coupling of the WC with the moir\'e lattice, can be effectively modeled as a Sine-Gordon theory of fluctuations. 
\end{abstract}

\maketitle
\let\originalnewpage\newpage \let\newpage\relax \let\newpage\originalnewpage


\section{Introduction}

A strongly interacting dilute gas of electrons  minimizes  its energy by spontaneously breaking translation invariance to  form a Wigner crystal (WC)~\cite{Wigner34}.   Though this physics is a simple  and intuitive manifestation of a strongly interacting many-body phase,  experimental realizations of  quantum Wigner crystals have been far and few between. Thus far, they have been seen in a two dimensional electron gas (2DEG) realized in semiconducting heterostructures~\cite{grimesadams} and liquid helium~\cite{MonarkhaRev}. Recently,  moir\'e materials, synthetic materials constituted from stacked monolayers with a mismatch in lattice size or orientation, have emerged as a highly tunable and experimentally accessible platform to study the physics of strong electronic correlations as well as topology \cite{Cao18Mott, Cao18SC,YoungDean, EfetovSC, PasupathySTM, Caltech19, CascadePrinceton, CascadeMIT, Efetov2Interplay, ReganWSeWS, xu2020abundance, jin2020stripe, ZurichSpacer, Allan2020simulation, TMDFengcheng, BSmoireTMD}.

\begin{figure}
\subfloat[]{\includegraphics[scale = 0.125]{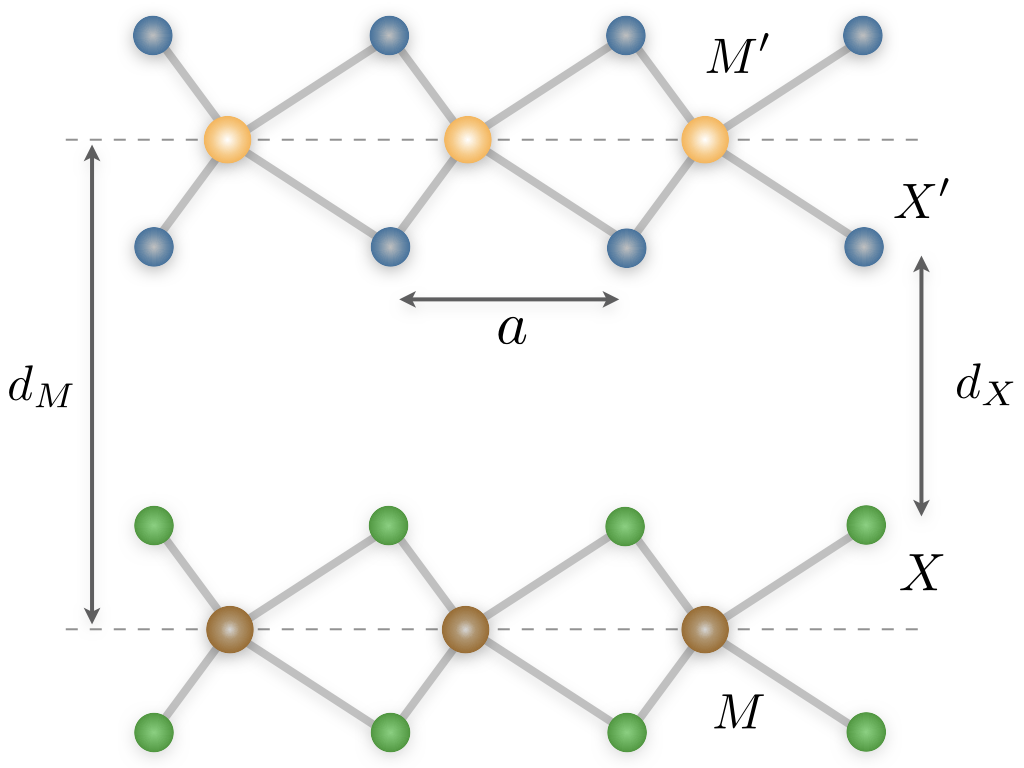}
\label{fig:sideview}}
\subfloat[]{\includegraphics[scale = 0.125]{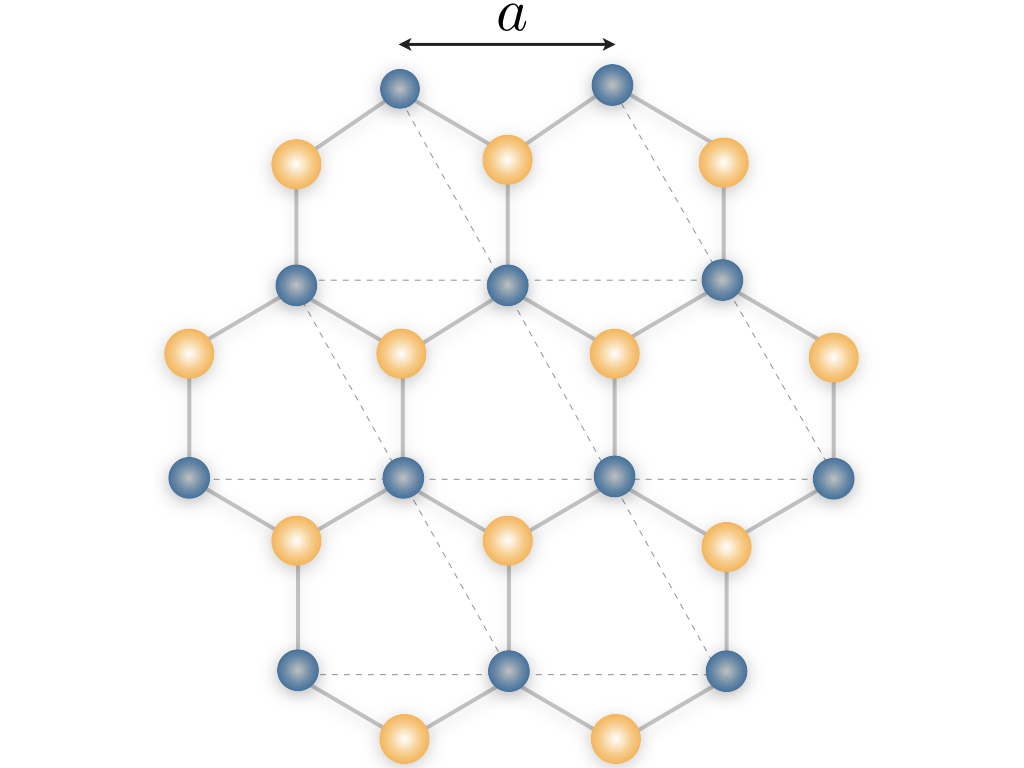}
\label{fig:topview}}
\caption{
Schematic of AA stacked TMD bilayer:
(a) the side-view shows the $MX_{2}$ layer, with trigonal prismatic (H) coordination, stacked on top of the $M'X'_{2}$ layer. The large (yellow or brown) balls represent the metal ions, $M$ or $M'$, and the small (blue or green) balls represent the chalcogens, $X$ or $X'$. The distance between the metal ions and the chalcogens are, respectively, denoted by $d_{M}$ and $d_{X}$.  
(b) The top-view is a honeycomb lattice of lattice of lattice constant $a$.
} 
\label{fig:TMDschemeatic}
\end{figure}

In particular, homobilayer moir\'e (HoM) materials or heterobilayer moir\'e (HeM) materials based on transition metal dichalcogenides (TMD), see Fig.~\ref{fig:TMDschemeatic}, have emerged as prime candidates for realizing WCs~\cite{ReganWSeWS, xu2020abundance, jin2020stripe}.  This can be largely attributed to the fact that the low energy moir\'e electrons in TMDs often reside in extremely narrow (quasi-flat) bands~\cite{MCDMoire, BSmoireTMD, TMDFengcheng, wang2019magic} or have very large effective masses, even compared to the traditional 2DEG systems~\cite{MonarkhaRev}. This makes them highly susceptible to charge localization. Such factors, coupled with the high controllability of TMDs for studying correlated phenomena~\cite{Allan2020simulation, TMDFengcheng, BSmoireTMD}, make them great candidates for studying Wigner crystallization. Given the plethora of TMDs, a primary goal in this article is to  explore material characteristics--lattice constant ($a$), dielectric constant ($\epsilon$), effective mass ($m_{\ast}$)--to characterize the ideal candidates for hosting a WC.

\begin{figure}
\includegraphics[scale = 0.20]{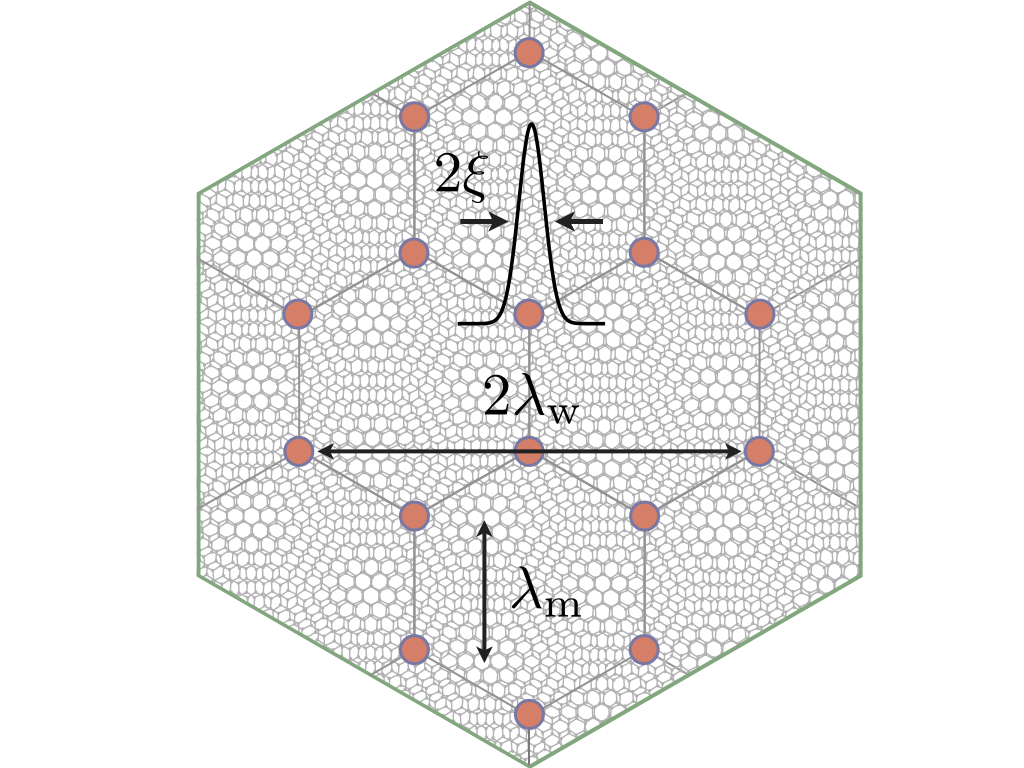}
\caption{
A cartoon rendition of a WC and the relevant length scales: A Wigner lattice is realized on a moir\'e superlattice (gray background) at filling fraction $2/3$. The distance between two nearest dark (or bright) spots is the moir\'e periodicity, $\lambda_{\mathrm{m}}$. The distance between the two nearest localized particles (red dots) is the Wigner lattice periodicity, $\lambda_{\mathrm{w}}$. The bell-shaped curve, representing the wavefunction of a localized particle, has a width of $2\xi$. Our discussion in this paper is confined to a crystal where the moir\'e electrons are highly localized, $\xi \ll \lambda_{\rm w}$.
} 
\label{fig:moireWC}
\end{figure}

Typically, a pure WC formed in  a 2DEG  slides  when subjected to a nonzero electric field due to the lack of a momentum relaxation mechanism.  A key signature of such a WC is its negative compressibility~\cite{Eisenstein92, LuLi11, SkinnerCapacitance, bello1981density, ChitraPRB01}. Disorder, however, pins the WC  and renders it incompressible as a result of the activation or pinning gap. 
A WC realized in moir\'e materials~\cite{OurPaper1,OurPaper2} is however,  ineluctably influenced by the underlying moir\'e lattice, which provides a uniform periodic background potential as illustrated in Fig.~\ref{fig:moireWC}.  This provides a pinning mechanism distinct from that induced by disorder  which will strongly influence  its  properties.  Such a crystal is often referred to as  a `generalized Wigner crystal' (GWC)~\cite{HubbardGWC}, see Fig.~\ref{fig:moireWC}.
Although  disorder-pinned-WCs have been studied widely~\cite{grimesadams,MonarkhaRev,MWRChen03, AshooriTunnelling17, hatke2015microwave, Monceau12, Delacretaz19}, an in-depth study of GWCs is still lacking.

In light of the recent experiments in TMD platforms exploring  the physics of strong correlations \cite{ReganWSeWS, xu2020abundance, jin2020stripe, ZurichSpacer, shimazaki2020optical},  a study of the properties of  the GWC is timely as it helps distinguish  the GWC from other density ordered gapped states that a lattice system may host alongside a GWC~\cite{ImadaPRL, jin2020stripe, slagle2020charge, shimazaki2020optical, pan2020quantum}. Insulating states observed in $\rm WSe_{2}/WS_{2}$ are at fractional fillings, $\nu = 1/3, 2/3$,~\cite{ReganWSeWS} and those in the twisted bilayer of graphene (TBLG) are at integer fillings~\cite{Cao18Mott,Cao18SC,YoungDean,EfetovSC,PasupathySTM,Caltech19,CascadePrinceton,CascadeMIT,Efetov2Interplay}. Simple observables like compressibility (or capacitance) are often misguiding and insufficient~\cite{Tomarken2019, GabiWCMott-Nature,Eisenstein94, QH-WCPuddle14}   to discern between  a   pinned WC  and a Mott state as these  exhibit  similar capacitive signatures~\cite{Eisenstein94, QH-WCPuddle14}. However,  in  the presence of the moir\'e lattice,  Mott states must preserve the underlying (moir\'e) lattice symmetry, and can only be observed at fillings for which a placement of the electrons preserves the underlying symmetry of the moir\'e lattice. While this is difficult for integer fillings exceeding unity~\cite{OurPaper1,OurPaper2}, it is impossible  at fractional filling observed in  $\rm WSe_{2}/WS_{2}$.  Consequently, the nature of the insulating states at integer fillings remains ambiguous. In  TBLG,  our earlier works~\cite{OurPaper1,OurPaper2},  based on  the Mott criterion, precluded  the interpretation of  the observed insulating states at integer  fillings as Mott states. Wigner crystallization~\cite{OurPaper1,OurPaper2} was envisaged  to be more favorable than Mott insulation at low charge densities in TBLG.

In this paper, we  explore from a materials perspective the viability of  both homo- and hetero-bilayer TMDs  for realizing WCs . Additionally,  we  study  the impact of the moir\'e lattice  on collective excitations~\cite{grimesadams,MonarkhaRev,MWRChen03} of GWCs and present estimates for the gap in the deep crystalline limit which can be directly accessed in transport experiments.   Our results are directly of relevance to a slew of recent experiments in these systems exploring the physics of the GWC~\cite{ReganWSeWS, ZurichSpacer}. We organize this article as follows. In Sec.~\ref{sec:Candidacy}, we analyze the material parameters of various HoM and HeM systems and assess their candidacy  for crystal formation  using several criteria. We identify a wide range of TMD materials that can support GWC phases and establish, broadly speaking, HeM to be better candidates than HoM for this purpose. 
In Sec.~\ref{sec:ElasticHam}, in the elastic limit~\cite{FluxLattice95, FluxLattice96, ChitraPRB01}, we obtain an effective Hamiltonian that describes harmonic fluctuations in a GWC pinned to a moir\'e lattice. We then move to obtaining the self-consistent equations for the pinning gaps corresponding to a GWC in Sec.~\ref{sec:GVM}. Finally, we conclude by connecting our results to the recent experiments in Sec.~\ref{sec:Conclude}.  Technical details are relegated to various appendices.

\section{TMD Candidacy For Wigner Crystallization}
\label{sec:Candidacy}

In this section, we discuss the key criteria for assessing the candidacy of various TMD bilayers, both HoM and HeM for Wigner crystallization. 
Generally, a material with low carrier density and a high degree of correlation can be susceptible to forming a WC. A natural way to measure correlation is to compare the strength of electronic interaction ($U$) with the kinetic energy ($W$) of the  relevant charge carriers.  Assuming the mean separation between the moir\'e particles to be of the order of the moir\'e periodicity, $\lambda_{\mathrm m}$, we set the scale of the Coulomb  repulsion  to $U= {e^2}/{\epsilon \lambda_{m}}$, where, $e$ is electronic charge and $\epsilon$ is the dielectric constant. In principle, one can also use a more realistic interaction potential for TMDs that can account for the encapsulating environment (such as the $\rm hBN/ SiO_{2}$ surroundings)~\cite{Rubio-RKpotential, Drummond-Falko, ScharfScreening}. However, at long distances, such a potential distills to a Coulomb-type potential~\cite{CoulombType}. Therefore, our assumption remains useful for discussing the low energy physics of TMDs. Another simplifying assumption we make is to ignore the full details of the TMD bandstructure~\cite{BSmoireTMD, TMDFengcheng}. We simply set $W = {\hbar^2 k^2}/{2 m^e_\ast} $ with $k \simeq \pi/\lambda_{\mathrm m}$. $m^e_\ast$ ($m^h_\ast$) is the effective mass of the electrons (holes) in the conduction (valence) band. We will later see that  typically, $W \sim \mathcal O(1 \, \text{meV})$ and $U \sim \mathcal O(10 \,\mathrm{meV})$.

Here, we reiterate that the important (in-plane) length scales in the problem, as shown in Figs.~\ref{fig:TMDschemeatic} and \ref{fig:moireWC}, are -- the monolayer lattice constant ($a$), the moir\'e periodicity ($\lambda_{\rm m}$), the Wigner lattice periodicity ($\lambda_{\rm w}$), and the localization length of the moir\'e particles ($\xi$).  $a$ is the smallest scale and can be
neglected  in a low energy theory. $\lambda_{\rm m}$ is a geometric scale which is fixed for a given TMD device. Unlike these two lengths, $\xi$ and $\lambda_{\rm w}$ are dynamically generated. By working in the deep crystalline limit where $\xi \ll \lambda_{\rm m}, \lambda_{\rm w}$ we can drop $\xi$. Thus, the most important scale in our problem is $\lambda_{\rm w}$, and its interplay with $\lambda_{\rm m}$.  
Since $\lambda_{\rm w}= 1/\sqrt{\pi n_{e}}$ is a function of  electronic density $n_e$ (or hole density $n_{h}$), it allows us to study crystallization as a function of doping levels.  Using this, the kinetic $(W)$ and potential $(U)$ energies can be recast as $W^{-1}=2 m \pi n_e$ and $U^{-1}= \epsilon \sqrt{\pi n_e} $. The dimensionless ratio of these two parameters, also known as $r_{s}$, provides crucial insight into nature of a correlated state~\cite{DavidTanatar}. Ignoring the effect of the moir\'e potential on the energies,  we obtain
\begin{align} 
\label{eq:rs}
r_s = \frac{\rm g}{a_0 m_{0}} \frac{m^{e}_\ast}{\epsilon} \lambda_{\rm w} \quad , \quad
\lambda_{\rm w} =  \frac{1}{\sqrt{\pi n_e}} ,
\end{align}%
where $a_0 = \hbar^{2}/m_{0} e^{2} = 0.529 \,\rm{\AA}$ is the Bohr radius with $m_0$ as the bare electron mass. And, $\mathrm{g}=2$ is a valley degeneracy factor for TMDs. This valley degree of freedom can significantly alter the correlation properties and the threshold for Wigner crystallization. In a 2-valley 2DEG the crystallization threshold drops to $r_{s} = 29.5$~\cite{2v2DEG-2017} from $r_{s} = 37$ in a 1-valley system~\cite{DavidTanatar, rscrit31}. The further $r_s$ exceeds this threshold value, the easier it is to form a WC. For further discussion on a more fine-tuned definition of $r_{s}$, see~\cite{OurPaper2}. Note however, due to the availability of a set of potential minima facilitated by the underlying moir\'e lattice,  the threshold value for GWCs could be lower than $r_s=29.5$.  

Clearly, Eq.~\eqref{eq:rs} shows that the material parameters that favor Wigner crystallization (or enhance $r_{s}$) are a high effective mass, reduced screening or a small dielectric constant and low carrier density.  Firstly,  low energy carriers in TMDs or twisted bilayers of TMDs are particularly heavy.  Secondly, though the dielectric constant of a material is  fixed,  it can be  altered by introducing a spacer layer~\cite{ZurichSpacer}, such as a hexagonal boron nitride (hBN) monolayer.  Screening can then be reduced by a judicious choice of  spacer material , thereby favoring Wigner crystallization.

Evidently however, the moir\'e scale dependence of $r_{s}$ is not manifest in Eq.~\eqref{eq:rs}. This can be naturally restored by measuring the carrier density through the filling fraction of a moir\'e unit supercell. This can be understood as follows. The area of a (hexagonal) moir\'e unit supercell is given by $A_{s} = \sqrt{3} \lambda_{\mathrm{m}}^{2}/2$. If the full occupancy of the relevant low energy band is $N_{0}$, usually determined by the discrete symmetries of the system, then the supercell density is given by $n_{s} = N_{0}/A_{s} \sim 10^{11 \text{-} 12}$ cm$^{ \text{-} 2}$.   A state consisting of $N$ electrons in this band is observed at a filling fraction of $N/N_{0} \equiv \nu$, or at a density $n_{e} = \nu n_{s} $. Inserting this in Eq.~\eqref{eq:rs}, we observe that, for a given material, there exists a critical density, $n_{e}^{\rm max}$, or a maximal filling fraction, $\nu_{\rm max}$, above which a GWC cannot exist. Correspondingly, since $r_{s} \propto \lambda_{m}$ [replacing $\lambda_{\rm w}$ with $\lambda_{\rm m}$ in Eq.~\eqref{eq:rs}], there also exists a critical moir\'e length below which a material cannot host a GWC. It is worth noting here that the true advantage of moir\'e materials in realizing WC is this availability of large length scales that govern most of the physics.

Before proceeding further, we note the above discussions are pertinent for zero temperature WC (or quantum WC) only. As the temperature increases, one needs to confront the problem of crystal melting. Although an accurate estimation of this melting temperature can be a subtle issue~\cite{IllingMelting, KhrapakMelting, Shayegan2020}, for simplicity, we estimate it using the classical Lindemann criterion, $k_B T_{\mathrm L} \simeq 0.01 U$. Our discussions in this paper will be confined to the physics of a GWC at $T \ll T_{\mathrm L}$. In the subsections below, we will explicitly evaluate all the above mentioned parameters for several TMDs.

\subsection{Homobilayers}

\begin{table}[b!]
\caption{Wigner crystallization criteria for HoMs: The effective mass of the conduction band ($m_e^\ast$) is obtained using DFT-LDA in Ref.~\cite{MX2EffMass}. 
Monolayer and bilayer dielectric constants ($\epsilon_\perp, \epsilon_\parallel$ and $\epsilon_\perp^{(2)}, \epsilon_\parallel^{(2)}$) are adapted from Ref.~\cite{MX2dielectric}.
Experimental lattice constant ($a$) data and the distance between the TMD layers ($d_{X}$) are compiled in Ref.~\cite{MX2EffMass}.
For this table, we set the twist angle to $\theta = 1^{\circ}$.
Eq.~\eqref{eq:rs} reduces to $r_s =674 {m_e^{*}}/{m_{0} \epsilon}$ for $n_e = 10^{11}$\csq, which we also use as the unit for densities mentioned thereafter.
The critical density, or the closest rational filling fraction, $\nu_{\rm max} = n_{e}^{\rm max}/ n_{s} $, below which a HoM system can host GWC is obtained by setting $r_{s} = 29.5$ in Eq.~\eqref{eq:rs}. 
The Lindemann melting temperature ($T_{\rm L}$) of a GWC and the density ($n_{\rm  Mott}$) for which the Mott criterion is satisfied are obtained in the end.
We observe that due to the larger effective masses, the $\rm Mo$-based compounds are generally better suited to forming GWCs as compared to the $\rm W$-based compounds. In regard to $r_{s}$, or $U/W$, we conclude a twisted bilayer of $\rm MoSe_{2}$ to be the best candidate for Wigner crystallization.
}
\begin{ruledtabular}
\begin{tabular}{ c c c c c c c} 
{HoMs} & ${\rm MoS_2} $ & $ \rm MoSe_2$ & $\rm MoTe_2$ & $\rm WS_2$ & $\rm WSe_2$  & $\rm WTe_2$\\
\colrule
$m_\ast^{e}/m_{0}$ & $0.46$ & $0.56$ & $0.62$ & $0.26$ & $0.28$ & $0.26$ \\
$\epsilon_\perp (\epsilon_\parallel)$ 
& $4.8 (3.0)$ & $6.9 (3.8)$ & $8 (4.4)$ & $4.4 (2.9)$ & $4.5 (2.9)$ & $5.7 (3.3)$ \\
$\epsilon^{(2)}_\perp (\epsilon^{(2)}_\parallel)$ 
& $6.9 (4.4)$ & $7.9 (4.6)$ & $8.6 (5.5)$ & $6.1 (4.2)$ & $6.3 (4.3)$ & $8.4 (5.2)$ \\
$d_{X}$ [{\AA}] & $3.17$ & $3.33$ & $3.60$ & $3.14$ & $3.34$ & $3.60$ \\
$a$ [{\AA}] & $3.16$  & $3.29$  & $3.52$  & $3.15$  & $3.28$  & $3.50$  \\
$ \lambda_{\mathrm{m}} $  [nm] & $18.1$ & $18.8$ & $20.2$ & $18.0$ & $18.8$ & $20.0$ \\
$ U/W $ & $5.0$ & $5.8$ & $6.0$ & $3.1$ & $3.3$ & $2.6$  \\
$r_s \rvert_{10^{11} \text{cm}^{\text{-2}}} $ & $56.3$ & $62.6$ & $60.8$ & $34.7$ & $36.3$ & $26.5$ \\
$n_e^{\rm max}$ & $3.6$ & $4.5$ & $4.2$ & $1.4$ & $1.5$ & $0.8$ \\
$\nu_{\rm max}$ & $1.02$ & $1.38$ & $1.48$ & $0.40$ & $0.46$ & $0.28$ \\
$T_{\mathrm{L}} $ [K]  & $1.7$ & $1.5$ & $1.2$ & $1.8$ & $1.7$ & $1.3$ \\
$ n_{\rm Mott}10^{-3} $  & $2.5$ & $3.1$ & $2.9$ & $0.9$ & $1.0$ & $0.6$ \\
\end{tabular}
\end{ruledtabular}
\label{tab:Homo}
\end{table}

In a  HoM system, the top and the bottom layers  consist of the same TMD where each layer projects to a  2D honeycomb lattice (see Fig.~\ref{fig:topview}). This, therefore, is geometrically  equivalent to a twisted bilayer graphene system. The moir\'e periodicity in a HoM is thus given by~\cite{LopesPRL} $\lambda_{\mathrm m}(\theta) = \frac{a}{2\sin(\theta/2)} \simeq a/\theta$. Here, $\theta$ is the twist angle between the two TMD layers.  ($\epsilon_\mypar^{\phantom{2}}, \epsilon_\perp^{\phantom{2}}$) and  ($\epsilon_\mypar^{2}, \epsilon_\perp^{2}$) denote the 
 in-plane  and out of plane dielectric constants of a monolayer and a homobilayer TMD  respectively.
We identify the geometric mean of these two constants, $\epsilon^{(2)} = \sqrt{\epsilon^{(2)}_\perp \epsilon^{(2)}_\mypar}$, as the dielectric constant of the bilayer system~\cite{MoS2rs}. 

Using these parameters, we  summarize our results for crystallization criteria in different candidate HoMs in Table~\ref{tab:Homo}. For a typical twist angle $\theta=1^{\circ}$,  we  find that, $U/W > 1$ for all the homobilayers in Table~\ref{tab:Homo}, rendering them strongly interacting systems.  The corresponding $r_{s}$ computed using Eq.~\eqref{eq:rs} shows that all the HoMs in Table~\ref{tab:Homo} are susceptible to forming GWCs since they all have $r_{s}$ fairly above the crystallization threshold.  The critical density  for  crystallization is  found to be nearly the order of $n_{s}$. The filling fraction ($\nu_{\rm max}$) below which the GWCs can be observed are also evaluated along with it.  Based on the Lindemann criteria, our results predict that the   GWCs  should be stable in the  range of $1$K-$3$K. 
Our simple analysis  shows that  $\rm Mo$-based HoMs  are more viable than $\rm W$-based compounds for the realization of GWCs.

Finally, we evaluate the Mott criterion, , $n_e^{1/2}a_0^\ast \approx O(1)$, which a system needs to satisfy in order to host Mott insulating states~\cite{MottDavis}. The effective Bohr radius, $a_0^\ast = \hbar^{2}/m^{e}_{\ast} e_{\ast}^{2}$ and $e_{\ast} = e^{2}/\epsilon$.  Evalutating this for HoMs,  we find that  for experimentally relevant densities (that is near the fractional fillings of a moir\'e unit supercell) the Mott criterion is far from being met, $n_e^{1/2}a_0^\ast \sim \mathcal{O} (10^{-2}) \ll 1$.  Satisfying  the Mott criterion, $n_{\rm Mott}^{1/2} a_{\ast} = 1$ requires densities  nearly three orders of magnitude larger than the  maximal density for  GWC formation. This indicates that for the typical density of operation, HoM systems are far from the Mott insulating regime.



\begin{table*}[ht]
\caption{Wigner crystallization criteria for a nearly-aligned HeMs: 
Any experimental study of the pertaining heterostructure is referenced here.
The electron (hole) effective mass $m_{e} (m_{h})$ are adapted from Ref.~\cite{HeteroMass}.
The moir\'e length is evaluated following Eq.~\eqref{eq:MoireHeM}. Like before, we set the twist angle to $\theta = 0.5^{\circ}$ and the particle density to $n_{e} = 10^{12} \rm cm^{-2}$.
Asterisked values in the $r_{s}$ row indicate values crossing the crystallization threshold.
Note that the Lindemann temperature is the same for both electron and hole crystals since $U$, under our assumptions, simply depends on the geometry and not on the effective mass.
Since the hole bands have higher effective mass they display larger correlation than the electron bands. In particular, among the heterobilayers listed here, $\rm MoSe_2/WSe_2$ and $\rm MoS_2/ WS_2$ seem to be the most susceptible to forming an electronic and hole GWC, respectively.
 }
\begin{ruledtabular}
\begin{tabular}{ c c c c c c c c} 
HeMs & $\rm WSe_2/WS_2$ & $\rm MoSe_2/ MoS_2$ & $\rm MoTe_2/MoSe_2$ & $\rm MoSe_2/WS_2$ & $\rm MoSe_2/WSe_2$ & $\rm MoS_2/ WS_2$ & $\rm MoTe_{2}/WSe_{2}$  
\\ \colrule
Refs. & \cite{ReganWSeWS,FengWangWSeWS} & \cite{MSeMS} & \cite{MSeWS1, MSeWS2} & \cite{MoSe2/WSe2} & \cite{MSWS1,MSWS2} & \cite{MoS2/WS2} & \cite{MoTe2/WSe2} \\
$m_{e (h)}^{*}/m_{0}$ & $0.28\, (0.46)$ & $0.42 \, (0.71)$ & $0.46 \, (1.37)$ & $0.28 \, (0.71)$ & $0.54\, (0.44)$ & $0.46 \, (1.70)$ &  $0.30 \, (1.33)$ \\
$2/(\epsilon_{1}^{-1} + \epsilon_{2}^{-1} )$ & $2.9$ & $3.35$ & $4.08$ & $3.29$ & $3.29$ & $2.95$ & $3.5$ \\
$ \lambda_{\mathrm{m}} $  [nm]  & $8.1$ & $8.1$ & $5.3$ & $7.6$ & $35.6$ & $34.0$ & $5.1$ \\
$U/W \rvert_{e (h)}$ & 
$3.0 \, (4.9)$ & $3.9 \, (6.6)$ & $2.3 \, (6.9)$ & $2.5 \, (6.3)$ & $22.4 \, (18.2)$ & $20.3 \,  (75.1)$ & $1.7 \, (7.5)$ \\
$ r_s^{e (h)} \rvert_{10^{12} \text{cm}^{\text{-2}}} $ & $20.6 \, (33.8^{*})$ & $26.7 \, (45.2^{*})$ & $24.0 \, (71.6^{*})$ & $18.2 \, (46.0^{*})$ & $35.0^{*} \, (28.5)$ & $33.3^{*} \, (122.9^{*})$ & $18.3 \, (81.1^{*})$ \\
$T_{\mathrm{L}}$ [K] & $7.1$ & $6.1$ & $7.6$ & $6.7$ & $1.4$ & $1.7$ & $9.3$ \\
$n_{e (h)}^{\rm max} \, [10^{12}$ \csq] &
$0.5 \, (1.3)$ & $0.8 \,(2.3)$ & $0.7 \, (5.9)$ & $0.4 \, (2.4)$ & $1.4 \, (0.9)$ & $1.3 \, (17.4)$ & $0.4 \, (7.5)$ \\
$\nu_{\rm max}^{e (h)}$ &
$0.28 \, (0.73)$ & $0.46\, (1.31)$ & $0.17 \, (1.46)$ & $0.2 \, (1.19)$ & $15.56 \, (10.0)$ & $13.0 \, (174.0)$ & $0.09 \, (1.7)$ \\
\end{tabular}
\end{ruledtabular}
\label{tab:Hetero}
\end{table*}

\subsection{Heterobilayers}

In HeM materials, the top and bottom layers contain different TMDs.  We now explore the potential for GWCs in HeMs  in the manner done in the preceding section for homobilayers. Although the planar projection of each layer is a honeycomb lattice with different periodicities, a moir\'e pattern emerges even without introducing any twist angle (`near-aligned sample'). Twisting  alters the moir\'e periodicity; in particular, it reduces with increasing twist angle and often approaches the original lattice constant at  `large-twist-angles'. For example, in a HeM with a small difference in lattice constants  ~\cite{AnderiMoireconstant},
the moir\'e periodicity  is ~\cite{FengWangWSeWS, ruiztijerina2020theory} 
\begin{align}
\lambda_{m} \simeq \frac{a_{>}}{\sqrt{\delta_{a}^{2} + 4 \sin^{2}(\theta/2)}} \quad , \quad \delta_{a} = 1-\frac{a_{<}}{a_{>}} .
\label{eq:MoireHeM}
\end{align}
Here $a_{> (<)}$ is the largest (smallest) lattice constant among the two layers.  We see that $\lambda_{m}$ is strongly influenced by the twist angle for samples with small $\delta_a$.  As shown inTable~\ref{tab:Homo},  this is the case of  HeMs with differing metal ions [$\mathit{MX_2/ M'X_2} $]  which have $\delta_a \lesssim 0.1\%$. HeMs with differing chalcogens [${ \mathit{MX_2/ MX'_2}}$] tend to have large $\delta_a$, i.e. around $4\%$ and are less sensitive to small angle twists.   Motivated by  the experiment of Ref.~\cite{ReganWSeWS} which concern $\theta \lesssim 1^{\circ}$~\cite{FengWangWSeWS}, we  confine our discussion to nearly-aligned heterobilayers.

The effective dielectric constant of the HeM system  is obtained by treating the two layers as two dielectrics (or capacitors) in series, 
\begin{align}
\frac{d_1 + d_2}{\epsilon} = \frac{d_1}{\epsilon_1} + \frac{d_2}{\epsilon_2}  ,
\end{align}
where $\epsilon_{i}$ and $d_{i}$ are the dielectric constants and the thickness of the top and bottom layers, respectively.   We assume   $d_{1} = d_{2}$  and as  the two layers are different and stacked along the direction that is normal to the dielectric plane, we set $\epsilon_{i}$ to be the in-plane monolayer dielectric constants, $\epsilon_{\mypar, i}$.
For near-aligned samples with $\theta = 0.5^{\circ}$,  using Eq.~\eqref{eq:MoireHeM} and Eq.~\eqref{eq:rs} we evaluate  $U/W$ and  $r_{s}$ for different HeMs.  Our results are summarized in Table~\ref{tab:Hetero}.   We find generically  that hole carriers  have larger  $r_{s}$ due to their larger effective masses. Almost all the  HeMs  considered in Table~\ref{tab:Hetero} can Wigner crystallize for a hole density of $10^{12}$ \csq or less. However, except for a few, most of the electronic carriers do not crystallize. 

Since $\rm MoSe_2/WSe_2$ and $\rm MoS_2/ WS_2$ share the same chalcogens, they are quite sensitive to twist angle. For $\theta \sim 0^{\circ}$, the moir\'e length can be as large as a micrometer and it gradually reduces to about a deca-nanometer by $5^{\circ}$ of twisting. The correlation factor $U/W$, therefore, also reduces by nearly two orders of magnitude. For the remainder of the HoMs, though the above mentioned trend is still valid, however, quantitatively, no significant change is observed in the correlation factor since the moir\'e length scale remains largely insensitive to small changes in the twist angle. In particular, for $\rm WSe_2/WS_2$, we find that at filling fraction $\nu=1/3$,  $r_{s} = 44.0$ ($26.8$) for holes (electrons), and at $\nu = 2/3$, it is $31.1$ ($19.0$) for holes (electrons). This thus explains why Regan \textit{et al.}~\cite{ReganWSeWS} observe GWC states on the hole side but not on the electronic side. This is one of our key results as it bares directly on the experiments.

Lastly, we evaluate the critical density, or filling fraction, above which the heterostructure will be unable to host GWCs. In particular, for $\rm WSe_2/WS_2$ we observe that no hole-crystal can exist above a filling fraction of $0.73 \, (\approx 3/4)$. States at any filling fraction below this, even other than those at $1/3$ and $2/3$~\cite{slagle2020charge}, are perfectly allowed. Similarly, on the electron side, GWC can exist up to $\nu =  0.28 \, (\approx 1/4)$.

To summarize, based on the best estimates of the system parameters, we find a variety of homobilayer and heterobilayer TMDs to be excellent candidates for realizing WCs. In particular, our analysis based on $r_{s}$ indicates that $\rm MoSe_{2}$ (among the homobilayers) and  $\rm MoSe_2/WSe_2$ or $\rm MoS_2/ WS_2$ (among the heterobilayers) are the best candidates for realizing WCs. We also establish that due to larger effective masses of the valence bands,  hole-crystals in general,  are easier to realize than electron-crystals, an observation consistent with experiments.  In the remainder of the paper, we focus on the properties of a GWC.

\section{Effective Theory of GWC}
\label{sec:ElasticHam}

Understanding the collective excitations of a GWC  is critical in distinguishing them from the other density ordered states observed in the lattice system.  Here, we will focus on the vibrational modes of the GWC in absence of an external magnetic field. For analytical tractability we will confine our discussion to the limit when the GWC is deep in the crystalline regime. 
We represent the particle density of the system using a lattice of Gaussian wave-packets of size $2\xi$ (see Fig.~(\ref{fig:moireWC})), 
\begin{align}
\begin{split}
\rho(\bs x) &= \sum_i  | \psi \left( \bs x - \bs R_i \right) |^2
, \\
| \psi(\bs x) |^2 &= \frac{1}{2 \pi \xi} \exp \left( -|\bs x^2|/{4 \xi^2} \right) .
\end{split}
\end{align}
where 
$\bs R_i  = \bs R_i^0 + u_i(t) $,
describes fluctuations around the mean  lattice sites  $\bs R_i^0 $.
The GWC  we consider is far away from the phase boundary with the liquid phase so that we can treat the mean fluctuation in the position of the localized electrons, $\expect{\bs r^{2}} \sim \xi^{2}$, to be much smaller than the Wigner lattice periodicity, $\xi \ll \lambda_{\mathrm w}$, as in Fig.~\ref{fig:moireWC}. Since the field $\bs u_{i}(t)$ measures the fluctuation around the mean position of a particle, it is naturally $\mathcal{O}(\xi)$. For a GWC at $T=0$, $\xi$ (hence, $\bs u_{i}$) can be tuned by changing the density alone. A self-consistent solution of $\xi$ as a function of density is discussed in Ref.~\cite{ChitraEPJB05}. Finally, since $\xi$ increases with increasing temperature, we will restrict our discussion to low temperature, $T \ll T_{\mathrm L}$.

In this regime,  the above density functional  can be written in terms of  harmonics (see App.~\ref{app:HarmonicDensity} for a derivation)
\begin{align}
\rho(\bs x) \simeq \rho_0 \Big[ 1  -  \bs \nabla \cdot \bs u(\bs x) +  \sum_{l \neq 0}  e^{i \bs K_{l} \cdot \bs x} \rho_{l} (\bs x) \Big] .
\label{eq:expandDensity}
\end{align}
Here, $\rho_l (\bs x) =  e^{- i \bs K_{l} \cdot \bs u(\bs x) } $ and $\rho_0$ is the average density (over the entire sample). 
The second term  accounts for  long range density fluctuations over several $\lambda_{\mathrm w}$ and couples to  couples to the long-range (or $q \sim 0$) component of the Coulomb interaction.  
The remaining terms take care of the density fluctuations at a length scale comparable to or smaller than $\lambda_{\mathrm w}$ and hence can be referred to as un-smeared density.   The wave vectors $\bs K_{l} = \{ \pm l \bs \kappa_{n} \}$  denote the Brillouin zone (BZ) vectors of the undeformed GWC. Here, $l=1, 2, \cdots$ are simply `size multipliers' of the BZ. Formally, the $l=0$ term is nothing other than $\rho_{0}$ in Eq.~\eqref{eq:expandDensity}. The last term above also contains a summation over the index $n$ appearing through $K_{l}$. We perform this summation implicitly since it does not play any significant role in our analysis. 

The long wavelength theory describing the fluctuations of the crystal  is given by an elastic  Hamiltonian \begin{align}
H_{\mathrm{eff}} = \frac12 \sum_{\omega_{n}} \int \frac{d^2 \bs q}{ (2\pi)^2 } u_\alpha(\bs q, \omega_{n}) \, \Phi_{\alpha \beta} (\bs q, \omega_{n}) \, u_\beta(-\bs q, -\omega_{n}) ,
\end{align}%
where $\alpha, \beta = x,y$ are summed over, momenta $\{\bs q \}$ form the Fourier basis, and the kernel $\Phi_{\alpha \beta} (\bs q, \omega_{n})$ is the elastic matrix. Henceforth, we will express all the quantities after performing the frequency ($\omega_{n}$) summation. In case of a classical ($\omega_{n} = 0$) free theory, this matrix is  $\Phi_{\alpha \beta} = c q^2 \delta_{\alpha \beta}$, with the real space Hamiltonian $H_{\mathrm{eff}} = \frac c 2 \int d^2 \bs x \left[ \bs \nabla \cdot \bs u(\bs x) \right]^2$. Here $c$ is an elastic modulus.
The presence of  the moir\'e potential and the Coulomb interaction between the particles   generates the following  terms in the hamiltonian
\begin{align}
H_1 =  \Hel + \Hee,
\label{eq:HamilDensity} 
\end{align}
where electron-moir\'e lattice interaction and the electron-electron interaction terms, respectively, are
\begin{subequations}
\begin{gather}
\Hel = -  \int_{\bs x} \, V(\bs x) \rho(\bs x) ,
\label{eq:ManyBodyHel} \\ 
\Hee = \frac12 \int_{\bs x , \bs x'} \, U(\bs x - \bs x') [\rho(\bs x) -\rho_0][\rho(\bs x')-\rho_0] \, . 
\label{eq:ManyBodyHee}
\end{gather}
\end{subequations}
We will approximate the (triangular) moir\'e potential, $V(\bs x)$, by~\cite{BSmoireTMD, LiangMoireChem}
\begin{align}
V(\bs x) = 2 \tilde V \sum_{m=1}^3 \cos \left(\bs x \cdot \bs g_m +  \phi \right) ,
\label{eq:MoirePot}
\end{align}%
where $\tilde V \sim \mathcal O(10 \,\text{meV})$ sets the depth of the moir\'e potential and $\phi$ determines the shape of the potential. These two (intrinsic) parameters can be fixed for a given TMD using methods developed in Ref.~\cite{LiangMoireChem}. Lastly, the unit vectors of the moir\'e Brillouin zone (MBZ) are given by $\bs g_m = \frac{4 \pi}{\sqrt{3} \lambda_{\mathrm m} } \left( \cos \frac{2\pi m}{3}, \sin \frac{2 \pi m}{3} \right)$.

\begin{figure}
\includegraphics[scale = 0.18]{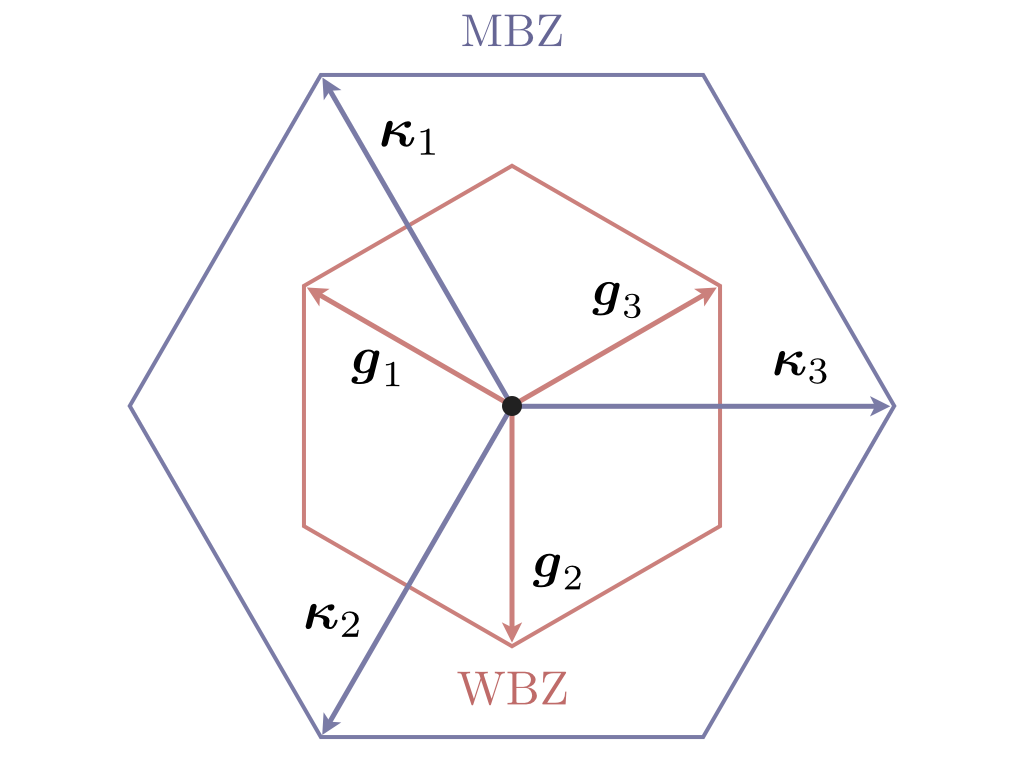}
\caption{
Schematic of a moir\'e BZ (MBZ, blue) and a Wigner BZ (WBZ, red). The BZ vectors are, $g_{n} = \frac{4\pi}{\sqrt{3} \lambda_{\rm m}} \left( \cos \frac{2\pi n}{3}, \sin \frac{2\pi n}{3} \right) e^{i \pi/6}$ and $\kappa_{n} = \frac{4\pi}{\sqrt{3} \lambda_{\rm w}} \left( \cos \frac{2\pi n}{3}, \sin \frac{2\pi n}{3} \right)$. In general, since $\lambda_{\rm w} > \lambda_{\rm m}$, the WBZ is smaller than the MBZ. For the particular case drawn above, $\lambda_{\rm w} = 3 \lambda_{\rm m}$. In other words, the third WBZ is the same as the first MBZ ($|\boldsymbol{\kappa}_n| = 3|\boldsymbol{g}_n| $).
} 
\label{fig:BZvecs}
\end{figure}%

\subsection{Interaction with the moir\'e potential}
\label{sub:Hel}

We now focus on the  moir\'e potential given by the first term term in Eq.~\eqref{eq:HamilDensity}.    In terms of a reciprocal vector of the MBZ,  $\bs G_{m} = \{ \pm m \bs g_{n} \}$, where $\bs g_{n}$ are the primitive MBZ vectors, see Fig.~\ref{fig:BZvecs}, the  periodic  moir\'e potential is~\cite{BSmoireTMD, FengWangWSeWS}
\begin{align}
V(\bs x) =  \sum_{m} V_m \, e^{i \bs G_{m} \cdot \bs x} .
\label{eq:GenericPot}
\end{align}%
As before, $m$ is a size multiplier for the principal MBZ and a summation over the index $n$ is made implicit. We assume the potential to be an even function in position space and set the $m=0$ mode to zero.  For the potential in Eq.~\eqref{eq:MoirePot}, we obtain $V(\bs G_{m}) = \tilde V e^{i \mathrm{sgn}(m) \phi}$. Substituting Eq.~\eqref{eq:GenericPot} in Eq.~\eqref{eq:ManyBodyHel}, we obtain the following moir\'e term
\begin{align}
\label{eq:EffectiveHel}
\Hel &= - \rho_0 \sum_{l,m}  \,  V_{m} \, \int d\bs x \,  e^{i (\bs K_l - \bs G_m ) \cdot \bs x } \, \rho_l(\bs x) .
\end{align}%

In writing the above expression, we have set the energy of the moir\'e lattice, $\sim \int_{\bs x} V(\bs x)$, to zero and neglected the gradient term in the density 
  as this term represents an external source term (linear in $\bs u$) and does not  contribute to the physics of the pinning gap. 
Note that  
the integrand here involves both the WBZ and the MBZ vectors. This term plays a critical role in imposing a certain set of commensuration constraints.
In general, a GWC need not conform to the lattice symmetries of a background (e.g., moir\'e) lattice. With changing density, one often anticipates the GWC to go through a large set of commensurate-incommensurate transitions, also known as the devil's staircase~\cite{BakDevil82, BakqFK80}, where the incommensurate structures may also have a completely different lattice symmetry~\cite{Rademaker}  and associated stability issues.  These states and the accompanying transitions  cannot be described by  the  elastic (linear harmonic) theory  developed here. 

In this paper,  we focus exclusively on the case where  the GWC and the background lattice share the same lattice symmetry, such as in the experiment of Regan \textit{et al.}~\cite{ReganWSeWS}.   As we will show below,  this leads to a geometrical  condition $r \bs G_1 = s \bs K_1$, where $(r, s)$ are co-primes and the subscript $1$ refers to the principal BZ vectors.  Finally, since usually $\lambda_\mathrm{w} \geq \lambda_{\mathrm{m}}$, hence $|\bs G_{ 1}| \geq |\bs K_{ 1}|$.  As a result, $r \leq s$. For instance, the WC observed in~\cite{ReganWSeWS} at $1/3$-filling, or a state  at $\nu = 1/3^{n}$ in general, simply has its BZ shrunk (without any rotation) by a factor of $2^{n}$. This state of affairs obtains because the GWC at $1/3^{n}$-filling has a unit cell  that is $2^{n}$ times larger than that of the moir\'e lattice. Therefore, for $\nu = 1/3^{n}$, $r=1$ and $s=2^{n}$. 

\subsection{Electronic interaction}

Using the underlying translation invariance, we write the interaction term in, Eq.~\eqref{eq:HamilDensity} as 
\begin{align} 
\label{eq:HeeExpand}
\Hee &= \frac{\rho_0^2}{2} \int_{\bs x, \bs x'} U(\bs x - \bs x') \left[\bs \nabla \cdot \bs u(\bs x) \right] \left[ \bs \nabla \cdot \bs u (\bs x') \right] +
 \nonumber  \\  & \frac{\rho_0^2}{2} \int_{\bs x, \bs x'} \sum_{l} U(\bs x - \bs x')  e^{i \bs K_l \cdot (\bs x - \bs x')} \rho_{l}(\bs x - \bs x')  .
\end{align}
Note that terms with $K_l \neq K_m$ have been discarded as they are  highly oscillatory.

We now switch from the cartesian basis $u_x, u_y$ to one  described by the longitudinal ($u_\mypar$) and transverse ($u_\perp$) components with respect to the momentum vectors ($\bs q$) 
\begin{align}
\label{eq:BasisRot}
u_\alpha(\bs q) = u_\mypar(\bs q) \hat{\bs q}_\alpha +  u_\perp(\bs q) \epsilon_{\alpha \beta} \hat{\bs q}_\beta ,
\end{align}%
where $\alpha, \beta = \{ x,y \}$, and $\epsilon_{\alpha \beta}$ is an antisymmetric tensor, $\epsilon_{xy} = 1 = -\epsilon_{yx}$.  Note that  $u_\mypar$  and  $u_\perp$ are  the bulk compression  and shear  modes respectively.  In this basis, the first term, $\Hee^{(1)}$, in Eq.~\eqref{eq:ManyBodyHee} becomes
\begin{align}
\Hee^{(1)} 
& = \frac{\rho_0^2}{2} \int_{\bs x, \bs x'} U(\bs x - \bs x') \left[\bs \nabla \cdot \bs u(\bs x) \right] \left[ \bs \nabla \cdot \bs u (\bs x') \right]  
\nonumber \\ 
& = \frac{ d_\mypar}{2 } \sum_{\bs q} \, q \, u_\mypar (\bs q) u_{\mypar} (-\bs q) \quad , \quad
\frac{\rho^2_0 e^2}{\epsilon} \equiv d_\mypar.
\label{eq:HeeLongDistance}
\end{align}
We see that 
the  $q=\vert {\bs q}\vert$ term results from the long-range (in 2D) nature of the interaction, $U(\bs q) \sim 1/q$. Had we considered a shorter-range interaction of the form $U(\bs q) \sim 1/q^\gamma$, the proportionality above would have been modified to $q^{2-\gamma}$.
The transverse modes do not change  the local density and remain unaffected by the Coulomb interaction.  Typically,  long wavelength electrostatic fluctuations, namely the plasma modes, are always longitudinal in the absence of a magnetic field (since $\bs q \times \bs E = 0$, where $\bs E$ is an electric field).

In the elastic limit $|\bs u(\bs x) - \bs u(\bs x')| \ll \lambda_{\mathrm{w}}$, we  Taylor expand the second term, $\Hee^{(2)}$, in Eq.~\eqref{eq:HeeExpand}.
 The  first-order  term vanishes because the undeformed GWC has   an energy minimum at $\bs u=0$  and the second-order term gives the correction
\begin{align}
\Hee^{(2)} 
\simeq \frac{\rho_0^2}{ 2} & \sum_{l}  \int_{\bs x, \bs x'} V(\bs x - \bs x') e^{i \bs K_{l} \cdot (\bs x - \bs x')}  K_{l,\alpha} K_{l,\beta}  \nonumber \\ 
& \times \left[ u_\alpha(\bs x) - u_\alpha(\bs x') \right] \left[ u_\beta(\bs x) - u_\beta(\bs x') \right] \, .
\label{eq:Hee2Taylor}
\end{align}
Here, $\bs K_{l,\alpha}$ denote the $\alpha = x, y$ components of $\bs K_{l}$. Henceforth, unless mentioned, we will set $\rho_{0} = 1$. 

\begin{figure}[t]
\includegraphics[scale = 0.3]{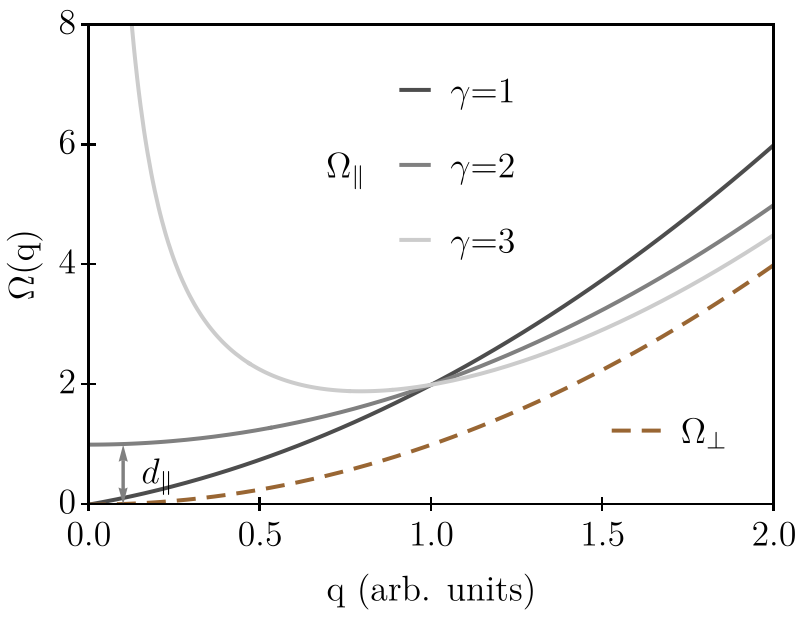}
\caption{
Dispersion of the longitudinal (solid) and transverse (dashed) modes of 2D WCs. Here, we have set $c_{a} =1 =d_{\parallel}$. $\gamma=1$ corresponds to the long-range Coulomb interaction. With increasing $\gamma$, the interaction becomes increasingly short-range. For $\gamma=2$, as can be seen from Eq.~\eqref{eq:EffectiveHee}, a gap of size $d_{\parallel}$ appears in the longitudinal mode. With further increase in $\gamma$, this gap diverges.
} 
\label{fig:dispersion}
\end{figure}

As shown in App.~\ref{app:SimplifyHee}, this term  can be absorbed into a redefinition of the elastic coefficients ~\cite{BonsallMaradudin,MakiZotos83,ChitraPRB01}. 
We note that we have considered these elastic constants to be $q$-independent, which is a feature of the local elastic theory. One can also extend this analysis to non-local elastic theories where these constants can be considered to be $q$-dependent.  Generalizing to an interaction of the form $U(\bs x) \sim 1/|\bs x|^{\gamma}$, we find that the full Hamiltonian defining the
low energy fluctuations of the GWC can be expressed as 
\begin{gather}
H_{\rm eff} = \int_{\bs q}
u_\mypar(\bs q) \, \Omega_\mypar u_\mypar(-\bs q) + 
u_\perp (\bs q)  \, \Omega_\perp u_\perp (-\bs q)  + \Hel; \nonumber \\
\Omega_\mypar(\bs q)  = c_\mypar q^2 + d_\mypar q^{2-\gamma} \quad, \quad
\Omega_\perp(\bs q)  = c_\perp q^{2}  .
\label{eq:EffectiveHee}
\end{gather}
$\Hel $ is given by \eqref{eq:EffectiveHel}. $\Omega_{a}$ are the dispersions of the longitudinal and the transverse modes. As discussed previously, it is only the longitudinal mode whose dispersion is affected by $\gamma$, see Fig.~\ref{fig:dispersion}.  Secondly, as discussed in App.~\ref{app:SimplifyHee}, these elastic constants follow $c_{a} \propto {\lambda_{\rm w}^{\gamma}}/{\epsilon}$. Notably, the elastic modulus $d_\mypar$ is a density-independent constant only in the low density limit  far away from WC  melting.
Also, as screening ($\epsilon$) increases, the WC becomes loosely bound due to reduced interaction. This makes a WC less rigid, or $c_a$ decreases with increasing $\epsilon$.

\section{Gaussian Variational Minimization}
\label{sec:GVM}

In this section, we treat the effective Hamiltonian obtained in the previous section using the Gaussian variational method (GVM) developed in Refs.~\cite{FluxLattice95, FluxLattice96, FukuyamaLee78}. This allows us to obtain the dispersion of the vibrational modes of the GWC and the associated pinning gap arising from the interaction between the Wigner lattice and the moir\'e lattice. Motivated by the experiments, we assume the GWCs to be weakly coupled to the moir\'e lattice. This allows us to treat the vibrations of the localized particles as harmonic fluctuations. This is formalized by the GVM as follows. Consider a Hamiltonian $H=\frac12 \int_q u(q)  \,  \Omega (q) u(-q) + H'$, where the kernel $ \Omega (q)$ is known \textit{a priori} and $H'$ can contain non-linear or polynomial terms in the field $u(q)$. For a vector field $\bs u(\bs q)$, this kernel becomes a matrix.  The goal is to approximate the Hamiltonian $H$ by the following quadratic form,
\begin{align}
H_0 = \frac12 \int_{q} u (q) \, {\mathcal G}^{-1} (q) \, u (-q) \, .
\end{align}
The optimal function  $\mathcal G(q)$ is then obtained by minimizing the  variational free energy of the theory $H$, $F_{\rm var} = F_0 + \langle H - H_0 \rangle_0$ where $\expect{\mathcal{..}}_{0}$ is the expectation value evaluated with  $H_0$  with respect to   $\mathcal G(q)$.  
In App.~\ref{app:GVM-SG} we provide a pedagogical discussion on using this GVM method for the simple case of a Sine-Gordon (SG) interaction as  the hamiltonian in \eqref{eq:EffectiveHee}  closely resembles the SG  problem.

\subsection{Applying GVM to GWC}
\label{sub:GVM-GWC}

We use the GVM to obtain the gap opened by the moir\'e lattice. Since the displacement is a two component field we have both $ \Omega_\mypar(\bs q) $ and 
$  \Omega_\perp(\bs q) $. The variational free energy becomes
\begin{align}
\label{eq:var}
F_{\mathrm{var}} 
 =  \frac T2 \int_q \sum_{a = \parallel, \perp} \Big \{ \Omega_a(\bs q) \mathcal G_a(\bs q) - \log [ T \mathcal G_a (\bs q) ]  \Big \} - \nonumber \\ 
\sum_{l,m } V_{m} \, \delta\left( \bs K_l - \bs G_m \right) \,  \exp \left[ - \frac T2  \sum_{b = \parallel, \perp}   K_{l,b}^2 \,   \int_{\bs q} \mathcal G_b (\bs q)\right] .
\end{align}
Note that in the absence of a magnetic field  there is no  admixture of the longitudinal and transverse modes.  

The Green function that minimizes the free energy in Eq.~\eqref{eq:var} can be approximated by $\mathcal G^{(0)}_{a}(\bs q) \simeq \frac{1}{\Omega_{a} (\bs q) + \Delta_{a}}$, where the gaps $ \Delta_{a}$ satisfy the following self-consistent equations (SCE)
\begin{align}
\Delta_{a} =  \sum_{m \in \mathcal M} V_m^{\phantom{2}}  G_{m,a}^2 \exp \left( - \frac T2  \int_{\bs q} \sum_{b=\perp, \mypar} \frac{G_{m,b}^2}{\Omega_{b}(\bs q) + \Delta_{b} }  \right) .
\label{eq:cons1}  
\end{align}
Here $a$ is not in the Cartesian basis but in the orthonormal basis discussed in Eq.~\eqref{eq:BasisRot}. Though at first glance  Eq.~\eqref{eq:cons1} seems  independent of $\bs K_{l}$ (or $\lambda_{\mathrm{w}}$), we note that  the conservation of momentum imposed through the delta function in Eq.~\eqref{eq:var},  restricts the set of  $\bs G_{m}$ to  those satisfying $r \bs G_{1} = s \bs K_{1}$. The set of such restricted (momentum conserving) values of $\bs G_{m}$ is denoted by $\mathcal M$. For instance, for the WC at $1/3$-filling, since, as explained previously, $r=1$, $\mathcal M$ is trivially the first MBZ.  After integrating, we find that  the gap equations take the form
\begin{align}
\label{eq:SCEgap}
\Delta_{a} \simeq&  \sum_{m \in \mathcal M} V_m^{\phantom{2}}  G_{m,a}^2 
\left( \frac{\Delta_\perp} {c_\perp \Lambda^2} \right)^{ \frac{T G_{m,\perp}^2}{8\pi c_\perp} }
\left( \frac{\Delta_\mypar} {c_\mypar \Lambda^2} \right)^{ \frac{T G_{m,\mypar}^2}{8\pi c_\mypar} }  \nonumber \\
&\times \exp \left[ {\frac{T G_{m,\mypar}^2}{8\pi c_\mypar} \tilde \Delta_\mypar \left( \pi  + 2 \tan ^{-1}  \tilde \Delta_\mypar  \right)} \right] .
\end{align}
Here, ${d_\mypar}/{\sqrt{4 c_\mypar \Delta_\mypar -d_\mypar^2}} \equiv \tilde \Delta_\mypar  $ and $\Lambda$ is a UV cutoff for the momentum space integration. The zero temperature limit for the gap above is $\sum_{m \in \mathcal M} V_m^{\phantom{2}}  G_{m,a}^2 \equiv \Delta_{a}^{0}$. And, a low temperature expansion is obtained to be
\begin{gather}
\Delta_\perp 
	= A_{\perp} +  B_{\perp} \log \Delta_\perp  ;
\nonumber \\
	 A_{\perp} = \Delta_\perp^0 + T \sum_{m \in \mathcal M} V_m^{\phantom{2}} G_{m,\perp}^2 \left[  D_{m}^{\mypar}  - \frac{ G_{m,\perp}^2}{8\pi c_\perp} \log {c_\perp \Lambda^2} \right]  ,
	\nonumber \\
	D_{m}^{\mypar} = \frac{ G_{m,\mypar}^2}{8\pi c_\mypar} \log \frac{\Delta_\mypar} {c_\mypar \Lambda^2}  + \frac{ G_{m,\mypar}^2}{8\pi c_\mypar} \tilde \Delta_\mypar \left( \pi + 2  \tan ^{-1}  \tilde \Delta_\mypar  \right) ,
	\nonumber \\ 
	B_{\perp} = T \sum_{m \in \mathcal M} V_m^{\phantom{2}}\frac{ G_{m,\perp}^4}{8\pi c_\perp}.
\label{eq:ExcplicitPerp}
\end{gather}
Here $A_{\perp}$ is dependent on $\Delta_{\mypar}$, and $B_{\perp}$ is a geometric constant. From this, we obtain a closed-form expression for $\Delta_{\perp}$ in terms of $\Delta_{\mypar}$. By bringing the above equation to the form $we^w=z$, we obtain the solution $w=W_k(z)$, where $W_{k}(z)$ is the (multivalued) Lambert $W$ function with its branch indexed by the integer $k$. In fact, when $w<0$ (for us, $w=- \Delta_\perp/B_\perp$), the solution has two branches, $W_{0}(z)$ and $W_{-1}(z)$. We will drop the latter solution since it is not a regular function at $\Delta_\perp=0$. Therefore,
\begin{align}
\label{eq:DeltaPerp}
\Delta_\perp = - B_{\perp} W_0 \left(- \frac{e^{- A_\perp/ B_\perp} }{B_\perp} \right) .
\end{align}%
This is the explicit dependence of $\Delta_{\perp}$ on $\Delta_{\mypar}$ (through $A_{\perp}$ only). Similarly, an SCE for the $\Delta_{\mypar}$ component is
\begin{align}
\Delta_{\mypar}^{\phantom{0}} = \Delta_{\mypar}^{0} + T \sum_{m \in \mathcal M} V_m^{\phantom{2}} G_{m, \mypar}^2 \frac{ G_{m,\perp}^2}{8\pi c_\perp}  \log  \frac{\Delta_\perp} {c_\perp \Lambda^2}  ,
\label{eq:DeltaPar}
\end{align}%
where $\Delta_{\perp}$ is given by Eq.~\eqref{eq:DeltaPerp}. In the next subsection we discuss the solutions obtained here, especially in conjunction with the recent experiments.

\subsection{Discussions}
\label{sub:GapDis}

Note that the last term in Eq.~\eqref{eq:SCEgap} is an artifact of the long-range interaction which vanishes if $d_{\parallel}=0$. This term, which is the compression term,  purely accounts for the elastic contribution to the gap. For $d_{\mypar}=0$, the gap equation is equivalent to the vector SG potential, see Eq.~\eqref{eq:VectorSG}. 

Secondly, since $\Lambda$ appears in the denominator of Eq.~\eqref{eq:SCEgap}, the gap vanishes for temperatures larger than a characteristic temperature, $\mathrm{min} \left(\frac{8\pi c_\mypar}{G_{\mypar}^2}, \frac{8\pi c_\perp}{G_{\perp}^{2} } \right) \equiv T_\ast$. This is a feature of the equivalence of the effective interaction Hamiltonian to that with the SG potential, see discussions in the App.~\ref{app:GVM-SG}.  The analysis is valid only if $T_{\ast}$ is much smaller than  the melting temperature (such as $T_{\rm L}$) of a GWC. Note that since $T_{\ast} \sim {c_{a}}{\lambda^{2}_{\rm w}}$,  this temperature scale can be  controlled  by means of the twist angle. 

The pinning frequency  is related to the zero temperature  gap as~\cite{FukuyamaLee78} $\omega_{p}^{a} = \sqrt{\Delta^{0}_{a}/\rho_{0}}$. Notably, since the pinning frequency scales with the size of the WBZ, $\omega_{p}^{a} \propto G_{a}$, it becomes increasingly difficult to de-pin a WC of smaller unit cell. This is since a WC with large unit cell (or small $\bs G$) will be loosely bound compared to one with smaller unit cell (as the particles are more tightly packed). Therefore, the former can be easily de-pinned by an external electric field. Similarly, for a deeper moir\'e potential the pinning frequency increases ($\omega_{p}^{a} \propto \tilde V^{1/2}$) since the particles get tightly bound to the potential minima. Introduction of a spacer layer can further modulate this frequency. Geometrical factors aside, the pinning gap thus becomes $\mathcal O(\lesssim \rm meV)$. With increasing temperature, as seen in Eq.~\eqref{eq:ExcplicitPerp}, this gap softens as the increasing thermal fluctuation facilitates de-pinning. The extent to which this gap decreases depends on various coefficients appearing in Eq.~\eqref{eq:ExcplicitPerp}. Most notably, via the elastic constants, $c_{\alpha}$, the logarithm term has a coefficient that is directly proportional to the dielectric term. Thus, the larger the screening, the smaller the pinning gap. Therefore, although the geometrical constants associated with various HoM or HeM TMDs may not affect the pinning gap of a GWC, the dielectric constant can however alter the physics. This gap translates into determining which state is a stronger insulator.

\section{Conclusion}
\label{sec:Conclude}

We have addressed the feasibility of realizing Wigner crystals in a host of  HoM and HeM  systems.  Note however,  that our results are based on  estimated material parameters  of the TMD moir\'e materials.  Corrections to these results might arise principally from three sources.  The first is from the full band structure of the TMD heterostructures~\cite{OurPaper2}.  Second, a material correction arising from twist-angle inhomogeneity across a sample~\cite{Uri2020, PRRtwistdis} which may cause additional pinning or de-pinning of the WC could also affect the physics. Similar effects may also arise from atomic relaxations~\cite{NamKoshino, Corrugation14}. Third, the presence or absence of a spacer layer~\cite{ZurichSpacer}, such as a monolayer hBN, may also affect the correlation energy, thereby affecting Wigner crystallization. A first principles calculation of the elastic coefficients of the GWC is also important to obtain good qualitative and quantitative estimates for the pinning gap and the phonon spectrum. All of these aspects merit  further studies  as this will help narrow  the density and temperature regimes where WC is feasible.

Due to the presence of a pinning gap, transport measurements to confirm the existence of WC states can be misleading as there can be many other kinds of insulating states with similar transport characteristics. Although observation of such states at fractional occupancy increases their likelihood of being Wigner states, especially for those observed at incommensurate fillings, however, the possibility of  other density ordered states cannot be ruled out, particularly for commensurate fractional occupancies. Devising smoking gun evidence for various density ordered states may be an interesting task for theorists and experimentalists alike.

As was mentioned before, once a system meets the material constraints to realize a GWC, there exists a plethora of crystalline states below the filling fraction $\nu^{\rm max}$. These states constitute a devil's staircase and have a rich physics of commensurate-incommensurate transitions~\cite{BakDevil82, AubryKAM, PokrovskyLR83}. Due to various stability criteria, only a few such states might display clear experimental signatures. However, with careful analysis or improvements in experimental conditions, one may gain insight into the other states as well. In fact, a theoretical framework to understand these commensurate-incommensurate transitions in presence of an underlying lattice  is an interesting theoretical task and is left for future work.

B.P. and P.W.P. thank the NSF under grant DMR19-19143 for partial funding of this project.

\section*{Appendix}

\appendix
\renewcommand{\theequation}{\thesection \arabic{equation}}
\setcounter{equation}{0}

\section{Harmonic Expansion of Density}
\label{app:HarmonicDensity} 

Following~\cite{FluxLattice95}, we derive the elastic limit of the density written in Eq.~\eqref{eq:expandDensity}. A continuum limit can be easily obtained if we treat the equilibrium GWC configuration, $\bs R_i^0 = \bs R_i - \bs u(\bs R_i^0)$, as a slowly varying smooth vector field, $\bs \varphi(\bs x)$, over the position of the particles
\begin{align}
\bs \varphi(\bs x) = \bs x - \bs u \left(\bs \varphi(\bs x) \right) \, .
\label{newfield}
\end{align}
Clearly a solution of $\bs \varphi(\bs x)$ is given by, $\bs \varphi(\bs R_i) = \bs \varphi \left(\bs R_i^0 + \bs u(\bs R_i^0) \right) =  \bs R_i^0 $. Using the above equality we can rewrite the density in terms of this new field as
\begin{subequations}
\begin{align}
\rho(\bs x) =& \sum_i  \delta^{(2)} \left[ \bs R_i - \bs \varphi(\bs x) - \bs u(\bs \varphi(\bs x)) \right] \\
\simeq& \det [\bs \partial_\alpha \bs \varphi_\beta (\bs x)]  \sum_i  \delta^{(2)} \left( \bs R_i - \bs \varphi(\bs x) \right) \\
=& \det [\bs \partial_\alpha \bs \varphi_\beta (\bs x)]  \int \frac{d \bs q}{(2\pi)^2} \rho_0 (\bs q) e^{i \bs q \cdot \bs \varphi(\bs x)} \, . \label{integraldensity}
\end{align}
\end{subequations}
The first simplification was done using the elastic limit, $\bs \partial_\alpha \bs u_\beta \ll 1$. In the last line, we have used the integral representation of the delta function. In the presence of an undeformed GWC, we can introduce its reciprocal vectors, $e^{i \bs K_{l} \cdot \bs R_i} = 1$, to write 
\begin{align}
 \rho_0(\bs q) = \sum_i e^{i \bs q \cdot \bs R_i}  = \rho_0 (2 \pi)^2 \sum_{l} \delta^{(2)} (\bs q - \bs K_l) \, .
\end{align}
Here $\rho_0$ is the average number density. Introducing the above simplification in Eq.~\eqref{integraldensity} and using Eq.~\eqref{newfield}, we obtain
\begin{align}
\rho(\bs x) =& \rho_0 \det [\bs \partial_\alpha \bs \varphi_\beta (\bs x)]  \sum_{l} e^{i \bs K_l \cdot \bs \varphi(\bs x) }  \nonumber \\
=& \rho_0 \det [1 - \bs \partial_\alpha \bs u_\beta \left( \bs \varphi (\bs x) \right) ]  \sum_{l} e^{i \bs K_l \cdot \left[ \bs x - \bs u(\bs \varphi(\bs x)) \right] } \nonumber \\
\simeq &  \rho_0 - \rho_0 \bs \nabla \cdot \bs u (\bs x) + \rho_0  \sum_{l}  e^{i \bs K_l \cdot \left[ \bs x - \bs u(\bs x) \right] } \, .
\end{align}
We again used the elastic limit by first Taylor-expanding the determinant operator,  $\det$, and then substituting $\bs u(\bs \varphi(\bs x))  \approx \bs u(\bs x)$ which works for $\bs x$ close to the equilibrium position and in the elastic limit. This leads us to Eq.~\eqref{eq:expandDensity}. Note there is complete decoupling between the gradient term and the terms with $\bs K_l$. This occurs because $\bs u(\bs x)$ has negligible Fourier components outside the WBZ.

\section{Elastic Interaction Hamiltonian} 
\label{app:SimplifyHee}

In this Appendix, we clarify the derivation of Eq.~\eqref{eq:Hee2Taylor}. First, we Fourier transform the second part of Eq.~\eqref{eq:HeeExpand},
\begin{align}
\Hee^{(2)} 
&=  \frac{1}{ 2} \sum_{l}\int_{\bs x, \bs x'} U(\bs x - \bs x') e^{i \bs K_{l} \cdot (\bs x - \bs x')} K_{l,\alpha} K_{l,\beta} \, \times \nonumber \\
& \int_{\bs q, \bs q'} u_{\alpha}(\bs q) \, u_{\beta}(\bs q') \, \left( e^{i \bs q \cdot \bs x} - e^{i \bs q \cdot \bs x'} \right) \left( e^{i \bs q' \cdot \bs x} - e^{i \bs q' \cdot \bs x'} \right).
\end{align}
In order to simplify it further, we introduce the center of mass coordinate, $2\bs X = \bs x + \bs x'$ and the relative coordinate $2\bs \delta = \bs x - \bs x'$ to obtain
\begin{align}
\Hee^{(2)}  
 = &   \sum_{l} K_{l,\alpha} K_{l,\beta} \int_{\bs q} u_\alpha(\bs q)  u_\beta (- \bs q) \, \times \nonumber \\
& \int_{\bs \delta} d\bs \delta \, U(\bs \delta) 
 \left[ 1 - \cos(\bs q \cdot \bs \delta)  \right]  e^{i \bs K_{l} \cdot \bs \delta } .
\end{align}
In coming to this line, we have also integrated out $\bs q'$, which introduced a delta function, $\delta^{(2)} (\bs q + \bs q')$. Next, we perform the last integration for a generic potential of the form, $U(\bs x) = e^{2}/ \epsilon |\bs x|^{\gamma}$. One can obtain the long-range Coulomb potential by setting $\gamma = 1$, and with increasing $\gamma$ the potential becomes increasingly short-range. For such a $U(\bs \delta)$ we find that
\begin{align}
\Hee^{(2)} 
= &  \sum_{l} K_{l,\alpha} K_{l,\beta} \int_{\bs q} u_\alpha(\bs q)  u_\beta (- \bs q) \, \times \nonumber \\
& \frac{e^{2}}{\epsilon} \left( \frac{2}{|\bs K_{l}|^{\gamma}} - \frac{1}{|\bs K_{l} - \bs q|^{\gamma} } - \frac{1}{|\bs K_{l} + \bs q|^{\gamma} }  \right) .
\label{eq:inverseK}
\end{align}

For further simplification, we confine our discussion to the low-energy limit. This allows us to Taylor-expand the last term in Eq.~\eqref{eq:inverseK} for the limit $|\bs q| \ll |\bs K_{l}|$. The first term in this expansion, which is linear in $\bs q$, vanishes because it involves integrating over a $\cos \theta_{l}$ term. Here, $\theta_{l}$ are the angles between the $\bs q$ vector and $\bs K_{l}$. Therefore, retaining up to the $\mathcal{O}(q^{2})$ term we obtain, 
\begin{align}
\Hee^{(2)} 
\simeq  \gamma \frac{e^{2}}{\epsilon} \, & \sum_{l} \frac{K_{l,\alpha} K_{l,\beta}}{| \bs K_{l} |^{2+\gamma}} \int_{\bs q} q^{2} u_\alpha(\bs q)  u_\beta (- \bs q)  \, \times \nonumber \\
& \left[ (2+\gamma) \cos^{2} \theta_{l} - 1 \right].
\end{align}%
Note that unlike the long-distance term, $\Hee^{(1)}$ in Eq.~\eqref{eq:HeeLongDistance}, the leading dispersion corresponding to $\Hee^{(2)}$ remains quadratic regardless of the choice of $\gamma$. 
\begin{align}
\Hee^{(2)}  = \int_{\bs q} c_\mypar q^2 u_\mypar(\bs q)  u_\mypar(- \bs q) +  c_\perp q^2 u_\perp(\bs q)  u_\perp(- \bs q).
\end{align}%

\section{GVM for Sine-Gordon Potential}
\label{app:GVM-SG}

In this Appendix, we demonstrate the GVM method discussed in the main text for a Sine-Gordon (SG) potential,
\begin{align}
H = \frac12 c \int dx \, \left[ \nabla \phi(x) \right]^2 - g \int dx \cos[2 \phi(x) ] .
\end{align}%
Here $c$ and $g$ are free parameters. Using the simplifications discussed in the main text [and using $\Omega(q) = cq^2$], we obtain the variational free energy to be
\begin{align}
F_{\mathrm{var}} &= - \frac T2 \int_q \log [ T \mathcal G(q) ] + \frac T2 \int_q c q^2 \mathcal G(q) - 
\nonumber \\
& g \left. \exp \left[ \frac T2 \int_q \mathcal G(q) \frac{\partial^2}{\partial \phi^2} \right]  \int dx \cos(2 \phi ) \right \rvert_{\phi=0} .
\end{align}%
Further simplifications of the last term leads us to 
\begin{align}
F_{\mathrm{var}} =&  - \frac T2 \int_q \log [ T \mathcal G(q) ] + \frac T2 \int_q c q^2 \mathcal G(q) 
- g e^{  - 2 \int_q T \mathcal G(q) } .
\end{align}%
In these equations, we fixed the sample area to $\int dx = 1$. The saddle point solution of the above free energy is
\begin{align}
\mathcal G^{-1} = cq^2 + 4 g e^{  - 2 \int_q T \mathcal G(q) }.
\end{align}%
We now set $\mathcal G^{-1}(q) = cq^2 + m$ and solve $m$ self-consistently,
\begin{align}
\label{eq:SGmass}
m = 4 g e^{-2 T \int_q^\Lambda \frac{1}{c q^2 + m} }  \simeq 4 g \left( \frac{m}{c \Lambda^2} \right)^{T/2\pi c} .
\end{align}
Here, $\Lambda$ is a UV cutoff in the momentum-space. A notable feature of this solution is that beyond a certain temperature maximum, $T > 2\pi c$, the SG mass must vanish simply due to the presence of the cutoff in the denominator above. Such a maximal temperature will also appear in our discussion in Sec.~\ref{sub:GapDis}. Additionally, from Eq.~\eqref{eq:SGmass} one can also deduce the scaling behavior of the mass, $m \sim g^{1/(1-\tau)}$, where $\tau = T/2\pi c$.

Pertaining to our discussion of GVM in the context of GWC, we extend the previous solutions for a SG potential to an $n$-component vector SG system. The interaction term here becomes $H' \sim \int dx \cos( \sum_n p_n \phi_n)$. The kernel corresponding to the field $\phi_n$ is $c_n q^2$. As before, we obtain the variational free energy 
\begin{subequations}
\label{eq:VectorSG}
\begin{align}
F_{\mathrm{var}} =&  - \frac T2  \sum_n \int_q \left \{ \log [ T \mathcal G_n(q) ] - c_n q^2 \mathcal G_n(q) \right \} \nonumber \\
& - g e^{  - \frac{T}{2}   \sum_n a_n^2 \int_q \mathcal G(q) } .
\end{align}%
Since, due to the vanishing average of cosine functions, there are no cross terms such as $\cos \phi_m \cos \phi_n$ (with $m\neq n$), the saddle-point equation (setting $n=1$ and $a_n =2$ goes back to the original case)
\begin{align}
\mathcal G_n^{-1} &= c_nq^2 + g a_n^2 e^{  - \frac T2 \sum_n a^2_n \int_q \mathcal G_n(q) } , \\
\therefore \quad m_n & = g a_n^2 \exp \left[  - \sum_n \frac {T a^2_n }{8\pi c_n} \log \left( \frac{c_n \Lambda^2}{m_n} \right) \right].
\end{align}%
\end{subequations}
We can solve this SCE exactly and, in this case as well, there exists a similar temperature window where gap vanishes, $ {T a^2_n }/{8\pi c_n} \equiv \tau_n > 1$. And, like before, the scaling of $m_{n}$ with the coupling constant becomes, $m_n \propto g^{ \left(1 - \sum_n \tau_n \right) ^{-1} } $. These solutions are not exactly transferable for our discussions in the main text since there the kernel has a $d_{\mypar} q^{2-\gamma}$ part. See Sec.~\ref{sub:GapDis} for the case when $d_{\mypar}=0$, where the above results are perfectly applicable.


\begin{thebibliography}{85}%
\makeatletter
\providecommand \@ifxundefined [1]{%
 \@ifx{#1\undefined}
}%
\providecommand \@ifnum [1]{%
 \ifnum #1\expandafter \@firstoftwo
 \else \expandafter \@secondoftwo
 \fi
}%
\providecommand \@ifx [1]{%
 \ifx #1\expandafter \@firstoftwo
 \else \expandafter \@secondoftwo
 \fi
}%
\providecommand \natexlab [1]{#1}%
\providecommand \enquote  [1]{``#1''}%
\providecommand \bibnamefont  [1]{#1}%
\providecommand \bibfnamefont [1]{#1}%
\providecommand \citenamefont [1]{#1}%
\providecommand \href@noop [0]{\@secondoftwo}%
\providecommand \href [0]{\begingroup \@sanitize@url \@href}%
\providecommand \@href[1]{\@@startlink{#1}\@@href}%
\providecommand \@@href[1]{\endgroup#1\@@endlink}%
\providecommand \@sanitize@url [0]{\catcode `\\12\catcode `\$12\catcode
  `\&12\catcode `\#12\catcode `\^12\catcode `\_12\catcode `\%12\relax}%
\providecommand \@@startlink[1]{}%
\providecommand \@@endlink[0]{}%
\providecommand \url  [0]{\begingroup\@sanitize@url \@url }%
\providecommand \@url [1]{\endgroup\@href {#1}{\urlprefix }}%
\providecommand \urlprefix  [0]{URL }%
\providecommand \Eprint [0]{\href }%
\providecommand \doibase [0]{http://dx.doi.org/}%
\providecommand \selectlanguage [0]{\@gobble}%
\providecommand \bibinfo  [0]{\@secondoftwo}%
\providecommand \bibfield  [0]{\@secondoftwo}%
\providecommand \translation [1]{[#1]}%
\providecommand \BibitemOpen [0]{}%
\providecommand \bibitemStop [0]{}%
\providecommand \bibitemNoStop [0]{.\EOS\space}%
\providecommand \EOS [0]{\spacefactor3000\relax}%
\providecommand \BibitemShut  [1]{\csname bibitem#1\endcsname}%
\let\auto@bib@innerbib\@empty
\bibitem [{\citenamefont {Wigner}(1934)}]{Wigner34}%
  \BibitemOpen
  \bibfield  {author} {\bibinfo {author} {\bibfnamefont {E.}~\bibnamefont
  {Wigner}},\ }\bibfield  {title} {\enquote {\bibinfo {title} {On the
  interaction of electrons in metals},}\ }\href {\doibase
  10.1103/PhysRev.46.1002} {\bibfield  {journal} {\bibinfo  {journal} {Phys.
  Rev.}\ }\textbf {\bibinfo {volume} {46}},\ \bibinfo {pages} {1002--1011}
  (\bibinfo {year} {1934})}\BibitemShut {NoStop}%
\bibitem [{\citenamefont {Grimes}\ and\ \citenamefont
  {Adams}(1979)}]{grimesadams}%
  \BibitemOpen
  \bibfield  {author} {\bibinfo {author} {\bibfnamefont {C.~C.}\ \bibnamefont
  {Grimes}}\ and\ \bibinfo {author} {\bibfnamefont {G.}~\bibnamefont {Adams}},\
  }\bibfield  {title} {\enquote {\bibinfo {title} {Evidence for a
  liquid-to-crystal phase transition in a classical, two-dimensional sheet of
  electrons},}\ }\href {\doibase 10.1103/PhysRevLett.42.795} {\bibfield
  {journal} {\bibinfo  {journal} {Phys. Rev. Lett.}\ }\textbf {\bibinfo
  {volume} {42}},\ \bibinfo {pages} {795--798} (\bibinfo {year}
  {1979})}\BibitemShut {NoStop}%
\bibitem [{\citenamefont {Monarkhaa}\ and\ \citenamefont
  {V.~E.~Syvokon}(2012)}]{MonarkhaRev}%
  \BibitemOpen
  \bibfield  {author} {\bibinfo {author} {\bibfnamefont {Y.~P.}\ \bibnamefont
  {Monarkhaa}}\ and\ \bibinfo {author} {\bibfnamefont {V.~E.}\ \bibnamefont
  {V.~E.~Syvokon}},\ }\bibfield  {title} {\enquote {\bibinfo {title} {A
  two-dimensional wigner crystal (review article)},}\ }\href@noop {} {\bibfield
   {journal} {\bibinfo  {journal} {Low Temperature Physics}\ }\textbf {\bibinfo
  {volume} {38}},\ \bibinfo {pages} {1067} (\bibinfo {year}
  {2012})}\BibitemShut {NoStop}%
\bibitem [{\citenamefont {Cao}\ \emph {et~al.}(2018{\natexlab{a}})\citenamefont
  {Cao}, \citenamefont {Fatemi}, \citenamefont {Demir}, \citenamefont {Fang},
  \citenamefont {Tomarken}, \citenamefont {Luo}, \citenamefont
  {Sanchez-Yamagishi}, \citenamefont {Watanabe}, \citenamefont {Taniguchi},
  \citenamefont {Kaxiras}, \citenamefont {Ashoori},\ and\ \citenamefont
  {Jarillo-Herrero}}]{Cao18Mott}%
  \BibitemOpen
  \bibfield  {author} {\bibinfo {author} {\bibfnamefont {Y.}~\bibnamefont
  {Cao}}, \bibinfo {author} {\bibfnamefont {V.}~\bibnamefont {Fatemi}},
  \bibinfo {author} {\bibfnamefont {A.}~\bibnamefont {Demir}}, \bibinfo
  {author} {\bibfnamefont {S.}~\bibnamefont {Fang}}, \bibinfo {author}
  {\bibfnamefont {S.~L.}\ \bibnamefont {Tomarken}}, \bibinfo {author}
  {\bibfnamefont {J.~Y.}\ \bibnamefont {Luo}}, \bibinfo {author} {\bibfnamefont
  {J.~D.}\ \bibnamefont {Sanchez-Yamagishi}}, \bibinfo {author} {\bibfnamefont
  {K.}~\bibnamefont {Watanabe}}, \bibinfo {author} {\bibfnamefont
  {T.}~\bibnamefont {Taniguchi}}, \bibinfo {author} {\bibfnamefont
  {E.}~\bibnamefont {Kaxiras}}, \bibinfo {author} {\bibfnamefont {R.~C.}\
  \bibnamefont {Ashoori}}, \ and\ \bibinfo {author} {\bibfnamefont
  {P.}~\bibnamefont {Jarillo-Herrero}},\ }\bibfield  {title} {\enquote
  {\bibinfo {title} {Correlated insulator behaviour at half-filling in
  magic-angle graphene superlattices},}\ }\href
  {https://www.nature.com/articles/nature26154} {\bibfield  {journal} {\bibinfo
   {journal} {Nature}\ }\textbf {\bibinfo {volume} {556}},\ \bibinfo {pages}
  {80} (\bibinfo {year} {2018}{\natexlab{a}})}\BibitemShut {NoStop}%
\bibitem [{\citenamefont {Cao}\ \emph {et~al.}(2018{\natexlab{b}})\citenamefont
  {Cao}, \citenamefont {Fatemi}, \citenamefont {Fang}, \citenamefont
  {Watanabe}, \citenamefont {Taniguchi}, \citenamefont {Kaxiras},\ and\
  \citenamefont {Jarillo-Herrero}}]{Cao18SC}%
  \BibitemOpen
  \bibfield  {author} {\bibinfo {author} {\bibfnamefont {Y.}~\bibnamefont
  {Cao}}, \bibinfo {author} {\bibfnamefont {V.}~\bibnamefont {Fatemi}},
  \bibinfo {author} {\bibfnamefont {S.}~\bibnamefont {Fang}}, \bibinfo {author}
  {\bibfnamefont {K.}~\bibnamefont {Watanabe}}, \bibinfo {author}
  {\bibfnamefont {T.}~\bibnamefont {Taniguchi}}, \bibinfo {author}
  {\bibfnamefont {E.}~\bibnamefont {Kaxiras}}, \ and\ \bibinfo {author}
  {\bibfnamefont {P.}~\bibnamefont {Jarillo-Herrero}},\ }\bibfield  {title}
  {\enquote {\bibinfo {title} {Unconventional superconductivity in magic-angle
  graphene superlattices},}\ }\href
  {https://www.nature.com/articles/nature26160} {\bibfield  {journal} {\bibinfo
   {journal} {Nature}\ }\textbf {\bibinfo {volume} {556}},\ \bibinfo {pages}
  {43} (\bibinfo {year} {2018}{\natexlab{b}})}\BibitemShut {NoStop}%
\bibitem [{\citenamefont {Yankowitz}\ \emph {et~al.}(2019)\citenamefont
  {Yankowitz}, \citenamefont {Chen}, \citenamefont {Polshyn}, \citenamefont
  {Zhang}, \citenamefont {Watanabe}, \citenamefont {Taniguchi}, \citenamefont
  {Graf}, \citenamefont {Young},\ and\ \citenamefont {Dean}}]{YoungDean}%
  \BibitemOpen
  \bibfield  {author} {\bibinfo {author} {\bibfnamefont {M.}~\bibnamefont
  {Yankowitz}}, \bibinfo {author} {\bibfnamefont {S.}~\bibnamefont {Chen}},
  \bibinfo {author} {\bibfnamefont {H.}~\bibnamefont {Polshyn}}, \bibinfo
  {author} {\bibfnamefont {Y.}~\bibnamefont {Zhang}}, \bibinfo {author}
  {\bibfnamefont {K.}~\bibnamefont {Watanabe}}, \bibinfo {author}
  {\bibfnamefont {T.}~\bibnamefont {Taniguchi}}, \bibinfo {author}
  {\bibfnamefont {D.}~\bibnamefont {Graf}}, \bibinfo {author} {\bibfnamefont
  {A.~F.}\ \bibnamefont {Young}}, \ and\ \bibinfo {author} {\bibfnamefont
  {C.~R.}\ \bibnamefont {Dean}},\ }\bibfield  {title} {\enquote {\bibinfo
  {title} {Tuning superconductivity in twisted bilayer graphene},}\ }\href
  {\doibase 10.1126/science.aav1910} {\bibfield  {journal} {\bibinfo  {journal}
  {Science}\ } (\bibinfo {year} {2019}),\ 10.1126/science.aav1910}\BibitemShut
  {NoStop}%
\bibitem [{\citenamefont {Lu}\ \emph {et~al.}(2019)\citenamefont {Lu},
  \citenamefont {Stepanov}, \citenamefont {Yang}, \citenamefont {Xie},
  \citenamefont {Aamir}, \citenamefont {Das}, \citenamefont {Urgell},
  \citenamefont {Watanabe}, \citenamefont {Taniguchi}, \citenamefont {Zhang},
  \citenamefont {Bachtold}, \citenamefont {MacDonald},\ and\ \citenamefont
  {Efetov}}]{EfetovSC}%
  \BibitemOpen
  \bibfield  {author} {\bibinfo {author} {\bibfnamefont {X.}~\bibnamefont
  {Lu}}, \bibinfo {author} {\bibfnamefont {P.}~\bibnamefont {Stepanov}},
  \bibinfo {author} {\bibfnamefont {W.}~\bibnamefont {Yang}}, \bibinfo {author}
  {\bibfnamefont {M.}~\bibnamefont {Xie}}, \bibinfo {author} {\bibfnamefont
  {M.~A.}\ \bibnamefont {Aamir}}, \bibinfo {author} {\bibfnamefont
  {I.}~\bibnamefont {Das}}, \bibinfo {author} {\bibfnamefont {C.}~\bibnamefont
  {Urgell}}, \bibinfo {author} {\bibfnamefont {K.}~\bibnamefont {Watanabe}},
  \bibinfo {author} {\bibfnamefont {T.}~\bibnamefont {Taniguchi}}, \bibinfo
  {author} {\bibfnamefont {G.}~\bibnamefont {Zhang}}, \bibinfo {author}
  {\bibfnamefont {A.}~\bibnamefont {Bachtold}}, \bibinfo {author}
  {\bibfnamefont {A.~H.}\ \bibnamefont {MacDonald}}, \ and\ \bibinfo {author}
  {\bibfnamefont {D.~K.}\ \bibnamefont {Efetov}},\ }\bibfield  {title}
  {\enquote {\bibinfo {title} {Superconductors, orbital magnets and correlated
  states in magic-angle bilayer graphene},}\ }\href {\doibase
  10.1038/s41586-019-1695-0} {\bibfield  {journal} {\bibinfo  {journal}
  {Nature}\ }\textbf {\bibinfo {volume} {574}},\ \bibinfo {pages} {653--657}
  (\bibinfo {year} {2019})}\BibitemShut {NoStop}%
\bibitem [{\citenamefont {{Kerelsky}}\ \emph {et~al.}(2018)\citenamefont
  {{Kerelsky}}, \citenamefont {{McGilly}}, \citenamefont {{Kennes}},
  \citenamefont {{Xian}}, \citenamefont {{Yankowitz}}, \citenamefont {{Chen}},
  \citenamefont {{Watanabe}}, \citenamefont {{Taniguchi}}, \citenamefont
  {{Hone}}, \citenamefont {{Dean}}, \citenamefont {{Rubio}},\ and\
  \citenamefont {{Pasupathy}}}]{PasupathySTM}%
  \BibitemOpen
  \bibfield  {author} {\bibinfo {author} {\bibfnamefont {A.}~\bibnamefont
  {{Kerelsky}}}, \bibinfo {author} {\bibfnamefont {L.}~\bibnamefont
  {{McGilly}}}, \bibinfo {author} {\bibfnamefont {D.~M.}\ \bibnamefont
  {{Kennes}}}, \bibinfo {author} {\bibfnamefont {L.}~\bibnamefont {{Xian}}},
  \bibinfo {author} {\bibfnamefont {M.}~\bibnamefont {{Yankowitz}}}, \bibinfo
  {author} {\bibfnamefont {S.}~\bibnamefont {{Chen}}}, \bibinfo {author}
  {\bibfnamefont {K.}~\bibnamefont {{Watanabe}}}, \bibinfo {author}
  {\bibfnamefont {T.}~\bibnamefont {{Taniguchi}}}, \bibinfo {author}
  {\bibfnamefont {J.}~\bibnamefont {{Hone}}}, \bibinfo {author} {\bibfnamefont
  {C.}~\bibnamefont {{Dean}}}, \bibinfo {author} {\bibfnamefont
  {A.}~\bibnamefont {{Rubio}}}, \ and\ \bibinfo {author} {\bibfnamefont
  {A.~N.}\ \bibnamefont {{Pasupathy}}},\ }\bibfield  {title} {\enquote
  {\bibinfo {title} {{Magic Angle Spectroscopy}},}\ }\href@noop {} {\bibfield
  {journal} {\bibinfo  {journal} {arXiv e-prints}\ ,\ \bibinfo {eid}
  {arXiv:1812.08776}} (\bibinfo {year} {2018})},\ \Eprint
  {http://arxiv.org/abs/1812.08776} {arXiv:1812.08776 [cond-mat.mes-hall]}
  \BibitemShut {NoStop}%
\bibitem [{\citenamefont {{Choi}}\ \emph {et~al.}(2019)\citenamefont {{Choi}},
  \citenamefont {{Kemmer}}, \citenamefont {{Peng}}, \citenamefont {{Thomson}},
  \citenamefont {{Arora}}, \citenamefont {{Polski}}, \citenamefont {{Zhang}},
  \citenamefont {{Ren}}, \citenamefont {{Alicea}}, \citenamefont {{Refael}},
  \citenamefont {{von Oppen}}, \citenamefont {{Watanabe}}, \citenamefont
  {{Taniguchi}},\ and\ \citenamefont {{Nadj-Perge}}}]{Caltech19}%
  \BibitemOpen
  \bibfield  {author} {\bibinfo {author} {\bibfnamefont {Y.}~\bibnamefont
  {{Choi}}}, \bibinfo {author} {\bibfnamefont {J.}~\bibnamefont {{Kemmer}}},
  \bibinfo {author} {\bibfnamefont {Y.}~\bibnamefont {{Peng}}}, \bibinfo
  {author} {\bibfnamefont {A.}~\bibnamefont {{Thomson}}}, \bibinfo {author}
  {\bibfnamefont {H.}~\bibnamefont {{Arora}}}, \bibinfo {author} {\bibfnamefont
  {R.}~\bibnamefont {{Polski}}}, \bibinfo {author} {\bibfnamefont
  {Y.}~\bibnamefont {{Zhang}}}, \bibinfo {author} {\bibfnamefont
  {H.}~\bibnamefont {{Ren}}}, \bibinfo {author} {\bibfnamefont
  {J.}~\bibnamefont {{Alicea}}}, \bibinfo {author} {\bibfnamefont
  {G.}~\bibnamefont {{Refael}}}, \bibinfo {author} {\bibfnamefont
  {F.}~\bibnamefont {{von Oppen}}}, \bibinfo {author} {\bibfnamefont
  {K.}~\bibnamefont {{Watanabe}}}, \bibinfo {author} {\bibfnamefont
  {T.}~\bibnamefont {{Taniguchi}}}, \ and\ \bibinfo {author} {\bibfnamefont
  {S.}~\bibnamefont {{Nadj-Perge}}},\ }\bibfield  {title} {\enquote {\bibinfo
  {title} {{Imaging Electronic Correlations in Twisted Bilayer Graphene near
  the Magic Angle}},}\ }\href@noop {} {\bibfield  {journal} {\bibinfo
  {journal} {arXiv e-prints}\ ,\ \bibinfo {eid} {arXiv:1901.02997}} (\bibinfo
  {year} {2019})},\ \Eprint {http://arxiv.org/abs/1901.02997} {arXiv:1901.02997
  [cond-mat.mes-hall]} \BibitemShut {NoStop}%
\bibitem [{\citenamefont {Wong}\ \emph {et~al.}(2020)\citenamefont {Wong},
  \citenamefont {Nuckolls}, \citenamefont {Oh}, \citenamefont {Lian},
  \citenamefont {Xie}, \citenamefont {Jeon}, \citenamefont {Watanabe},
  \citenamefont {Taniguchi}, \citenamefont {Bernevig},\ and\ \citenamefont
  {Yazdani}}]{CascadePrinceton}%
  \BibitemOpen
  \bibfield  {author} {\bibinfo {author} {\bibfnamefont {D.}~\bibnamefont
  {Wong}}, \bibinfo {author} {\bibfnamefont {K.~P.}\ \bibnamefont {Nuckolls}},
  \bibinfo {author} {\bibfnamefont {M.}~\bibnamefont {Oh}}, \bibinfo {author}
  {\bibfnamefont {B.}~\bibnamefont {Lian}}, \bibinfo {author} {\bibfnamefont
  {Y.}~\bibnamefont {Xie}}, \bibinfo {author} {\bibfnamefont {S.}~\bibnamefont
  {Jeon}}, \bibinfo {author} {\bibfnamefont {K.}~\bibnamefont {Watanabe}},
  \bibinfo {author} {\bibfnamefont {T.}~\bibnamefont {Taniguchi}}, \bibinfo
  {author} {\bibfnamefont {B.~A.}\ \bibnamefont {Bernevig}}, \ and\ \bibinfo
  {author} {\bibfnamefont {A.}~\bibnamefont {Yazdani}},\ }\bibfield  {title}
  {\enquote {\bibinfo {title} {Cascade of electronic transitions in magic-angle
  twisted bilayer graphene},}\ }\href {\doibase 10.1038/s41586-020-2339-0}
  {\bibfield  {journal} {\bibinfo  {journal} {Nature}\ }\textbf {\bibinfo
  {volume} {582}},\ \bibinfo {pages} {198--202} (\bibinfo {year}
  {2020})}\BibitemShut {NoStop}%
\bibitem [{\citenamefont {Zondiner}\ \emph {et~al.}(2020)\citenamefont
  {Zondiner}, \citenamefont {Rozen}, \citenamefont {Rodan-Legrain},
  \citenamefont {Cao}, \citenamefont {Queiroz}, \citenamefont {Taniguchi},
  \citenamefont {Watanabe}, \citenamefont {Oreg}, \citenamefont {von Oppen},
  \citenamefont {Stern},\ and\ \citenamefont {et~al.}}]{CascadeMIT}%
  \BibitemOpen
  \bibfield  {author} {\bibinfo {author} {\bibfnamefont {U.}~\bibnamefont
  {Zondiner}}, \bibinfo {author} {\bibfnamefont {A.}~\bibnamefont {Rozen}},
  \bibinfo {author} {\bibfnamefont {D.}~\bibnamefont {Rodan-Legrain}}, \bibinfo
  {author} {\bibfnamefont {Y.}~\bibnamefont {Cao}}, \bibinfo {author}
  {\bibfnamefont {R.}~\bibnamefont {Queiroz}}, \bibinfo {author} {\bibfnamefont
  {T.}~\bibnamefont {Taniguchi}}, \bibinfo {author} {\bibfnamefont
  {K.}~\bibnamefont {Watanabe}}, \bibinfo {author} {\bibfnamefont
  {Y.}~\bibnamefont {Oreg}}, \bibinfo {author} {\bibfnamefont {F.}~\bibnamefont
  {von Oppen}}, \bibinfo {author} {\bibfnamefont {A.}~\bibnamefont {Stern}}, \
  and\ \bibinfo {author} {\bibnamefont {et~al.}},\ }\bibfield  {title}
  {\enquote {\bibinfo {title} {Cascade of phase transitions and dirac revivals
  in magic-angle graphene},}\ }\href {\doibase 10.1038/s41586-020-2373-y}
  {\bibfield  {journal} {\bibinfo  {journal} {Nature}\ }\textbf {\bibinfo
  {volume} {582}},\ \bibinfo {pages} {203--208} (\bibinfo {year}
  {2020})}\BibitemShut {NoStop}%
\bibitem [{\citenamefont {Stepanov}\ \emph {et~al.}(2019)\citenamefont
  {Stepanov}, \citenamefont {Das}, \citenamefont {Lu}, \citenamefont
  {Fahimniya}, \citenamefont {Watanabe}, \citenamefont {Taniguchi},
  \citenamefont {Koppens}, \citenamefont {Lischner}, \citenamefont {Levitov},\
  and\ \citenamefont {Efetov}}]{Efetov2Interplay}%
  \BibitemOpen
  \bibfield  {author} {\bibinfo {author} {\bibfnamefont {P.}~\bibnamefont
  {Stepanov}}, \bibinfo {author} {\bibfnamefont {I.}~\bibnamefont {Das}},
  \bibinfo {author} {\bibfnamefont {X.}~\bibnamefont {Lu}}, \bibinfo {author}
  {\bibfnamefont {A.}~\bibnamefont {Fahimniya}}, \bibinfo {author}
  {\bibfnamefont {K.}~\bibnamefont {Watanabe}}, \bibinfo {author}
  {\bibfnamefont {T.}~\bibnamefont {Taniguchi}}, \bibinfo {author}
  {\bibfnamefont {F.~H.~L.}\ \bibnamefont {Koppens}}, \bibinfo {author}
  {\bibfnamefont {J.}~\bibnamefont {Lischner}}, \bibinfo {author}
  {\bibfnamefont {L.}~\bibnamefont {Levitov}}, \ and\ \bibinfo {author}
  {\bibfnamefont {D.~K.}\ \bibnamefont {Efetov}},\ }\bibfield  {title}
  {\enquote {\bibinfo {title} {The interplay of insulating and superconducting
  orders in magic-angle graphene bilayers},}\ }\href@noop {} {\  (\bibinfo
  {year} {2019})},\ \Eprint {http://arxiv.org/abs/1911.09198} {arXiv:1911.09198
  [cond-mat.supr-con]} \BibitemShut {NoStop}%
\bibitem [{\citenamefont {Regan}\ \emph {et~al.}(2020)\citenamefont {Regan},
  \citenamefont {Wang}, \citenamefont {Jin}, \citenamefont {Bakti~Utama},
  \citenamefont {Gao}, \citenamefont {Wei}, \citenamefont {Zhao}, \citenamefont
  {Zhao}, \citenamefont {Zhang}, \citenamefont {Yumigeta},\ and\ \citenamefont
  {et~al.}}]{ReganWSeWS}%
  \BibitemOpen
  \bibfield  {author} {\bibinfo {author} {\bibfnamefont {E.~C.}\ \bibnamefont
  {Regan}}, \bibinfo {author} {\bibfnamefont {D.}~\bibnamefont {Wang}},
  \bibinfo {author} {\bibfnamefont {C.}~\bibnamefont {Jin}}, \bibinfo {author}
  {\bibfnamefont {M.~I.}\ \bibnamefont {Bakti~Utama}}, \bibinfo {author}
  {\bibfnamefont {B.}~\bibnamefont {Gao}}, \bibinfo {author} {\bibfnamefont
  {X.}~\bibnamefont {Wei}}, \bibinfo {author} {\bibfnamefont {S.}~\bibnamefont
  {Zhao}}, \bibinfo {author} {\bibfnamefont {W.}~\bibnamefont {Zhao}}, \bibinfo
  {author} {\bibfnamefont {Z.}~\bibnamefont {Zhang}}, \bibinfo {author}
  {\bibfnamefont {K.}~\bibnamefont {Yumigeta}}, \ and\ \bibinfo {author}
  {\bibnamefont {et~al.}},\ }\bibfield  {title} {\enquote {\bibinfo {title}
  {{Mott and generalized Wigner crystal states in WSe2/WS2 moir\'e
  superlattices}},}\ }\href {\doibase 10.1038/s41586-020-2092-4} {\bibfield
  {journal} {\bibinfo  {journal} {Nature}\ }\textbf {\bibinfo {volume} {579}},\
  \bibinfo {pages} {359--363} (\bibinfo {year} {2020})}\BibitemShut {NoStop}%
\bibitem [{\citenamefont {Xu}\ \emph {et~al.}(2020)\citenamefont {Xu},
  \citenamefont {Liu}, \citenamefont {Rhodes}, \citenamefont {Watanabe},
  \citenamefont {Taniguchi}, \citenamefont {Hone}, \citenamefont {Elser},
  \citenamefont {Mak},\ and\ \citenamefont {Shan}}]{xu2020abundance}%
  \BibitemOpen
  \bibfield  {author} {\bibinfo {author} {\bibfnamefont {Y.}~\bibnamefont
  {Xu}}, \bibinfo {author} {\bibfnamefont {S.}~\bibnamefont {Liu}}, \bibinfo
  {author} {\bibfnamefont {D.~A.}\ \bibnamefont {Rhodes}}, \bibinfo {author}
  {\bibfnamefont {K.}~\bibnamefont {Watanabe}}, \bibinfo {author}
  {\bibfnamefont {T.}~\bibnamefont {Taniguchi}}, \bibinfo {author}
  {\bibfnamefont {J.}~\bibnamefont {Hone}}, \bibinfo {author} {\bibfnamefont
  {V.}~\bibnamefont {Elser}}, \bibinfo {author} {\bibfnamefont {K.~F.}\
  \bibnamefont {Mak}}, \ and\ \bibinfo {author} {\bibfnamefont
  {J.}~\bibnamefont {Shan}},\ }\bibfield  {title} {\enquote {\bibinfo {title}
  {Abundance of correlated insulating states at fractional fillings of
  wse$_{2}$/ws$_{2}$ moiré superlattices},}\ }\href
  {https://arxiv.org/abs/2007.11128} {\  (\bibinfo {year} {2020})},\ \Eprint
  {http://arxiv.org/abs/2007.11128} {arXiv:2007.11128 [cond-mat.str-el]}
  \BibitemShut {NoStop}%
\bibitem [{\citenamefont {Jin}\ \emph {et~al.}(2020)\citenamefont {Jin},
  \citenamefont {Tao}, \citenamefont {Li}, \citenamefont {Xu}, \citenamefont
  {Tang}, \citenamefont {Zhu}, \citenamefont {Liu}, \citenamefont {Watanabe},
  \citenamefont {Taniguchi}, \citenamefont {Hone}, \citenamefont {Fu},
  \citenamefont {Shan},\ and\ \citenamefont {Mak}}]{jin2020stripe}%
  \BibitemOpen
  \bibfield  {author} {\bibinfo {author} {\bibfnamefont {C.}~\bibnamefont
  {Jin}}, \bibinfo {author} {\bibfnamefont {Z.}~\bibnamefont {Tao}}, \bibinfo
  {author} {\bibfnamefont {T.}~\bibnamefont {Li}}, \bibinfo {author}
  {\bibfnamefont {Y.}~\bibnamefont {Xu}}, \bibinfo {author} {\bibfnamefont
  {Y.}~\bibnamefont {Tang}}, \bibinfo {author} {\bibfnamefont {J.}~\bibnamefont
  {Zhu}}, \bibinfo {author} {\bibfnamefont {S.}~\bibnamefont {Liu}}, \bibinfo
  {author} {\bibfnamefont {K.}~\bibnamefont {Watanabe}}, \bibinfo {author}
  {\bibfnamefont {T.}~\bibnamefont {Taniguchi}}, \bibinfo {author}
  {\bibfnamefont {J.~C.}\ \bibnamefont {Hone}}, \bibinfo {author}
  {\bibfnamefont {L.}~\bibnamefont {Fu}}, \bibinfo {author} {\bibfnamefont
  {J.}~\bibnamefont {Shan}}, \ and\ \bibinfo {author} {\bibfnamefont {K.~F.}\
  \bibnamefont {Mak}},\ }\bibfield  {title} {\enquote {\bibinfo {title} {Stripe
  phases in wse2/ws2 moiré superlattices},}\ }\href
  {https://arxiv.org/abs/2007.12068} {\  (\bibinfo {year} {2020})},\ \Eprint
  {http://arxiv.org/abs/2007.12068} {arXiv:2007.12068 [cond-mat.mes-hall]}
  \BibitemShut {NoStop}%
\bibitem [{\citenamefont {Shimazaki}\ \emph
  {et~al.}(2020{\natexlab{a}})\citenamefont {Shimazaki}, \citenamefont
  {Schwartz}, \citenamefont {Watanabe}, \citenamefont {Taniguchi},
  \citenamefont {Kroner},\ and\ \citenamefont {Imamo\u{g}lu}}]{ZurichSpacer}%
  \BibitemOpen
  \bibfield  {author} {\bibinfo {author} {\bibfnamefont {Y.}~\bibnamefont
  {Shimazaki}}, \bibinfo {author} {\bibfnamefont {I.}~\bibnamefont {Schwartz}},
  \bibinfo {author} {\bibfnamefont {K.}~\bibnamefont {Watanabe}}, \bibinfo
  {author} {\bibfnamefont {T.}~\bibnamefont {Taniguchi}}, \bibinfo {author}
  {\bibfnamefont {M.}~\bibnamefont {Kroner}}, \ and\ \bibinfo {author}
  {\bibfnamefont {A.}~\bibnamefont {Imamo\u{g}lu}},\ }\bibfield  {title}
  {\enquote {\bibinfo {title} {Strongly correlated electrons and hybrid
  excitons in a moir\'e heterostructure},}\ }\href {\doibase
  10.1038/s41586-020-2191-2} {\bibfield  {journal} {\bibinfo  {journal}
  {Nature}\ }\textbf {\bibinfo {volume} {580}},\ \bibinfo {pages} {472--477}
  (\bibinfo {year} {2020}{\natexlab{a}})}\BibitemShut {NoStop}%
\bibitem [{\citenamefont {Tang}\ \emph {et~al.}(2020)\citenamefont {Tang},
  \citenamefont {Li}, \citenamefont {Li}, \citenamefont {Xu}, \citenamefont
  {Liu}, \citenamefont {Barmak}, \citenamefont {Watanabe}, \citenamefont
  {Taniguchi}, \citenamefont {MacDonald}, \citenamefont {Shan} \emph
  {et~al.}}]{Allan2020simulation}%
  \BibitemOpen
  \bibfield  {author} {\bibinfo {author} {\bibfnamefont {Y.}~\bibnamefont
  {Tang}}, \bibinfo {author} {\bibfnamefont {L.}~\bibnamefont {Li}}, \bibinfo
  {author} {\bibfnamefont {T.}~\bibnamefont {Li}}, \bibinfo {author}
  {\bibfnamefont {Y.}~\bibnamefont {Xu}}, \bibinfo {author} {\bibfnamefont
  {S.}~\bibnamefont {Liu}}, \bibinfo {author} {\bibfnamefont {K.}~\bibnamefont
  {Barmak}}, \bibinfo {author} {\bibfnamefont {K.}~\bibnamefont {Watanabe}},
  \bibinfo {author} {\bibfnamefont {T.}~\bibnamefont {Taniguchi}}, \bibinfo
  {author} {\bibfnamefont {A.~H.}\ \bibnamefont {MacDonald}}, \bibinfo {author}
  {\bibfnamefont {J.}~\bibnamefont {Shan}},  \emph {et~al.},\ }\bibfield
  {title} {\enquote {\bibinfo {title} {Simulation of hubbard model physics in
  wse 2/ws 2 moir{\'e} superlattices},}\ }\href
  {https://www.nature.com/articles/s41586-020-2085-3} {\bibfield  {journal}
  {\bibinfo  {journal} {Nature}\ }\textbf {\bibinfo {volume} {579}},\ \bibinfo
  {pages} {353--358} (\bibinfo {year} {2020})}\BibitemShut {NoStop}%
\bibitem [{\citenamefont {Wu}\ \emph {et~al.}(2018)\citenamefont {Wu},
  \citenamefont {Lovorn}, \citenamefont {Tutuc},\ and\ \citenamefont
  {MacDonald}}]{TMDFengcheng}%
  \BibitemOpen
  \bibfield  {author} {\bibinfo {author} {\bibfnamefont {F.}~\bibnamefont
  {Wu}}, \bibinfo {author} {\bibfnamefont {T.}~\bibnamefont {Lovorn}}, \bibinfo
  {author} {\bibfnamefont {E.}~\bibnamefont {Tutuc}}, \ and\ \bibinfo {author}
  {\bibfnamefont {A.~H.}\ \bibnamefont {MacDonald}},\ }\bibfield  {title}
  {\enquote {\bibinfo {title} {Hubbard model physics in transition metal
  dichalcogenide moir\'e bands},}\ }\href {\doibase
  10.1103/PhysRevLett.121.026402} {\bibfield  {journal} {\bibinfo  {journal}
  {Phys. Rev. Lett.}\ }\textbf {\bibinfo {volume} {121}},\ \bibinfo {pages}
  {026402} (\bibinfo {year} {2018})}\BibitemShut {NoStop}%
\bibitem [{\citenamefont {Wu}\ \emph {et~al.}(2019)\citenamefont {Wu},
  \citenamefont {Lovorn}, \citenamefont {Tutuc}, \citenamefont {Martin},\ and\
  \citenamefont {MacDonald}}]{BSmoireTMD}%
  \BibitemOpen
  \bibfield  {author} {\bibinfo {author} {\bibfnamefont {F.}~\bibnamefont
  {Wu}}, \bibinfo {author} {\bibfnamefont {T.}~\bibnamefont {Lovorn}}, \bibinfo
  {author} {\bibfnamefont {E.}~\bibnamefont {Tutuc}}, \bibinfo {author}
  {\bibfnamefont {I.}~\bibnamefont {Martin}}, \ and\ \bibinfo {author}
  {\bibfnamefont {A.~H.}\ \bibnamefont {MacDonald}},\ }\bibfield  {title}
  {\enquote {\bibinfo {title} {Topological insulators in twisted transition
  metal dichalcogenide homobilayers},}\ }\href {\doibase
  10.1103/PhysRevLett.122.086402} {\bibfield  {journal} {\bibinfo  {journal}
  {Phys. Rev. Lett.}\ }\textbf {\bibinfo {volume} {122}},\ \bibinfo {pages}
  {086402} (\bibinfo {year} {2019})}\BibitemShut {NoStop}%
\bibitem [{\citenamefont {Bistritzer}\ and\ \citenamefont
  {MacDonald}(2011)}]{MCDMoire}%
  \BibitemOpen
  \bibfield  {author} {\bibinfo {author} {\bibfnamefont {R.}~\bibnamefont
  {Bistritzer}}\ and\ \bibinfo {author} {\bibfnamefont {A.~H.}\ \bibnamefont
  {MacDonald}},\ }\bibfield  {title} {\enquote {\bibinfo {title} {Moir{\'e}
  bands in twisted double-layer graphene},}\ }\href
  {http://www.pnas.org/content/108/30/12233} {\bibfield  {journal} {\bibinfo
  {journal} {Proceedings of the National Academy of Sciences}\ }\textbf
  {\bibinfo {volume} {108}},\ \bibinfo {pages} {12233--12237} (\bibinfo {year}
  {2011})}\BibitemShut {NoStop}%
\bibitem [{\citenamefont {Wang}\ \emph {et~al.}(2019)\citenamefont {Wang},
  \citenamefont {Shih}, \citenamefont {Ghiotto}, \citenamefont {Xian},
  \citenamefont {Rhodes}, \citenamefont {Tan}, \citenamefont {Claassen},
  \citenamefont {Kennes}, \citenamefont {Bai}, \citenamefont {Kim},
  \citenamefont {Watanabe}, \citenamefont {Taniguchi}, \citenamefont {Zhu},
  \citenamefont {Hone}, \citenamefont {Rubio}, \citenamefont {Pasupathy},\ and\
  \citenamefont {Dean}}]{wang2019magic}%
  \BibitemOpen
  \bibfield  {author} {\bibinfo {author} {\bibfnamefont {L.}~\bibnamefont
  {Wang}}, \bibinfo {author} {\bibfnamefont {E.-M.}\ \bibnamefont {Shih}},
  \bibinfo {author} {\bibfnamefont {A.}~\bibnamefont {Ghiotto}}, \bibinfo
  {author} {\bibfnamefont {L.}~\bibnamefont {Xian}}, \bibinfo {author}
  {\bibfnamefont {D.~A.}\ \bibnamefont {Rhodes}}, \bibinfo {author}
  {\bibfnamefont {C.}~\bibnamefont {Tan}}, \bibinfo {author} {\bibfnamefont
  {M.}~\bibnamefont {Claassen}}, \bibinfo {author} {\bibfnamefont {D.~M.}\
  \bibnamefont {Kennes}}, \bibinfo {author} {\bibfnamefont {Y.}~\bibnamefont
  {Bai}}, \bibinfo {author} {\bibfnamefont {B.}~\bibnamefont {Kim}}, \bibinfo
  {author} {\bibfnamefont {K.}~\bibnamefont {Watanabe}}, \bibinfo {author}
  {\bibfnamefont {T.}~\bibnamefont {Taniguchi}}, \bibinfo {author}
  {\bibfnamefont {X.}~\bibnamefont {Zhu}}, \bibinfo {author} {\bibfnamefont
  {J.}~\bibnamefont {Hone}}, \bibinfo {author} {\bibfnamefont {A.}~\bibnamefont
  {Rubio}}, \bibinfo {author} {\bibfnamefont {A.}~\bibnamefont {Pasupathy}}, \
  and\ \bibinfo {author} {\bibfnamefont {C.~R.}\ \bibnamefont {Dean}},\ }\href
  {https://arxiv.org/abs/1910.12147} {\enquote {\bibinfo {title} {Magic
  continuum in twisted bilayer wse2},}\ } (\bibinfo {year} {2019}),\ \Eprint
  {http://arxiv.org/abs/1910.12147} {arXiv:1910.12147 [cond-mat.mes-hall]}
  \BibitemShut {NoStop}%
\bibitem [{\citenamefont {Eisenstein}\ \emph {et~al.}(1992)\citenamefont
  {Eisenstein}, \citenamefont {Pfeiffer},\ and\ \citenamefont
  {West}}]{Eisenstein92}%
  \BibitemOpen
  \bibfield  {author} {\bibinfo {author} {\bibfnamefont {J.~P.}\ \bibnamefont
  {Eisenstein}}, \bibinfo {author} {\bibfnamefont {L.~N.}\ \bibnamefont
  {Pfeiffer}}, \ and\ \bibinfo {author} {\bibfnamefont {K.~W.}\ \bibnamefont
  {West}},\ }\bibfield  {title} {\enquote {\bibinfo {title} {Negative
  compressibility of interacting two-dimensional electron and quasiparticle
  gases},}\ }\href {\doibase 10.1103/PhysRevLett.68.674} {\bibfield  {journal}
  {\bibinfo  {journal} {Phys. Rev. Lett.}\ }\textbf {\bibinfo {volume} {68}},\
  \bibinfo {pages} {674--677} (\bibinfo {year} {1992})}\BibitemShut {NoStop}%
\bibitem [{\citenamefont {Li}\ \emph {et~al.}(2011)\citenamefont {Li},
  \citenamefont {Richter}, \citenamefont {Paetel}, \citenamefont {Kopp},
  \citenamefont {Mannhart},\ and\ \citenamefont {Ashoori}}]{LuLi11}%
  \BibitemOpen
  \bibfield  {author} {\bibinfo {author} {\bibfnamefont {L.}~\bibnamefont
  {Li}}, \bibinfo {author} {\bibfnamefont {C.}~\bibnamefont {Richter}},
  \bibinfo {author} {\bibfnamefont {S.}~\bibnamefont {Paetel}}, \bibinfo
  {author} {\bibfnamefont {T.}~\bibnamefont {Kopp}}, \bibinfo {author}
  {\bibfnamefont {J.}~\bibnamefont {Mannhart}}, \ and\ \bibinfo {author}
  {\bibfnamefont {R.}~\bibnamefont {Ashoori}},\ }\bibfield  {title} {\enquote
  {\bibinfo {title} {Very large capacitance enhancement in a two-dimensional
  electron system},}\ }\href@noop {} {\bibfield  {journal} {\bibinfo  {journal}
  {Science}\ }\textbf {\bibinfo {volume} {332}},\ \bibinfo {pages} {825--828}
  (\bibinfo {year} {2011})}\BibitemShut {NoStop}%
\bibitem [{\citenamefont {Skinner}\ and\ \citenamefont
  {Shklovskii}(2010)}]{SkinnerCapacitance}%
  \BibitemOpen
  \bibfield  {author} {\bibinfo {author} {\bibfnamefont {B.}~\bibnamefont
  {Skinner}}\ and\ \bibinfo {author} {\bibfnamefont {B.~I.}\ \bibnamefont
  {Shklovskii}},\ }\bibfield  {title} {\enquote {\bibinfo {title} {Anomalously
  large capacitance of a plane capacitor with a two-dimensional electron
  gas},}\ }\href {\doibase 10.1103/PhysRevB.82.155111} {\bibfield  {journal}
  {\bibinfo  {journal} {Phys. Rev. B}\ }\textbf {\bibinfo {volume} {82}},\
  \bibinfo {pages} {155111} (\bibinfo {year} {2010})}\BibitemShut {NoStop}%
\bibitem [{\citenamefont {Bello}\ \emph {et~al.}(1981)\citenamefont {Bello},
  \citenamefont {Levin}, \citenamefont {Shklovskii},\ and\ \citenamefont
  {Efros}}]{bello1981density}%
  \BibitemOpen
  \bibfield  {author} {\bibinfo {author} {\bibfnamefont {M.}~\bibnamefont
  {Bello}}, \bibinfo {author} {\bibfnamefont {E.}~\bibnamefont {Levin}},
  \bibinfo {author} {\bibfnamefont {B.}~\bibnamefont {Shklovskii}}, \ and\
  \bibinfo {author} {\bibfnamefont {A.}~\bibnamefont {Efros}},\ }\bibfield
  {title} {\enquote {\bibinfo {title} {Density of localized states in the
  surface impurity band of a metal--insulator--semiconductor structure},}\
  }\href {http://www.jetp.ac.ru/cgi-bin/e/index/e/53/4/p822?a=list} {\bibfield
  {journal} {\bibinfo  {journal} {Sov. Phys. JETP,}\ }\textbf {\bibinfo
  {volume} {80}},\ \bibinfo {pages} {822--829} (\bibinfo {year}
  {1981})}\BibitemShut {NoStop}%
\bibitem [{\citenamefont {Chitra}\ \emph {et~al.}(2001)\citenamefont {Chitra},
  \citenamefont {Giamarchi},\ and\ \citenamefont {Le~Doussal}}]{ChitraPRB01}%
  \BibitemOpen
  \bibfield  {author} {\bibinfo {author} {\bibfnamefont {R.}~\bibnamefont
  {Chitra}}, \bibinfo {author} {\bibfnamefont {T.}~\bibnamefont {Giamarchi}}, \
  and\ \bibinfo {author} {\bibfnamefont {P.}~\bibnamefont {Le~Doussal}},\
  }\bibfield  {title} {\enquote {\bibinfo {title} {Pinned wigner crystals},}\
  }\href {\doibase 10.1103/PhysRevB.65.035312} {\bibfield  {journal} {\bibinfo
  {journal} {Phys. Rev. B}\ }\textbf {\bibinfo {volume} {65}},\ \bibinfo
  {pages} {035312} (\bibinfo {year} {2001})}\BibitemShut {NoStop}%
\bibitem [{\citenamefont {Padhi}\ \emph {et~al.}(2018)\citenamefont {Padhi},
  \citenamefont {Setty},\ and\ \citenamefont {Phillips}}]{OurPaper1}%
  \BibitemOpen
  \bibfield  {author} {\bibinfo {author} {\bibfnamefont {B.}~\bibnamefont
  {Padhi}}, \bibinfo {author} {\bibfnamefont {C.}~\bibnamefont {Setty}}, \ and\
  \bibinfo {author} {\bibfnamefont {P.~W.}\ \bibnamefont {Phillips}},\
  }\bibfield  {title} {\enquote {\bibinfo {title} {Doped twisted bilayer
  graphene near magic angles: Proximity to wigner crystallization, not mott
  insulation},}\ }\href {\doibase 10.1021/acs.nanolett.8b02033} {\bibfield
  {journal} {\bibinfo  {journal} {Nano Letters}\ }\textbf {\bibinfo {volume}
  {18}},\ \bibinfo {pages} {6175--6180} (\bibinfo {year} {2018})}\BibitemShut
  {NoStop}%
\bibitem [{\citenamefont {Padhi}\ and\ \citenamefont
  {Phillips}(2019)}]{OurPaper2}%
  \BibitemOpen
  \bibfield  {author} {\bibinfo {author} {\bibfnamefont {B.}~\bibnamefont
  {Padhi}}\ and\ \bibinfo {author} {\bibfnamefont {P.~W.}\ \bibnamefont
  {Phillips}},\ }\bibfield  {title} {\enquote {\bibinfo {title}
  {Pressure-induced metal-insulator transition in twisted bilayer graphene},}\
  }\href {\doibase 10.1103/PhysRevB.99.205141} {\bibfield  {journal} {\bibinfo
  {journal} {Phys. Rev. B}\ }\textbf {\bibinfo {volume} {99}},\ \bibinfo
  {pages} {205141} (\bibinfo {year} {2019})}\BibitemShut {NoStop}%
\bibitem [{\citenamefont {Hubbard}(1978)}]{HubbardGWC}%
  \BibitemOpen
  \bibfield  {author} {\bibinfo {author} {\bibfnamefont {J.}~\bibnamefont
  {Hubbard}},\ }\bibfield  {title} {\enquote {\bibinfo {title} {Generalized
  wigner lattices in one dimension and some applications to
  tetracyanoquinodimethane (tcnq) salts},}\ }\href {\doibase
  10.1103/PhysRevB.17.494} {\bibfield  {journal} {\bibinfo  {journal} {Phys.
  Rev. B}\ }\textbf {\bibinfo {volume} {17}},\ \bibinfo {pages} {494--505}
  (\bibinfo {year} {1978})}\BibitemShut {NoStop}%
\bibitem [{\citenamefont {Chen}\ \emph {et~al.}(2003)\citenamefont {Chen},
  \citenamefont {Lewis}, \citenamefont {Engel}, \citenamefont {Tsui},
  \citenamefont {Ye}, \citenamefont {Pfeiffer},\ and\ \citenamefont
  {West}}]{MWRChen03}%
  \BibitemOpen
  \bibfield  {author} {\bibinfo {author} {\bibfnamefont {Y.}~\bibnamefont
  {Chen}}, \bibinfo {author} {\bibfnamefont {R.~M.}\ \bibnamefont {Lewis}},
  \bibinfo {author} {\bibfnamefont {L.~W.}\ \bibnamefont {Engel}}, \bibinfo
  {author} {\bibfnamefont {D.~C.}\ \bibnamefont {Tsui}}, \bibinfo {author}
  {\bibfnamefont {P.~D.}\ \bibnamefont {Ye}}, \bibinfo {author} {\bibfnamefont
  {L.~N.}\ \bibnamefont {Pfeiffer}}, \ and\ \bibinfo {author} {\bibfnamefont
  {K.~W.}\ \bibnamefont {West}},\ }\bibfield  {title} {\enquote {\bibinfo
  {title} {Microwave resonance of the 2d wigner crystal around integer landau
  fillings},}\ }\href {\doibase 10.1103/PhysRevLett.91.016801} {\bibfield
  {journal} {\bibinfo  {journal} {Phys. Rev. Lett.}\ }\textbf {\bibinfo
  {volume} {91}},\ \bibinfo {pages} {016801} (\bibinfo {year}
  {2003})}\BibitemShut {NoStop}%
\bibitem [{\citenamefont {{Jang}}\ \emph {et~al.}(2017)\citenamefont {{Jang}},
  \citenamefont {{Hunt}}, \citenamefont {{Pfeiffer}}, \citenamefont {{West}},\
  and\ \citenamefont {{Ashoori}}}]{AshooriTunnelling17}%
  \BibitemOpen
  \bibfield  {author} {\bibinfo {author} {\bibfnamefont {J.}~\bibnamefont
  {{Jang}}}, \bibinfo {author} {\bibfnamefont {B.~M.}\ \bibnamefont {{Hunt}}},
  \bibinfo {author} {\bibfnamefont {L.~N.}\ \bibnamefont {{Pfeiffer}}},
  \bibinfo {author} {\bibfnamefont {K.~W.}\ \bibnamefont {{West}}}, \ and\
  \bibinfo {author} {\bibfnamefont {R.~C.}\ \bibnamefont {{Ashoori}}},\
  }\bibfield  {title} {\enquote {\bibinfo {title} {{Sharp tunneling resonance
  from the vibrations of an electronic {Wigner crystal}}},}\ }\href {\doibase
  10.1038/nphys3979} {\bibfield  {journal} {\bibinfo  {journal} {Nature
  Physics}\ }\textbf {\bibinfo {volume} {13}},\ \bibinfo {pages} {340--344}
  (\bibinfo {year} {2017})},\ \Eprint {http://arxiv.org/abs/1604.06220}
  {arXiv:1604.06220 [cond-mat.str-el]} \BibitemShut {NoStop}%
\bibitem [{\citenamefont {Hatke}\ \emph {et~al.}(2015)\citenamefont {Hatke},
  \citenamefont {Liu}, \citenamefont {Engel}, \citenamefont {Shayegan},
  \citenamefont {Pfeiffer}, \citenamefont {West},\ and\ \citenamefont
  {Baldwin}}]{hatke2015microwave}%
  \BibitemOpen
  \bibfield  {author} {\bibinfo {author} {\bibfnamefont {A.}~\bibnamefont
  {Hatke}}, \bibinfo {author} {\bibfnamefont {Y.}~\bibnamefont {Liu}}, \bibinfo
  {author} {\bibfnamefont {L.}~\bibnamefont {Engel}}, \bibinfo {author}
  {\bibfnamefont {M.}~\bibnamefont {Shayegan}}, \bibinfo {author}
  {\bibfnamefont {L.}~\bibnamefont {Pfeiffer}}, \bibinfo {author}
  {\bibfnamefont {K.}~\bibnamefont {West}}, \ and\ \bibinfo {author}
  {\bibfnamefont {K.}~\bibnamefont {Baldwin}},\ }\bibfield  {title} {\enquote
  {\bibinfo {title} {Microwave spectroscopy of the low-filling-factor bilayer
  electron solid in a wide quantum well},}\ }\href
  {https://www.nature.com/articles/ncomms8071} {\bibfield  {journal} {\bibinfo
  {journal} {Nature communications}\ }\textbf {\bibinfo {volume} {6}},\
  \bibinfo {pages} {1--6} (\bibinfo {year} {2015})}\BibitemShut {NoStop}%
\bibitem [{\citenamefont {Monceau}(2012)}]{Monceau12}%
  \BibitemOpen
  \bibfield  {author} {\bibinfo {author} {\bibfnamefont {P.}~\bibnamefont
  {Monceau}},\ }\bibfield  {title} {\enquote {\bibinfo {title} {Electronic
  crystals: an experimental overview},}\ }\href {\doibase
  10.1080/00018732.2012.719674} {\bibfield  {journal} {\bibinfo  {journal}
  {Advances in Physics}\ }\textbf {\bibinfo {volume} {61}},\ \bibinfo {pages}
  {325--581} (\bibinfo {year} {2012})}\BibitemShut {NoStop}%
\bibitem [{\citenamefont {Delacr\'etaz}\ \emph {et~al.}(2019)\citenamefont
  {Delacr\'etaz}, \citenamefont {Gout\'eraux}, \citenamefont {Hartnoll},\ and\
  \citenamefont {Karlsson}}]{Delacretaz19}%
  \BibitemOpen
  \bibfield  {author} {\bibinfo {author} {\bibfnamefont {L.~V.}\ \bibnamefont
  {Delacr\'etaz}}, \bibinfo {author} {\bibfnamefont {B.}~\bibnamefont
  {Gout\'eraux}}, \bibinfo {author} {\bibfnamefont {S.~A.}\ \bibnamefont
  {Hartnoll}}, \ and\ \bibinfo {author} {\bibfnamefont {A.}~\bibnamefont
  {Karlsson}},\ }\bibfield  {title} {\enquote {\bibinfo {title} {Theory of
  collective magnetophonon resonance and melting of a field-induced wigner
  solid},}\ }\href {\doibase 10.1103/PhysRevB.100.085140} {\bibfield  {journal}
  {\bibinfo  {journal} {Phys. Rev. B}\ }\textbf {\bibinfo {volume} {100}},\
  \bibinfo {pages} {085140} (\bibinfo {year} {2019})}\BibitemShut {NoStop}%
\bibitem [{\citenamefont {Shimazaki}\ \emph
  {et~al.}(2020{\natexlab{b}})\citenamefont {Shimazaki}, \citenamefont
  {Kuhlenkamp}, \citenamefont {Schwartz}, \citenamefont {Smolenski},
  \citenamefont {Watanabe}, \citenamefont {Taniguchi}, \citenamefont {Kroner},
  \citenamefont {Schmidt}, \citenamefont {Knap},\ and\ \citenamefont
  {Imamoglu}}]{shimazaki2020optical}%
  \BibitemOpen
  \bibfield  {author} {\bibinfo {author} {\bibfnamefont {Y.}~\bibnamefont
  {Shimazaki}}, \bibinfo {author} {\bibfnamefont {C.}~\bibnamefont
  {Kuhlenkamp}}, \bibinfo {author} {\bibfnamefont {I.}~\bibnamefont
  {Schwartz}}, \bibinfo {author} {\bibfnamefont {T.}~\bibnamefont {Smolenski}},
  \bibinfo {author} {\bibfnamefont {K.}~\bibnamefont {Watanabe}}, \bibinfo
  {author} {\bibfnamefont {T.}~\bibnamefont {Taniguchi}}, \bibinfo {author}
  {\bibfnamefont {M.}~\bibnamefont {Kroner}}, \bibinfo {author} {\bibfnamefont
  {R.}~\bibnamefont {Schmidt}}, \bibinfo {author} {\bibfnamefont
  {M.}~\bibnamefont {Knap}}, \ and\ \bibinfo {author} {\bibfnamefont
  {A.}~\bibnamefont {Imamoglu}},\ }\href {https://arxiv.org/abs/2008.04156}
  {\enquote {\bibinfo {title} {Optical signatures of charge order in a
  mott-wigner state},}\ } (\bibinfo {year} {2020}{\natexlab{b}}),\ \Eprint
  {http://arxiv.org/abs/2008.04156} {arXiv:2008.04156 [cond-mat.mes-hall]}
  \BibitemShut {NoStop}%
\bibitem [{\citenamefont {Noda}\ and\ \citenamefont {Imada}(2002)}]{ImadaPRL}%
  \BibitemOpen
  \bibfield  {author} {\bibinfo {author} {\bibfnamefont {Y.}~\bibnamefont
  {Noda}}\ and\ \bibinfo {author} {\bibfnamefont {M.}~\bibnamefont {Imada}},\
  }\bibfield  {title} {\enquote {\bibinfo {title} {Quantum phase transitions to
  charge-ordered and wigner-crystal states under the interplay of lattice
  commensurability and long-range coulomb interactions},}\ }\href {\doibase
  10.1103/PhysRevLett.89.176803} {\bibfield  {journal} {\bibinfo  {journal}
  {Phys. Rev. Lett.}\ }\textbf {\bibinfo {volume} {89}},\ \bibinfo {pages}
  {176803} (\bibinfo {year} {2002})}\BibitemShut {NoStop}%
\bibitem [{\citenamefont {Slagle}\ and\ \citenamefont
  {Fu}(2020)}]{slagle2020charge}%
  \BibitemOpen
  \bibfield  {author} {\bibinfo {author} {\bibfnamefont {K.}~\bibnamefont
  {Slagle}}\ and\ \bibinfo {author} {\bibfnamefont {L.}~\bibnamefont {Fu}},\
  }\href@noop {} {\enquote {\bibinfo {title} {{Charge Transfer Excitations,
  Pair Density Waves, and Superconductivity in Moiré Materials}},}\ } (\bibinfo
  {year} {2020}),\ \Eprint {http://arxiv.org/abs/2003.13690} {arXiv:2003.13690
  [cond-mat.str-el]} \BibitemShut {NoStop}%
\bibitem [{\citenamefont {Pan}\ \emph {et~al.}(2020)\citenamefont {Pan},
  \citenamefont {Wu},\ and\ \citenamefont {Sarma}}]{pan2020quantum}%
  \BibitemOpen
  \bibfield  {author} {\bibinfo {author} {\bibfnamefont {H.}~\bibnamefont
  {Pan}}, \bibinfo {author} {\bibfnamefont {F.}~\bibnamefont {Wu}}, \ and\
  \bibinfo {author} {\bibfnamefont {S.~D.}\ \bibnamefont {Sarma}},\ }\href
  {https://arxiv.org/abs/2008.08998} {\enquote {\bibinfo {title} {Quantum phase
  diagram of a moir\'e-hubbard model},}\ } (\bibinfo {year} {2020}),\ \Eprint
  {http://arxiv.org/abs/2008.08998} {arXiv:2008.08998 [cond-mat.str-el]}
  \BibitemShut {NoStop}%
\bibitem [{\citenamefont {Tomarken}\ \emph {et~al.}(2019)\citenamefont
  {Tomarken}, \citenamefont {Cao}, \citenamefont {Demir}, \citenamefont
  {Watanabe}, \citenamefont {Taniguchi}, \citenamefont {Jarillo-Herrero},\ and\
  \citenamefont {Ashoori}}]{Tomarken2019}%
  \BibitemOpen
  \bibfield  {author} {\bibinfo {author} {\bibfnamefont {S.}~\bibnamefont
  {Tomarken}}, \bibinfo {author} {\bibfnamefont {Y.}~\bibnamefont {Cao}},
  \bibinfo {author} {\bibfnamefont {A.}~\bibnamefont {Demir}}, \bibinfo
  {author} {\bibfnamefont {K.}~\bibnamefont {Watanabe}}, \bibinfo {author}
  {\bibfnamefont {T.}~\bibnamefont {Taniguchi}}, \bibinfo {author}
  {\bibfnamefont {P.}~\bibnamefont {Jarillo-Herrero}}, \ and\ \bibinfo {author}
  {\bibfnamefont {R.}~\bibnamefont {Ashoori}},\ }\bibfield  {title} {\enquote
  {\bibinfo {title} {Electronic compressibility of magic-angle graphene
  superlattices},}\ }\href {\doibase 10.1103/physrevlett.123.046601} {\bibfield
   {journal} {\bibinfo  {journal} {Physical Review Letters}\ }\textbf {\bibinfo
  {volume} {123}} (\bibinfo {year} {2019}),\
  10.1103/physrevlett.123.046601}\BibitemShut {NoStop}%
\bibitem [{\citenamefont {{Camjayi}}\ \emph {et~al.}(2008)\citenamefont
  {{Camjayi}}, \citenamefont {{Haule}}, \citenamefont {{Dobrosavljevi{\'c}}},\
  and\ \citenamefont {{Kotliar}}}]{GabiWCMott-Nature}%
  \BibitemOpen
  \bibfield  {author} {\bibinfo {author} {\bibfnamefont {A.}~\bibnamefont
  {{Camjayi}}}, \bibinfo {author} {\bibfnamefont {K.}~\bibnamefont {{Haule}}},
  \bibinfo {author} {\bibfnamefont {V.}~\bibnamefont {{Dobrosavljevi{\'c}}}}, \
  and\ \bibinfo {author} {\bibfnamefont {G.}~\bibnamefont {{Kotliar}}},\
  }\bibfield  {title} {\enquote {\bibinfo {title} {{Coulomb correlations and
  the Wigner-Mott transition}},}\ }\href {\doibase 10.1038/nphys1106}
  {\bibfield  {journal} {\bibinfo  {journal} {Nature Physics}\ }\textbf
  {\bibinfo {volume} {4}},\ \bibinfo {pages} {932--935} (\bibinfo {year}
  {2008})}\BibitemShut {NoStop}%
\bibitem [{\citenamefont {Eisenstein}\ \emph {et~al.}(1994)\citenamefont
  {Eisenstein}, \citenamefont {Pfeiffer},\ and\ \citenamefont
  {West}}]{Eisenstein94}%
  \BibitemOpen
  \bibfield  {author} {\bibinfo {author} {\bibfnamefont {J.~P.}\ \bibnamefont
  {Eisenstein}}, \bibinfo {author} {\bibfnamefont {L.~N.}\ \bibnamefont
  {Pfeiffer}}, \ and\ \bibinfo {author} {\bibfnamefont {K.~W.}\ \bibnamefont
  {West}},\ }\bibfield  {title} {\enquote {\bibinfo {title} {Compressibility of
  the two-dimensional electron gas: Measurements of the zero-field exchange
  energy and fractional quantum hall gap},}\ }\href {\doibase
  10.1103/PhysRevB.50.1760} {\bibfield  {journal} {\bibinfo  {journal} {Phys.
  Rev. B}\ }\textbf {\bibinfo {volume} {50}},\ \bibinfo {pages} {1760--1778}
  (\bibinfo {year} {1994})}\BibitemShut {NoStop}%
\bibitem [{\citenamefont {Zhang}\ \emph {et~al.}(2014)\citenamefont {Zhang},
  \citenamefont {Huang}, \citenamefont {Dietsche}, \citenamefont {von
  Klitzing},\ and\ \citenamefont {Smet}}]{QH-WCPuddle14}%
  \BibitemOpen
  \bibfield  {author} {\bibinfo {author} {\bibfnamefont {D.}~\bibnamefont
  {Zhang}}, \bibinfo {author} {\bibfnamefont {X.}~\bibnamefont {Huang}},
  \bibinfo {author} {\bibfnamefont {W.}~\bibnamefont {Dietsche}}, \bibinfo
  {author} {\bibfnamefont {K.}~\bibnamefont {von Klitzing}}, \ and\ \bibinfo
  {author} {\bibfnamefont {J.~H.}\ \bibnamefont {Smet}},\ }\bibfield  {title}
  {\enquote {\bibinfo {title} {Signatures for wigner crystal formation in the
  chemical potential of a two-dimensional electron system},}\ }\href {\doibase
  10.1103/PhysRevLett.113.076804} {\bibfield  {journal} {\bibinfo  {journal}
  {Phys. Rev. Lett.}\ }\textbf {\bibinfo {volume} {113}},\ \bibinfo {pages}
  {076804} (\bibinfo {year} {2014})}\BibitemShut {NoStop}%
\bibitem [{\citenamefont {Giamarchi}\ and\ \citenamefont
  {Le~Doussal}(1995)}]{FluxLattice95}%
  \BibitemOpen
  \bibfield  {author} {\bibinfo {author} {\bibfnamefont {T.}~\bibnamefont
  {Giamarchi}}\ and\ \bibinfo {author} {\bibfnamefont {P.}~\bibnamefont
  {Le~Doussal}},\ }\bibfield  {title} {\enquote {\bibinfo {title} {Elastic
  theory of flux lattices in the presence of weak disorder},}\ }\href {\doibase
  10.1103/PhysRevB.52.1242} {\bibfield  {journal} {\bibinfo  {journal} {Phys.
  Rev. B}\ }\textbf {\bibinfo {volume} {52}},\ \bibinfo {pages} {1242--1270}
  (\bibinfo {year} {1995})}\BibitemShut {NoStop}%
\bibitem [{\citenamefont {Giamarchi}\ and\ \citenamefont
  {Le~Doussal}(1997)}]{FluxLattice96}%
  \BibitemOpen
  \bibfield  {author} {\bibinfo {author} {\bibfnamefont {T.}~\bibnamefont
  {Giamarchi}}\ and\ \bibinfo {author} {\bibfnamefont {P.}~\bibnamefont
  {Le~Doussal}},\ }\bibfield  {title} {\enquote {\bibinfo {title} {Phase
  diagrams of flux lattices with disorder},}\ }\href {\doibase
  10.1103/PhysRevB.55.6577} {\bibfield  {journal} {\bibinfo  {journal} {Phys.
  Rev. B}\ }\textbf {\bibinfo {volume} {55}},\ \bibinfo {pages} {6577--6583}
  (\bibinfo {year} {1997})}\BibitemShut {NoStop}%
\bibitem [{\citenamefont {Cudazzo}\ \emph {et~al.}(2011)\citenamefont
  {Cudazzo}, \citenamefont {Tokatly},\ and\ \citenamefont
  {Rubio}}]{Rubio-RKpotential}%
  \BibitemOpen
  \bibfield  {author} {\bibinfo {author} {\bibfnamefont {P.}~\bibnamefont
  {Cudazzo}}, \bibinfo {author} {\bibfnamefont {I.~V.}\ \bibnamefont
  {Tokatly}}, \ and\ \bibinfo {author} {\bibfnamefont {A.}~\bibnamefont
  {Rubio}},\ }\bibfield  {title} {\enquote {\bibinfo {title} {Dielectric
  screening in two-dimensional insulators: Implications for excitonic and
  impurity states in graphane},}\ }\href {\doibase 10.1103/PhysRevB.84.085406}
  {\bibfield  {journal} {\bibinfo  {journal} {Phys. Rev. B}\ }\textbf {\bibinfo
  {volume} {84}},\ \bibinfo {pages} {085406} (\bibinfo {year}
  {2011})}\BibitemShut {NoStop}%
\bibitem [{\citenamefont {Danovich}\ \emph {et~al.}(2018)\citenamefont
  {Danovich}, \citenamefont {Ruiz-Tijerina}, \citenamefont {Hunt},
  \citenamefont {Szyniszewski}, \citenamefont {Drummond},\ and\ \citenamefont
  {Fal'ko}}]{Drummond-Falko}%
  \BibitemOpen
  \bibfield  {author} {\bibinfo {author} {\bibfnamefont {M.}~\bibnamefont
  {Danovich}}, \bibinfo {author} {\bibfnamefont {D.~A.}\ \bibnamefont
  {Ruiz-Tijerina}}, \bibinfo {author} {\bibfnamefont {R.~J.}\ \bibnamefont
  {Hunt}}, \bibinfo {author} {\bibfnamefont {M.}~\bibnamefont {Szyniszewski}},
  \bibinfo {author} {\bibfnamefont {N.~D.}\ \bibnamefont {Drummond}}, \ and\
  \bibinfo {author} {\bibfnamefont {V.~I.}\ \bibnamefont {Fal'ko}},\ }\bibfield
   {title} {\enquote {\bibinfo {title} {Localized interlayer complexes in
  heterobilayer transition metal dichalcogenides},}\ }\href {\doibase
  10.1103/PhysRevB.97.195452} {\bibfield  {journal} {\bibinfo  {journal} {Phys.
  Rev. B}\ }\textbf {\bibinfo {volume} {97}},\ \bibinfo {pages} {195452}
  (\bibinfo {year} {2018})}\BibitemShut {NoStop}%
\bibitem [{\citenamefont {Scharf}\ \emph {et~al.}(2019)\citenamefont {Scharf},
  \citenamefont {Van~Tuan}, \citenamefont {\v{Z}uti\'c},\ and\ \citenamefont
  {Dery}}]{ScharfScreening}%
  \BibitemOpen
  \bibfield  {author} {\bibinfo {author} {\bibfnamefont {B.}~\bibnamefont
  {Scharf}}, \bibinfo {author} {\bibfnamefont {D.}~\bibnamefont {Van~Tuan}},
  \bibinfo {author} {\bibfnamefont {I.}~\bibnamefont {\v{Z}uti\'c}}, \ and\ \bibinfo
  {author} {\bibfnamefont {H.}~\bibnamefont {Dery}},\ }\bibfield  {title}
  {\enquote {\bibinfo {title} {Dynamical screening in monolayer
  transition-metal dichalcogenides and its manifestations in the exciton
  spectrum},}\ }\href {\doibase 10.1088/1361-648x/ab071f} {\bibfield  {journal}
  {\bibinfo  {journal} {Journal of Physics: Condensed Matter}\ }\textbf
  {\bibinfo {volume} {31}},\ \bibinfo {pages} {203001} (\bibinfo {year}
  {2019})}\BibitemShut {NoStop}%
\bibitem [{\citenamefont {Yang}\ \emph {et~al.}(1991)\citenamefont {Yang},
  \citenamefont {Guo}, \citenamefont {Chan}, \citenamefont {Wong},\ and\
  \citenamefont {Ching}}]{CoulombType}%
  \BibitemOpen
  \bibfield  {author} {\bibinfo {author} {\bibfnamefont {X.~L.}\ \bibnamefont
  {Yang}}, \bibinfo {author} {\bibfnamefont {S.~H.}\ \bibnamefont {Guo}},
  \bibinfo {author} {\bibfnamefont {F.~T.}\ \bibnamefont {Chan}}, \bibinfo
  {author} {\bibfnamefont {K.~W.}\ \bibnamefont {Wong}}, \ and\ \bibinfo
  {author} {\bibfnamefont {W.~Y.}\ \bibnamefont {Ching}},\ }\bibfield  {title}
  {\enquote {\bibinfo {title} {Analytic solution of a two-dimensional hydrogen
  atom. i. nonrelativistic theory},}\ }\href {\doibase
  10.1103/PhysRevA.43.1186} {\bibfield  {journal} {\bibinfo  {journal} {Phys.
  Rev. A}\ }\textbf {\bibinfo {volume} {43}},\ \bibinfo {pages} {1186--1196}
  (\bibinfo {year} {1991})}\BibitemShut {NoStop}%
\bibitem [{\citenamefont {Tanatar}\ and\ \citenamefont
  {Ceperley}(1989)}]{DavidTanatar}%
  \BibitemOpen
  \bibfield  {author} {\bibinfo {author} {\bibfnamefont {B.}~\bibnamefont
  {Tanatar}}\ and\ \bibinfo {author} {\bibfnamefont {D.~M.}\ \bibnamefont
  {Ceperley}},\ }\bibfield  {title} {\enquote {\bibinfo {title} {Ground state
  of the two-dimensional electron gas},}\ }\href {\doibase
  10.1103/PhysRevB.39.5005} {\bibfield  {journal} {\bibinfo  {journal} {Phys.
  Rev. B}\ }\textbf {\bibinfo {volume} {39}},\ \bibinfo {pages} {5005--5016}
  (\bibinfo {year} {1989})}\BibitemShut {NoStop}%
\bibitem [{\citenamefont {Zarenia}\ \emph {et~al.}(2017)\citenamefont
  {Zarenia}, \citenamefont {Neilson}, \citenamefont {Partoens},\ and\
  \citenamefont {Peeters}}]{2v2DEG-2017}%
  \BibitemOpen
  \bibfield  {author} {\bibinfo {author} {\bibfnamefont {M.}~\bibnamefont
  {Zarenia}}, \bibinfo {author} {\bibfnamefont {D.}~\bibnamefont {Neilson}},
  \bibinfo {author} {\bibfnamefont {B.}~\bibnamefont {Partoens}}, \ and\
  \bibinfo {author} {\bibfnamefont {F.~M.}\ \bibnamefont {Peeters}},\
  }\bibfield  {title} {\enquote {\bibinfo {title} {Wigner crystallization in
  transition metal dichalcogenides: A new approach to correlation energy},}\
  }\href {\doibase 10.1103/PhysRevB.95.115438} {\bibfield  {journal} {\bibinfo
  {journal} {Phys. Rev. B}\ }\textbf {\bibinfo {volume} {95}},\ \bibinfo
  {pages} {115438} (\bibinfo {year} {2017})}\BibitemShut {NoStop}%
\bibitem [{\citenamefont {Drummond}\ and\ \citenamefont
  {Needs}(2009)}]{rscrit31}%
  \BibitemOpen
  \bibfield  {author} {\bibinfo {author} {\bibfnamefont {N.~D.}\ \bibnamefont
  {Drummond}}\ and\ \bibinfo {author} {\bibfnamefont {R.~J.}\ \bibnamefont
  {Needs}},\ }\bibfield  {title} {\enquote {\bibinfo {title} {Phase diagram of
  the low-density two-dimensional homogeneous electron gas},}\ }\href {\doibase
  10.1103/PhysRevLett.102.126402} {\bibfield  {journal} {\bibinfo  {journal}
  {Phys. Rev. Lett.}\ }\textbf {\bibinfo {volume} {102}},\ \bibinfo {pages}
  {126402} (\bibinfo {year} {2009})}\BibitemShut {NoStop}%
\bibitem [{\citenamefont {Illing}\ \emph {et~al.}(2017)\citenamefont {Illing},
  \citenamefont {Fritschi}, \citenamefont {Kaiser}, \citenamefont {Klix},
  \citenamefont {Maret},\ and\ \citenamefont {Keim}}]{IllingMelting}%
  \BibitemOpen
  \bibfield  {author} {\bibinfo {author} {\bibfnamefont {B.}~\bibnamefont
  {Illing}}, \bibinfo {author} {\bibfnamefont {S.}~\bibnamefont {Fritschi}},
  \bibinfo {author} {\bibfnamefont {H.}~\bibnamefont {Kaiser}}, \bibinfo
  {author} {\bibfnamefont {C.~L.}\ \bibnamefont {Klix}}, \bibinfo {author}
  {\bibfnamefont {G.}~\bibnamefont {Maret}}, \ and\ \bibinfo {author}
  {\bibfnamefont {P.}~\bibnamefont {Keim}},\ }\bibfield  {title} {\enquote
  {\bibinfo {title} {Mermin{\textendash}{Wagner} fluctuations in 2d amorphous
  solids},}\ }\href {\doibase 10.1073/pnas.1612964114} {\bibfield  {journal}
  {\bibinfo  {journal} {Proceedings of the National Academy of Sciences}\
  }\textbf {\bibinfo {volume} {114}},\ \bibinfo {pages} {1856--1861} (\bibinfo
  {year} {2017})}\BibitemShut {NoStop}%
\bibitem [{\citenamefont {Khrapak}(2020)}]{KhrapakMelting}%
  \BibitemOpen
  \bibfield  {author} {\bibinfo {author} {\bibfnamefont {S.~A.}\ \bibnamefont
  {Khrapak}},\ }\bibfield  {title} {\enquote {\bibinfo {title} {Lindemann
  melting criterion in two dimensions},}\ }\href {\doibase
  10.1103/PhysRevResearch.2.012040} {\bibfield  {journal} {\bibinfo  {journal}
  {Phys. Rev. Research}\ }\textbf {\bibinfo {volume} {2}},\ \bibinfo {pages}
  {012040} (\bibinfo {year} {2020})}\BibitemShut {NoStop}%
\bibitem [{\citenamefont {Ma}\ \emph {et~al.}(2020)\citenamefont {Ma},
  \citenamefont {Villegas~Rosales}, \citenamefont {Deng}, \citenamefont
  {Chung}, \citenamefont {Pfeiffer}, \citenamefont {West}, \citenamefont
  {Baldwin}, \citenamefont {Winkler},\ and\ \citenamefont
  {Shayegan}}]{Shayegan2020}%
  \BibitemOpen
  \bibfield  {author} {\bibinfo {author} {\bibfnamefont {M.~K.}\ \bibnamefont
  {Ma}}, \bibinfo {author} {\bibfnamefont {K.}~\bibnamefont
  {Villegas~Rosales}}, \bibinfo {author} {\bibfnamefont {H.}~\bibnamefont
  {Deng}}, \bibinfo {author} {\bibfnamefont {Y.}~\bibnamefont {Chung}},
  \bibinfo {author} {\bibfnamefont {L.}~\bibnamefont {Pfeiffer}}, \bibinfo
  {author} {\bibfnamefont {K.}~\bibnamefont {West}}, \bibinfo {author}
  {\bibfnamefont {K.}~\bibnamefont {Baldwin}}, \bibinfo {author} {\bibfnamefont
  {R.}~\bibnamefont {Winkler}}, \ and\ \bibinfo {author} {\bibfnamefont
  {M.}~\bibnamefont {Shayegan}},\ }\bibfield  {title} {\enquote {\bibinfo
  {title} {{Thermal and Quantum Melting Phase Diagrams for a
  Magnetic-Field-Induced Wigner Solid}},}\ }\href {\doibase
  10.1103/physrevlett.125.036601} {\bibfield  {journal} {\bibinfo  {journal}
  {Physical Review Letters}\ }\textbf {\bibinfo {volume} {125}} (\bibinfo
  {year} {2020}),\ 10.1103/physrevlett.125.036601}\BibitemShut {NoStop}%
\bibitem [{\citenamefont {Korm{\'{a}}nyos}\ \emph {et~al.}(2015)\citenamefont
  {Korm{\'{a}}nyos}, \citenamefont {Burkard}, \citenamefont {Gmitra},
  \citenamefont {Fabian}, \citenamefont {Z{\'{o}}lyomi}, \citenamefont
  {Drummond},\ and\ \citenamefont {Fal'ko}}]{MX2EffMass}%
  \BibitemOpen
  \bibfield  {author} {\bibinfo {author} {\bibfnamefont {A.}~\bibnamefont
  {Korm{\'{a}}nyos}}, \bibinfo {author} {\bibfnamefont {G.}~\bibnamefont
  {Burkard}}, \bibinfo {author} {\bibfnamefont {M.}~\bibnamefont {Gmitra}},
  \bibinfo {author} {\bibfnamefont {J.}~\bibnamefont {Fabian}}, \bibinfo
  {author} {\bibfnamefont {V.}~\bibnamefont {Z{\'{o}}lyomi}}, \bibinfo {author}
  {\bibfnamefont {N.~D.}\ \bibnamefont {Drummond}}, \ and\ \bibinfo {author}
  {\bibfnamefont {V.}~\bibnamefont {Fal'ko}},\ }\bibfield  {title} {\enquote
  {\bibinfo {title} {k $\cdotp$ p theory for two-dimensional transition metal
  dichalcogenide semiconductors},}\ }\href {\doibase
  10.1088/2053-1583/2/2/022001} {\bibfield  {journal} {\bibinfo  {journal} {2D
  Materials}\ }\textbf {\bibinfo {volume} {2}},\ \bibinfo {pages} {022001}
  (\bibinfo {year} {2015})}\BibitemShut {NoStop}%
\bibitem [{\citenamefont {Kumar}\ and\ \citenamefont
  {Ahluwalia}(2012)}]{MX2dielectric}%
  \BibitemOpen
  \bibfield  {author} {\bibinfo {author} {\bibfnamefont {A.}~\bibnamefont
  {Kumar}}\ and\ \bibinfo {author} {\bibfnamefont {P.}~\bibnamefont
  {Ahluwalia}},\ }\bibfield  {title} {\enquote {\bibinfo {title} {Tunable
  dielectric response of transition metals dichalcogenides {MX$_2$ (M=Mo, W;
  X=S, Se, Te)}: Effect of quantum confinement},}\ }\href {\doibase
  https://doi.org/10.1016/j.physb.2012.08.034} {\bibfield  {journal} {\bibinfo
  {journal} {Physica B: Condensed Matter}\ }\textbf {\bibinfo {volume} {407}},\
  \bibinfo {pages} {4627 -- 4634} (\bibinfo {year} {2012})}\BibitemShut
  {NoStop}%
\bibitem [{\citenamefont {Lopes~dos Santos}\ \emph {et~al.}(2007)\citenamefont
  {Lopes~dos Santos}, \citenamefont {Peres},\ and\ \citenamefont
  {Castro~Neto}}]{LopesPRL}%
  \BibitemOpen
  \bibfield  {author} {\bibinfo {author} {\bibfnamefont {J.~M.~B.}\
  \bibnamefont {Lopes~dos Santos}}, \bibinfo {author} {\bibfnamefont
  {N.~M.~R.}\ \bibnamefont {Peres}}, \ and\ \bibinfo {author} {\bibfnamefont
  {A.~H.}\ \bibnamefont {Castro~Neto}},\ }\bibfield  {title} {\enquote
  {\bibinfo {title} {Graphene bilayer with a twist: Electronic structure},}\
  }\href {\doibase 10.1103/PhysRevLett.99.256802} {\bibfield  {journal}
  {\bibinfo  {journal} {Phys. Rev. Lett.}\ }\textbf {\bibinfo {volume} {99}},\
  \bibinfo {pages} {256802} (\bibinfo {year} {2007})}\BibitemShut {NoStop}%
\bibitem [{\citenamefont {{Mak}}\ \emph {et~al.}(2013)\citenamefont {{Mak}},
  \citenamefont {{He}}, \citenamefont {{Lee}}, \citenamefont {{Lee}},
  \citenamefont {{Hone}}, \citenamefont {{Heinz}},\ and\ \citenamefont
  {{Shan}}}]{MoS2rs}%
  \BibitemOpen
  \bibfield  {author} {\bibinfo {author} {\bibfnamefont {K.~F.}\ \bibnamefont
  {{Mak}}}, \bibinfo {author} {\bibfnamefont {K.}~\bibnamefont {{He}}},
  \bibinfo {author} {\bibfnamefont {C.}~\bibnamefont {{Lee}}}, \bibinfo
  {author} {\bibfnamefont {G.~H.}\ \bibnamefont {{Lee}}}, \bibinfo {author}
  {\bibfnamefont {J.}~\bibnamefont {{Hone}}}, \bibinfo {author} {\bibfnamefont
  {T.~F.}\ \bibnamefont {{Heinz}}}, \ and\ \bibinfo {author} {\bibfnamefont
  {J.}~\bibnamefont {{Shan}}},\ }\bibfield  {title} {\enquote {\bibinfo {title}
  {{Tightly bound trions in monolayer MoS$_{2}$}},}\ }\href {\doibase
  10.1038/nmat3505} {\bibfield  {journal} {\bibinfo  {journal} {Nature
  Materials}\ }\textbf {\bibinfo {volume} {12}},\ \bibinfo {pages} {207--211}
  (\bibinfo {year} {2013})},\ \Eprint {http://arxiv.org/abs/1210.8226}
  {arXiv:1210.8226 [cond-mat.mtrl-sci]} \BibitemShut {NoStop}%
\bibitem [{\citenamefont {Mott}\ and\ \citenamefont {Davis}(2012)}]{MottDavis}%
  \BibitemOpen
  \bibfield  {author} {\bibinfo {author} {\bibfnamefont {N.~F.}\ \bibnamefont
  {Mott}}\ and\ \bibinfo {author} {\bibfnamefont {E.~A.}\ \bibnamefont
  {Davis}},\ }\href@noop {} {\emph {\bibinfo {title} {Electronic processes in
  non-crystalline materials}}}\ (\bibinfo  {publisher} {OUP Oxford},\ \bibinfo
  {year} {2012})\BibitemShut {NoStop}%
\bibitem [{\citenamefont {Xu}\ \emph {et~al.}(2018)\citenamefont {Xu},
  \citenamefont {Xu}, \citenamefont {Zhang}, \citenamefont {Peng},
  \citenamefont {Shao}, \citenamefont {Ni}, \citenamefont {Li}, \citenamefont
  {Yao}, \citenamefont {Lu}, \citenamefont {Zhu},\ and\ \citenamefont
  {et~al.}}]{HeteroMass}%
  \BibitemOpen
  \bibfield  {author} {\bibinfo {author} {\bibfnamefont {K.}~\bibnamefont
  {Xu}}, \bibinfo {author} {\bibfnamefont {Y.}~\bibnamefont {Xu}}, \bibinfo
  {author} {\bibfnamefont {H.}~\bibnamefont {Zhang}}, \bibinfo {author}
  {\bibfnamefont {B.}~\bibnamefont {Peng}}, \bibinfo {author} {\bibfnamefont
  {H.}~\bibnamefont {Shao}}, \bibinfo {author} {\bibfnamefont {G.}~\bibnamefont
  {Ni}}, \bibinfo {author} {\bibfnamefont {J.}~\bibnamefont {Li}}, \bibinfo
  {author} {\bibfnamefont {M.}~\bibnamefont {Yao}}, \bibinfo {author}
  {\bibfnamefont {H.}~\bibnamefont {Lu}}, \bibinfo {author} {\bibfnamefont
  {H.}~\bibnamefont {Zhu}}, \ and\ \bibinfo {author} {\bibnamefont {et~al.}},\
  }\bibfield  {title} {\enquote {\bibinfo {title} {The role of {A}nderson's
  rule in determining electronic, optical and transport properties of
  transition metal dichalcogenide heterostructures},}\ }\href {\doibase
  10.1039/c8cp05522j} {\bibfield  {journal} {\bibinfo  {journal} {Physical
  Chemistry Chemical Physics}\ }\textbf {\bibinfo {volume} {20}},\ \bibinfo
  {pages} {30351--30364} (\bibinfo {year} {2018})}\BibitemShut {NoStop}%
\bibitem [{\citenamefont {Jin}\ \emph {et~al.}(2019)\citenamefont {Jin},
  \citenamefont {Regan}, \citenamefont {Yan}, \citenamefont {Iqbal
  Bakti~Utama}, \citenamefont {Wang}, \citenamefont {Zhao}, \citenamefont
  {Qin}, \citenamefont {Yang}, \citenamefont {Zheng}, \citenamefont {Shi},\
  and\ \citenamefont {et~al.}}]{FengWangWSeWS}%
  \BibitemOpen
  \bibfield  {author} {\bibinfo {author} {\bibfnamefont {C.}~\bibnamefont
  {Jin}}, \bibinfo {author} {\bibfnamefont {E.~C.}\ \bibnamefont {Regan}},
  \bibinfo {author} {\bibfnamefont {A.}~\bibnamefont {Yan}}, \bibinfo {author}
  {\bibfnamefont {M.}~\bibnamefont {Iqbal Bakti~Utama}}, \bibinfo {author}
  {\bibfnamefont {D.}~\bibnamefont {Wang}}, \bibinfo {author} {\bibfnamefont
  {S.}~\bibnamefont {Zhao}}, \bibinfo {author} {\bibfnamefont {Y.}~\bibnamefont
  {Qin}}, \bibinfo {author} {\bibfnamefont {S.}~\bibnamefont {Yang}}, \bibinfo
  {author} {\bibfnamefont {Z.}~\bibnamefont {Zheng}}, \bibinfo {author}
  {\bibfnamefont {S.}~\bibnamefont {Shi}}, \ and\ \bibinfo {author}
  {\bibnamefont {et~al.}},\ }\bibfield  {title} {\enquote {\bibinfo {title}
  {Observation of moir\'e excitons in {WSe$_2$/WS$_2$} heterostructure
  superlattices},}\ }\href {\doibase 10.1038/s41586-019-0976-y} {\bibfield
  {journal} {\bibinfo  {journal} {Nature}\ }\textbf {\bibinfo {volume} {567}},\
  \bibinfo {pages} {76--80} (\bibinfo {year} {2019})}\BibitemShut {NoStop}%
\bibitem [{\citenamefont {Zhang}\ \emph {et~al.}(2018)\citenamefont {Zhang},
  \citenamefont {Surrente}, \citenamefont {Baranowski}, \citenamefont {Maude},
  \citenamefont {Gant}, \citenamefont {Castellanos-Gomez},\ and\ \citenamefont
  {Plochocka}}]{MSeMS}%
  \BibitemOpen
  \bibfield  {author} {\bibinfo {author} {\bibfnamefont {N.}~\bibnamefont
  {Zhang}}, \bibinfo {author} {\bibfnamefont {A.}~\bibnamefont {Surrente}},
  \bibinfo {author} {\bibfnamefont {M.}~\bibnamefont {Baranowski}}, \bibinfo
  {author} {\bibfnamefont {D.~K.}\ \bibnamefont {Maude}}, \bibinfo {author}
  {\bibfnamefont {P.}~\bibnamefont {Gant}}, \bibinfo {author} {\bibfnamefont
  {A.}~\bibnamefont {Castellanos-Gomez}}, \ and\ \bibinfo {author}
  {\bibfnamefont {P.}~\bibnamefont {Plochocka}},\ }\bibfield  {title} {\enquote
  {\bibinfo {title} {Moir\'e intralayer excitons in a mose2/mos2
  heterostructure},}\ }\href {\doibase 10.1021/acs.nanolett.8b03266} {\bibfield
   {journal} {\bibinfo  {journal} {Nano Letters}\ }\textbf {\bibinfo {volume}
  {18}},\ \bibinfo {pages} {7651--7657} (\bibinfo {year} {2018})}\BibitemShut
  {NoStop}%
\bibitem [{\citenamefont {Kozawa}\ \emph {et~al.}(2016)\citenamefont {Kozawa},
  \citenamefont {Carvalho}, \citenamefont {Verzhbitskiy}, \citenamefont
  {Giustiniano}, \citenamefont {Miyauchi}, \citenamefont {Mouri}, \citenamefont
  {Castro~Neto}, \citenamefont {Matsuda},\ and\ \citenamefont {Eda}}]{MSeWS1}%
  \BibitemOpen
  \bibfield  {author} {\bibinfo {author} {\bibfnamefont {D.}~\bibnamefont
  {Kozawa}}, \bibinfo {author} {\bibfnamefont {A.}~\bibnamefont {Carvalho}},
  \bibinfo {author} {\bibfnamefont {I.}~\bibnamefont {Verzhbitskiy}}, \bibinfo
  {author} {\bibfnamefont {F.}~\bibnamefont {Giustiniano}}, \bibinfo {author}
  {\bibfnamefont {Y.}~\bibnamefont {Miyauchi}}, \bibinfo {author}
  {\bibfnamefont {S.}~\bibnamefont {Mouri}}, \bibinfo {author} {\bibfnamefont
  {A.~H.}\ \bibnamefont {Castro~Neto}}, \bibinfo {author} {\bibfnamefont
  {K.}~\bibnamefont {Matsuda}}, \ and\ \bibinfo {author} {\bibfnamefont
  {G.}~\bibnamefont {Eda}},\ }\bibfield  {title} {\enquote {\bibinfo {title}
  {Evidence for fast interlayer energy transfer in mose2/ws2
  heterostructures},}\ }\href {\doibase 10.1021/acs.nanolett.6b00801}
  {\bibfield  {journal} {\bibinfo  {journal} {Nano Letters}\ }\textbf {\bibinfo
  {volume} {16}},\ \bibinfo {pages} {4087--4093} (\bibinfo {year}
  {2016})}\BibitemShut {NoStop}%
\bibitem [{\citenamefont {Alexeev}\ \emph {et~al.}(2019)\citenamefont
  {Alexeev}, \citenamefont {Ruiz-Tijerina}, \citenamefont {Danovich},
  \citenamefont {Hamer}, \citenamefont {Terry}, \citenamefont {Nayak},
  \citenamefont {Ahn}, \citenamefont {Pak}, \citenamefont {Lee}, \citenamefont
  {Sohn},\ and\ \citenamefont {et~al.}}]{MSeWS2}%
  \BibitemOpen
  \bibfield  {author} {\bibinfo {author} {\bibfnamefont {E.~M.}\ \bibnamefont
  {Alexeev}}, \bibinfo {author} {\bibfnamefont {D.~A.}\ \bibnamefont
  {Ruiz-Tijerina}}, \bibinfo {author} {\bibfnamefont {M.}~\bibnamefont
  {Danovich}}, \bibinfo {author} {\bibfnamefont {M.~J.}\ \bibnamefont {Hamer}},
  \bibinfo {author} {\bibfnamefont {D.~J.}\ \bibnamefont {Terry}}, \bibinfo
  {author} {\bibfnamefont {P.~K.}\ \bibnamefont {Nayak}}, \bibinfo {author}
  {\bibfnamefont {S.}~\bibnamefont {Ahn}}, \bibinfo {author} {\bibfnamefont
  {S.}~\bibnamefont {Pak}}, \bibinfo {author} {\bibfnamefont {J.}~\bibnamefont
  {Lee}}, \bibinfo {author} {\bibfnamefont {J.~I.}\ \bibnamefont {Sohn}}, \
  and\ \bibinfo {author} {\bibnamefont {et~al.}},\ }\bibfield  {title}
  {\enquote {\bibinfo {title} {Resonantly hybridized excitons in moir\'e
  superlattices in van der waals heterostructures},}\ }\href {\doibase
  10.1038/s41586-019-0986-9} {\bibfield  {journal} {\bibinfo  {journal}
  {Nature}\ }\textbf {\bibinfo {volume} {567}},\ \bibinfo {pages} {81--86}
  (\bibinfo {year} {2019})}\BibitemShut {NoStop}%
\bibitem [{\citenamefont {Seyler}\ \emph {et~al.}(2019)\citenamefont {Seyler},
  \citenamefont {Rivera}, \citenamefont {Yu}, \citenamefont {Wilson},
  \citenamefont {Ray}, \citenamefont {Mandrus}, \citenamefont {Yan},
  \citenamefont {Yao},\ and\ \citenamefont {Xu}}]{MoSe2/WSe2}%
  \BibitemOpen
  \bibfield  {author} {\bibinfo {author} {\bibfnamefont {K.~L.}\ \bibnamefont
  {Seyler}}, \bibinfo {author} {\bibfnamefont {P.}~\bibnamefont {Rivera}},
  \bibinfo {author} {\bibfnamefont {H.}~\bibnamefont {Yu}}, \bibinfo {author}
  {\bibfnamefont {N.~P.}\ \bibnamefont {Wilson}}, \bibinfo {author}
  {\bibfnamefont {E.~L.}\ \bibnamefont {Ray}}, \bibinfo {author} {\bibfnamefont
  {D.~G.}\ \bibnamefont {Mandrus}}, \bibinfo {author} {\bibfnamefont
  {J.}~\bibnamefont {Yan}}, \bibinfo {author} {\bibfnamefont {W.}~\bibnamefont
  {Yao}}, \ and\ \bibinfo {author} {\bibfnamefont {X.}~\bibnamefont {Xu}},\
  }\href {\doibase 10.1038/s41586-019-0957-1} {\bibfield  {journal} {\bibinfo
  {journal} {Nature}\ }\textbf {\bibinfo {volume} {567}},\ \bibinfo {pages}
  {66--70} (\bibinfo {year} {2019})}\BibitemShut {NoStop}%
\bibitem [{\citenamefont {Hong}\ \emph {et~al.}(2014)\citenamefont {Hong},
  \citenamefont {Kim}, \citenamefont {Shi}, \citenamefont {Zhang},
  \citenamefont {Jin}, \citenamefont {Sun}, \citenamefont {Tongay},
  \citenamefont {Wu}, \citenamefont {Zhang},\ and\ \citenamefont
  {Wang}}]{MSWS1}%
  \BibitemOpen
  \bibfield  {author} {\bibinfo {author} {\bibfnamefont {X.}~\bibnamefont
  {Hong}}, \bibinfo {author} {\bibfnamefont {J.}~\bibnamefont {Kim}}, \bibinfo
  {author} {\bibfnamefont {S.-F.}\ \bibnamefont {Shi}}, \bibinfo {author}
  {\bibfnamefont {Y.}~\bibnamefont {Zhang}}, \bibinfo {author} {\bibfnamefont
  {C.}~\bibnamefont {Jin}}, \bibinfo {author} {\bibfnamefont {Y.}~\bibnamefont
  {Sun}}, \bibinfo {author} {\bibfnamefont {S.}~\bibnamefont {Tongay}},
  \bibinfo {author} {\bibfnamefont {J.}~\bibnamefont {Wu}}, \bibinfo {author}
  {\bibfnamefont {Y.}~\bibnamefont {Zhang}}, \ and\ \bibinfo {author}
  {\bibfnamefont {F.}~\bibnamefont {Wang}},\ }\bibfield  {title} {\enquote
  {\bibinfo {title} {Ultrafast charge transfer in atomically thin mos2/ws2
  heterostructures},}\ }\href {\doibase 10.1038/nnano.2014.167} {\bibfield
  {journal} {\bibinfo  {journal} {Nature Nanotechnology}\ }\textbf {\bibinfo
  {volume} {9}},\ \bibinfo {pages} {682--686} (\bibinfo {year}
  {2014})}\BibitemShut {NoStop}%
\bibitem [{\citenamefont {Yang}\ \emph {et~al.}(2018)\citenamefont {Yang},
  \citenamefont {Kawai}, \citenamefont {Bosman}, \citenamefont {Tang},
  \citenamefont {Chai}, \citenamefont {Tay}, \citenamefont {Yang},
  \citenamefont {Seng}, \citenamefont {Zhu}, \citenamefont {Gong},
  \citenamefont {Liu}, \citenamefont {Goh}, \citenamefont {Wang},\ and\
  \citenamefont {Chi}}]{MSWS2}%
  \BibitemOpen
  \bibfield  {author} {\bibinfo {author} {\bibfnamefont {W.}~\bibnamefont
  {Yang}}, \bibinfo {author} {\bibfnamefont {H.}~\bibnamefont {Kawai}},
  \bibinfo {author} {\bibfnamefont {M.}~\bibnamefont {Bosman}}, \bibinfo
  {author} {\bibfnamefont {B.}~\bibnamefont {Tang}}, \bibinfo {author}
  {\bibfnamefont {J.}~\bibnamefont {Chai}}, \bibinfo {author} {\bibfnamefont
  {W.~L.}\ \bibnamefont {Tay}}, \bibinfo {author} {\bibfnamefont
  {J.}~\bibnamefont {Yang}}, \bibinfo {author} {\bibfnamefont {H.~L.}\
  \bibnamefont {Seng}}, \bibinfo {author} {\bibfnamefont {H.}~\bibnamefont
  {Zhu}}, \bibinfo {author} {\bibfnamefont {H.}~\bibnamefont {Gong}}, \bibinfo
  {author} {\bibfnamefont {H.}~\bibnamefont {Liu}}, \bibinfo {author}
  {\bibfnamefont {K.~E.~J.}\ \bibnamefont {Goh}}, \bibinfo {author}
  {\bibfnamefont {S.}~\bibnamefont {Wang}}, \ and\ \bibinfo {author}
  {\bibfnamefont {D.}~\bibnamefont {Chi}},\ }\bibfield  {title} {\enquote
  {\bibinfo {title} {Interlayer interactions in 2d ws2/mos2 heterostructures
  monolithically grown by in situ physical vapor deposition},}\ }\href
  {\doibase 10.1039/C8NR07498D} {\bibfield  {journal} {\bibinfo  {journal}
  {Nanoscale}\ }\textbf {\bibinfo {volume} {10}},\ \bibinfo {pages}
  {22927--22936} (\bibinfo {year} {2018})}\BibitemShut {NoStop}%
\bibitem [{\citenamefont {Gong}\ \emph {et~al.}(2014)\citenamefont {Gong},
  \citenamefont {Lin}, \citenamefont {Wang}, \citenamefont {Shi}, \citenamefont
  {Lei}, \citenamefont {Lin}, \citenamefont {Zou}, \citenamefont {Ye},
  \citenamefont {Vajtai}, \citenamefont {Yakobson} \emph {et~al.}}]{MoS2/WS2}%
  \BibitemOpen
  \bibfield  {author} {\bibinfo {author} {\bibfnamefont {Y.}~\bibnamefont
  {Gong}}, \bibinfo {author} {\bibfnamefont {J.}~\bibnamefont {Lin}}, \bibinfo
  {author} {\bibfnamefont {X.}~\bibnamefont {Wang}}, \bibinfo {author}
  {\bibfnamefont {G.}~\bibnamefont {Shi}}, \bibinfo {author} {\bibfnamefont
  {S.}~\bibnamefont {Lei}}, \bibinfo {author} {\bibfnamefont {Z.}~\bibnamefont
  {Lin}}, \bibinfo {author} {\bibfnamefont {X.}~\bibnamefont {Zou}}, \bibinfo
  {author} {\bibfnamefont {G.}~\bibnamefont {Ye}}, \bibinfo {author}
  {\bibfnamefont {R.}~\bibnamefont {Vajtai}}, \bibinfo {author} {\bibfnamefont
  {B.~I.}\ \bibnamefont {Yakobson}},  \emph {et~al.},\ }\bibfield  {title}
  {\enquote {\bibinfo {title} {Vertical and in-plane heterostructures from
  {WS$_{2}$/MoS$_{2}$} monolayers},}\ }\href
  {https://www.nature.com/articles/nmat4091?page=16} {\bibfield  {journal}
  {\bibinfo  {journal} {Nature materials}\ }\textbf {\bibinfo {volume} {13}},\
  \bibinfo {pages} {1135--1142} (\bibinfo {year} {2014})}\BibitemShut {NoStop}%
\bibitem [{\citenamefont {Yamaoka}\ \emph {et~al.}(2018)\citenamefont
  {Yamaoka}, \citenamefont {Lim}, \citenamefont {Koirala}, \citenamefont
  {Wang}, \citenamefont {Shinokita}, \citenamefont {Maruyama}, \citenamefont
  {Okada}, \citenamefont {Miyauchi},\ and\ \citenamefont
  {Matsuda}}]{MoTe2/WSe2}%
  \BibitemOpen
  \bibfield  {author} {\bibinfo {author} {\bibfnamefont {T.}~\bibnamefont
  {Yamaoka}}, \bibinfo {author} {\bibfnamefont {H.~E.}\ \bibnamefont {Lim}},
  \bibinfo {author} {\bibfnamefont {S.}~\bibnamefont {Koirala}}, \bibinfo
  {author} {\bibfnamefont {X.}~\bibnamefont {Wang}}, \bibinfo {author}
  {\bibfnamefont {K.}~\bibnamefont {Shinokita}}, \bibinfo {author}
  {\bibfnamefont {M.}~\bibnamefont {Maruyama}}, \bibinfo {author}
  {\bibfnamefont {S.}~\bibnamefont {Okada}}, \bibinfo {author} {\bibfnamefont
  {Y.}~\bibnamefont {Miyauchi}}, \ and\ \bibinfo {author} {\bibfnamefont
  {K.}~\bibnamefont {Matsuda}},\ }\bibfield  {title} {\enquote {\bibinfo
  {title} {Efficient photocarrier transfer and effective photoluminescence
  enhancement in {Type I Monolayer MoTe$_{2}$/WSe$_{2}$} heterostructure},}\
  }\href {\doibase 10.1002/adfm.201801021} {\bibfield  {journal} {\bibinfo
  {journal} {Advanced Functional Materials}\ }\textbf {\bibinfo {volume}
  {28}},\ \bibinfo {pages} {1801021} (\bibinfo {year} {2018})}\BibitemShut
  {NoStop}%
\bibitem [{\citenamefont {Lu}\ \emph {et~al.}(2014)\citenamefont {Lu},
  \citenamefont {Li}, \citenamefont {Watanabe}, \citenamefont {Taniguchi},\
  and\ \citenamefont {Andrei}}]{AnderiMoireconstant}%
  \BibitemOpen
  \bibfield  {author} {\bibinfo {author} {\bibfnamefont {C.-P.}\ \bibnamefont
  {Lu}}, \bibinfo {author} {\bibfnamefont {G.}~\bibnamefont {Li}}, \bibinfo
  {author} {\bibfnamefont {K.}~\bibnamefont {Watanabe}}, \bibinfo {author}
  {\bibfnamefont {T.}~\bibnamefont {Taniguchi}}, \ and\ \bibinfo {author}
  {\bibfnamefont {E.}~\bibnamefont {Andrei}},\ }\bibfield  {title} {\enquote
  {\bibinfo {title} {Mos2: Choice substrate for accessing and tuning the
  electronic properties of graphene},}\ }\href {\doibase
  10.1103/physrevlett.113.156804} {\bibfield  {journal} {\bibinfo  {journal}
  {Physical Review Letters}\ }\textbf {\bibinfo {volume} {113}} (\bibinfo
  {year} {2014}),\ 10.1103/physrevlett.113.156804}\BibitemShut {NoStop}%
\bibitem [{\citenamefont {Ruiz-Tijerina}\ \emph {et~al.}(2020)\citenamefont
  {Ruiz-Tijerina}, \citenamefont {Soltero},\ and\ \citenamefont
  {Mireles}}]{ruiztijerina2020theory}%
  \BibitemOpen
  \bibfield  {author} {\bibinfo {author} {\bibfnamefont {D.~A.}\ \bibnamefont
  {Ruiz-Tijerina}}, \bibinfo {author} {\bibfnamefont {I.}~\bibnamefont
  {Soltero}}, \ and\ \bibinfo {author} {\bibfnamefont {F.}~\bibnamefont
  {Mireles}},\ }\href {https://arxiv.org/abs/2007.03754} {\enquote {\bibinfo
  {title} {Theory of moiré localized excitons in transition-metal
  dichalcogenide heterobilayers},}\ } (\bibinfo {year} {2020}),\ \Eprint
  {http://arxiv.org/abs/2007.03754} {arXiv:2007.03754 [cond-mat.mes-hall]}
  \BibitemShut {NoStop}%
\bibitem [{\citenamefont {Chitra}\ and\ \citenamefont
  {Giamarchi}(2005)}]{ChitraEPJB05}%
  \BibitemOpen
  \bibfield  {author} {\bibinfo {author} {\bibfnamefont {R.}~\bibnamefont
  {Chitra}}\ and\ \bibinfo {author} {\bibfnamefont {T.}~\bibnamefont
  {Giamarchi}},\ }\bibfield  {title} {\enquote {\bibinfo {title} {Zero field
  wigner crystal},}\ }\href {\doibase 10.1140/epjb/e2005-00145-0} {\bibfield
  {journal} {\bibinfo  {journal} {The European Physical Journal B - Condensed
  Matter and Complex Systems}\ }\textbf {\bibinfo {volume} {44}},\ \bibinfo
  {pages} {455--467} (\bibinfo {year} {2005})}\BibitemShut {NoStop}%
\bibitem [{\citenamefont {Zhang}\ \emph {et~al.}(2019)\citenamefont {Zhang},
  \citenamefont {Yuan},\ and\ \citenamefont {Fu}}]{LiangMoireChem}%
  \BibitemOpen
  \bibfield  {author} {\bibinfo {author} {\bibfnamefont {Y.}~\bibnamefont
  {Zhang}}, \bibinfo {author} {\bibfnamefont {N.~F.~Q.}\ \bibnamefont {Yuan}},
  \ and\ \bibinfo {author} {\bibfnamefont {L.}~\bibnamefont {Fu}},\ }\href@noop
  {} {\enquote {\bibinfo {title} {Moir\'e quantum chemistry: charge transfer in
  transition metal dichalcogenide superlattices},}\ } (\bibinfo {year}
  {2019}),\ \Eprint {http://arxiv.org/abs/1910.14061} {arXiv:1910.14061
  [cond-mat.str-el]} \BibitemShut {NoStop}%
\bibitem [{\citenamefont {Bak}(1982)}]{BakDevil82}%
  \BibitemOpen
  \bibfield  {author} {\bibinfo {author} {\bibfnamefont {P.}~\bibnamefont
  {Bak}},\ }\bibfield  {title} {\enquote {\bibinfo {title} {Commensurate
  phases, incommensurate phases and the devil's staircase},}\ }\href {\doibase
  10.1088/0034-4885/45/6/001} {\bibfield  {journal} {\bibinfo  {journal}
  {Reports on Progress in Physics}\ }\textbf {\bibinfo {volume} {45}},\
  \bibinfo {pages} {587--629} (\bibinfo {year} {1982})}\BibitemShut {NoStop}%
\bibitem [{\citenamefont {Bak}\ and\ \citenamefont
  {Fukuyama}(1980)}]{BakqFK80}%
  \BibitemOpen
  \bibfield  {author} {\bibinfo {author} {\bibfnamefont {P.}~\bibnamefont
  {Bak}}\ and\ \bibinfo {author} {\bibfnamefont {H.}~\bibnamefont {Fukuyama}},\
  }\bibfield  {title} {\enquote {\bibinfo {title} {Destruction of "the devil's
  staircase" by quantum fluctuations},}\ }\href {\doibase
  10.1103/PhysRevB.21.3287} {\bibfield  {journal} {\bibinfo  {journal} {Phys.
  Rev. B}\ }\textbf {\bibinfo {volume} {21}},\ \bibinfo {pages} {3287--3289}
  (\bibinfo {year} {1980})}\BibitemShut {NoStop}%
\bibitem [{\citenamefont {Rademaker}\ \emph {et~al.}(2013)\citenamefont
  {Rademaker}, \citenamefont {Pramudya}, \citenamefont {Zaanen},\ and\
  \citenamefont {Dobrosavljevi\ifmmode~\acute{c}\else \'{c}\fi{}}}]{Rademaker}%
  \BibitemOpen
  \bibfield  {author} {\bibinfo {author} {\bibfnamefont {L.}~\bibnamefont
  {Rademaker}}, \bibinfo {author} {\bibfnamefont {Y.}~\bibnamefont {Pramudya}},
  \bibinfo {author} {\bibfnamefont {J.}~\bibnamefont {Zaanen}}, \ and\ \bibinfo
  {author} {\bibfnamefont {V.}~\bibnamefont
  {Dobrosavljevi\ifmmode~\acute{c}\else \'{c}\fi{}}},\ }\bibfield  {title}
  {\enquote {\bibinfo {title} {Influence of long-range interactions on charge
  ordering phenomena on a square lattice},}\ }\href {\doibase
  10.1103/PhysRevE.88.032121} {\bibfield  {journal} {\bibinfo  {journal} {Phys.
  Rev. E}\ }\textbf {\bibinfo {volume} {88}},\ \bibinfo {pages} {032121}
  (\bibinfo {year} {2013})}\BibitemShut {NoStop}%
\bibitem [{\citenamefont {Bonsall}\ and\ \citenamefont
  {Maradudin}(1977)}]{BonsallMaradudin}%
  \BibitemOpen
  \bibfield  {author} {\bibinfo {author} {\bibfnamefont {L.}~\bibnamefont
  {Bonsall}}\ and\ \bibinfo {author} {\bibfnamefont {A.~A.}\ \bibnamefont
  {Maradudin}},\ }\bibfield  {title} {\enquote {\bibinfo {title} {Some static
  and dynamical properties of a two-dimensional wigner crystal},}\ }\href
  {\doibase 10.1103/PhysRevB.15.1959} {\bibfield  {journal} {\bibinfo
  {journal} {Phys. Rev. B}\ }\textbf {\bibinfo {volume} {15}},\ \bibinfo
  {pages} {1959--1973} (\bibinfo {year} {1977})}\BibitemShut {NoStop}%
\bibitem [{\citenamefont {Maki}\ and\ \citenamefont
  {Zotos}(1983)}]{MakiZotos83}%
  \BibitemOpen
  \bibfield  {author} {\bibinfo {author} {\bibfnamefont {K.}~\bibnamefont
  {Maki}}\ and\ \bibinfo {author} {\bibfnamefont {X.}~\bibnamefont {Zotos}},\
  }\bibfield  {title} {\enquote {\bibinfo {title} {Static and dynamic
  properties of a two-dimensional wigner crystal in a strong magnetic field},}\
  }\href {\doibase 10.1103/PhysRevB.28.4349} {\bibfield  {journal} {\bibinfo
  {journal} {Phys. Rev. B}\ }\textbf {\bibinfo {volume} {28}},\ \bibinfo
  {pages} {4349--4356} (\bibinfo {year} {1983})}\BibitemShut {NoStop}%
\bibitem [{\citenamefont {Fukuyama}\ and\ \citenamefont
  {Lee}(1978)}]{FukuyamaLee78}%
  \BibitemOpen
  \bibfield  {author} {\bibinfo {author} {\bibfnamefont {H.}~\bibnamefont
  {Fukuyama}}\ and\ \bibinfo {author} {\bibfnamefont {P.~A.}\ \bibnamefont
  {Lee}},\ }\bibfield  {title} {\enquote {\bibinfo {title} {Dynamics of the
  charge-density wave. i. impurity pinning in a single chain},}\ }\href
  {\doibase 10.1103/PhysRevB.17.535} {\bibfield  {journal} {\bibinfo  {journal}
  {Phys. Rev. B}\ }\textbf {\bibinfo {volume} {17}},\ \bibinfo {pages}
  {535--541} (\bibinfo {year} {1978})}\BibitemShut {NoStop}%
\bibitem [{\citenamefont {Uri}\ \emph {et~al.}(2020)\citenamefont {Uri},
  \citenamefont {Grover}, \citenamefont {Cao}, \citenamefont {Crosse},
  \citenamefont {Bagani}, \citenamefont {Rodan-Legrain}, \citenamefont
  {Myasoedov}, \citenamefont {Watanabe}, \citenamefont {Taniguchi},
  \citenamefont {Moon},\ and\ \citenamefont {et~al.}}]{Uri2020}%
  \BibitemOpen
  \bibfield  {author} {\bibinfo {author} {\bibfnamefont {A.}~\bibnamefont
  {Uri}}, \bibinfo {author} {\bibfnamefont {S.}~\bibnamefont {Grover}},
  \bibinfo {author} {\bibfnamefont {Y.}~\bibnamefont {Cao}}, \bibinfo {author}
  {\bibfnamefont {J.}~\bibnamefont {Crosse}}, \bibinfo {author} {\bibfnamefont
  {K.}~\bibnamefont {Bagani}}, \bibinfo {author} {\bibfnamefont
  {D.}~\bibnamefont {Rodan-Legrain}}, \bibinfo {author} {\bibfnamefont
  {Y.}~\bibnamefont {Myasoedov}}, \bibinfo {author} {\bibfnamefont
  {K.}~\bibnamefont {Watanabe}}, \bibinfo {author} {\bibfnamefont
  {T.}~\bibnamefont {Taniguchi}}, \bibinfo {author} {\bibfnamefont
  {P.}~\bibnamefont {Moon}}, \ and\ \bibinfo {author} {\bibnamefont {et~al.}},\
  }\bibfield  {title} {\enquote {\bibinfo {title} {Mapping the twist-angle
  disorder and landau levels in magic-angle graphene},}\ }\href {\doibase
  10.1038/s41586-020-2255-3} {\bibfield  {journal} {\bibinfo  {journal}
  {Nature}\ }\textbf {\bibinfo {volume} {581}},\ \bibinfo {pages} {47--52}
  (\bibinfo {year} {2020})}\BibitemShut {NoStop}%
\bibitem [{\citenamefont {Padhi}\ \emph {et~al.}(2020)\citenamefont {Padhi},
  \citenamefont {Tiwari}, \citenamefont {Neupert},\ and\ \citenamefont
  {Ryu}}]{PRRtwistdis}%
  \BibitemOpen
  \bibfield  {author} {\bibinfo {author} {\bibfnamefont {B.}~\bibnamefont
  {Padhi}}, \bibinfo {author} {\bibfnamefont {A.}~\bibnamefont {Tiwari}},
  \bibinfo {author} {\bibfnamefont {T.}~\bibnamefont {Neupert}}, \ and\
  \bibinfo {author} {\bibfnamefont {S.}~\bibnamefont {Ryu}},\ }\href
  {https://arxiv.org/abs/2005.02406} {\enquote {\bibinfo {title} {Transport
  across twist angle domains in moir\'e graphene},}\ } (\bibinfo {year}
  {2020}),\ \Eprint {http://arxiv.org/abs/2005.02406} {arXiv:2005.02406
  [cond-mat.mes-hall]} \BibitemShut {NoStop}%
\bibitem [{\citenamefont {Nam}\ and\ \citenamefont
  {Koshino}(2017)}]{NamKoshino}%
  \BibitemOpen
  \bibfield  {author} {\bibinfo {author} {\bibfnamefont {N.~N.~T.}\
  \bibnamefont {Nam}}\ and\ \bibinfo {author} {\bibfnamefont {M.}~\bibnamefont
  {Koshino}},\ }\bibfield  {title} {\enquote {\bibinfo {title} {Lattice
  relaxation and energy band modulation in twisted bilayer graphene},}\ }\href
  {\doibase 10.1103/PhysRevB.96.075311} {\bibfield  {journal} {\bibinfo
  {journal} {Phys. Rev. B}\ }\textbf {\bibinfo {volume} {96}},\ \bibinfo
  {pages} {075311} (\bibinfo {year} {2017})}\BibitemShut {NoStop}%
\bibitem [{\citenamefont {Uchida}\ \emph {et~al.}(2014)\citenamefont {Uchida},
  \citenamefont {Furuya}, \citenamefont {Iwata},\ and\ \citenamefont
  {Oshiyama}}]{Corrugation14}%
  \BibitemOpen
  \bibfield  {author} {\bibinfo {author} {\bibfnamefont {K.}~\bibnamefont
  {Uchida}}, \bibinfo {author} {\bibfnamefont {S.}~\bibnamefont {Furuya}},
  \bibinfo {author} {\bibfnamefont {J.-I.}\ \bibnamefont {Iwata}}, \ and\
  \bibinfo {author} {\bibfnamefont {A.}~\bibnamefont {Oshiyama}},\ }\bibfield
  {title} {\enquote {\bibinfo {title} {Atomic corrugation and electron
  localization due to moir\'e patterns in twisted bilayer graphenes},}\ }\href
  {\doibase 10.1103/PhysRevB.90.155451} {\bibfield  {journal} {\bibinfo
  {journal} {Phys. Rev. B}\ }\textbf {\bibinfo {volume} {90}},\ \bibinfo
  {pages} {155451} (\bibinfo {year} {2014})}\BibitemShut {NoStop}%
\bibitem [{\citenamefont {Aubry}(1983)}]{AubryKAM}%
  \BibitemOpen
  \bibfield  {author} {\bibinfo {author} {\bibfnamefont {S.}~\bibnamefont
  {Aubry}},\ }\bibfield  {title} {\enquote {\bibinfo {title} {The twist map,
  the extended frenkel-kontorova model and the devil's staircase},}\ }\href
  {\doibase 10.1016/0167-2789(83)90129-X} {\bibfield  {journal} {\bibinfo
  {journal} {Physica D: Nonlinear Phenomena}\ }\textbf {\bibinfo {volume}
  {7}},\ \bibinfo {pages} {240 -- 258} (\bibinfo {year} {1983})}\BibitemShut
  {NoStop}%
\bibitem [{\citenamefont {Pokrovsky}\ and\ \citenamefont
  {Virosztek}(1983)}]{PokrovskyLR83}%
  \BibitemOpen
  \bibfield  {author} {\bibinfo {author} {\bibfnamefont {V.~L.}\ \bibnamefont
  {Pokrovsky}}\ and\ \bibinfo {author} {\bibfnamefont {A.}~\bibnamefont
  {Virosztek}},\ }\bibfield  {title} {\enquote {\bibinfo {title} {Long-range
  interactions in commensurate-incommensurate phase transition},}\ }\href
  {\doibase 10.1088/0022-3719/16/23/013} {\bibfield  {journal} {\bibinfo
  {journal} {Journal of Physics C: Solid State Physics}\ }\textbf {\bibinfo
  {volume} {16}},\ \bibinfo {pages} {4513--4525} (\bibinfo {year}
  {1983})}\BibitemShut {NoStop}%
\end{thebibliography}

%

\end{document}